\documentclass[]{aastex631}
 \pdfoutput=1 

\usepackage{graphicx}
\usepackage[caption=false]{subfig}
\DeclareCaptionLabelFormat{continued}{#1 ̃#2 (cont.)}
\usepackage[figure,figure*]{hypcap}

\accepted{to ApJ, May 23, 2022}

\def\aap{{A\&Ap}}
\def\apjl{{ApJL}}
\def\apjs{{ApJS}}

\newcommand{\aox}{\ifmmode{\alpha_{\mathrm{ox}}} \else $\alpha_{\mathrm{ox}}$\fi} 
\newcommand{\atoms}{\ifmmode{\mathrm{\,atoms~cm^{-2}}} \else \,atoms cm$^{-2}$\fi}
\newcommand{\ax}{\ifmmode{\alpha_x} \else $\alpha_x$\fi} 
\newcommand{\cmsq}{\ifmmode{\mathrm{cm^{-2}}} \else cm$^{-2}$\fi}
\newcommand{\degs}{\ifmmode ^{\circ}\else$^{\circ}$\fi}
\newcommand{\degsq}{\ifmmode {\mathrm{deg^2}} \else deg$^2$\fi}
\newcommand{\perdegsq}{\ifmmode {\mathrm{deg^{-2}}} \else deg$^{-2}$\fi}
\newcommand{\ew}{\ifmmode{W_{\lambda}} \else $W_{\lambda}$\fi}
\newcommand{\fbol}{\ifmmode f_{\mathrm{bol}} \else $f_{\mathrm{bol}}$\fi} 
\newcommand{\fcgs}{\ifmmode \mathrm{erg~cm^{-2}~s^{-1}}\else erg~cm$^{-2}$~s$^{-1}$\fi}
\newcommand{\flamcgs}{\ifmmode \mathrm{erg\,cm^{-2}\,s^{-1}\,\AA^{-1}}\else erg\,cm$^{-2}$\,s$^{-1}$\,\AA$^{-1}$\fi}
\newcommand{\fnucgs}{\ifmmode {\mathrm{erg~cm^{-2}~s^{-1}~Hz^{-1}}}\else erg~cm$^{-2}$~s$^{-1}$~Hz$^{-1}$\fi}
\newcommand{\gax }{{\lower0.8ex\hbox{$\buildrel >\over\sim$}}}
\newcommand\Ha{\ifmmode {\mathrm H}\alpha \else H$\alpha$\fi}
\newcommand\Hb{\ifmmode {\mathrm H}\beta \else H$\beta$\fi}
\newcommand\Hg{\ifmmode {\mathrm H}\gamma \else H$\gamma$\fi}
\newcommand\Nsigma{\ifmmode N_{\sigma}({\mathrm H}\beta) \else $N_{\sigma}({\rm H}\beta)$\fi}
\newcommand{\MgII}{\ifmmode {{\mathrm Mg\,II}}\else{Mg\,II}\fi}
\newcommand{\kms}{\ifmmode~{\mathrm{km~s}}^{-1}\else ~km~s$^{-1}~$\fi}
\newcommand{\lax }{{\lower0.8ex\hbox{$\buildrel <\over\sim$}}}
\newcommand{\lcgs}{\ifmmode \mathrm{erg~s^{-1}}\else erg~s$^{-1}$\fi}
\newcommand{\lnucgs}{\ifmmode erg~s^{-1}~Hz^{-1}\else erg~s$^{-1}$~Hz$^{-1}$\fi}

\newcommand{\logz}{\ifmmode{\mathrm{log}}~z \else log$~z$\fi}
\newcommand{\lo}{\ifmmode l_o \else $~l_o$\fi}
\newcommand{\Lo}{\ifmmode L_o \else $~L_o$\fi}
\newcommand{\lx}{\ifmmode l_x \else $~l_x$\fi}
\newcommand{\Lx}{\ifmmode L_x \else $~L_x$\fi}
\newcommand{\lbol}{\ifmmode L_{\mathrm{bol}} \else $L_{\mathrm{bol}}$\fi}
\newcommand{\Lbol}{\ifmmode L_{\mathrm{bol}} \else $L_{\mathrm{bol}}$\fi}
\newcommand{\LBol}{\ifmmode L_{\mathrm{bol}} \else $L_{\mathrm{bol}}$\fi}
\newcommand{\LEdd}{\ifmmode L_{\mathrm{Edd}} \else $L_{\mathrm{Edd}}$\fi}
\newcommand{\LxLbol}{\ifmmode L_x/L_{\mathrm{bol}} \else $L_x/L_{\mathrm{bol}}$\fi}
\newcommand{\rEdd}{\ifmmode L/L_{\mathrm{Edd}} \else $L/L_{\mathrm{Edd}}$\fi}
\newcommand{\REdd}{\ifmmode L/L_{\mathrm{Edd}} \else $L/L_{\mathrm{Edd}}$\fi}
\newcommand{\Rblr}{\ifmmode {R_{\mathrm BLR}} \else $R_{\mathrm BLR}$\fi}
\newcommand{\lamEdd}{\ifmmode \lambda_{\mathrm{Edd}} \else $\lambda_{\mathrm{Edd}}$\fi}
\newcommand{\mbh}{\ifmmode {M_{\rm BH}}\else${M_{\rm BH}}$\fi}
\newcommand{\Mbh}{\ifmmode {M_{\rm BH}}\else${M_{\rm BH}}$\fi}
\newcommand{\Msigma}{\ifmmode {M_{\rm BH}-\sigma_*}\else ${M_{\rm BH}-\sigma_*}$\fi}
\newcommand{\mdot}{\ifmmode \dot{m} \else $\dot{m}$\fi}
\newcommand{\mdote}{\ifmmode \dot{m}_{E} \else $\dot{m}_{E}$\fi}
\newcommand{\mone}{\ifmmode ^{-1}\else$^{-1}$\fi}
\newcommand{\msun}{\ifmmode {M_{\odot}}\else${M_{\odot}}$\fi}
\newcommand{\Msun}{\ifmmode {M_{\odot}}\else${M_{\odot}}$\fi}
\newcommand{\mtwo}{\ifmmode ^{-2}\else$^{-2}$\fi}
\newcommand{\Mvir}{\ifmmode {M_{\rm BH}^{\mathrm SE}}\else${M_{\rm BH}^{\mathrm SE}}$\fi}
\newcommand{\nhgal}{\ifmmode{ N_{H}^{Gal}} \else N$_{H}^{Gal}$\fi}
\newcommand{\nh}{\ifmmode{\mathrm N_{H}} \else N$_{H}$\fi}
\newcommand{\nhintr}{\ifmmode{ N_{H}^{intr}} \else N$_{H}^{intr}$\fi}
\newcommand{\nhtot}{\ifmmode{ N_{H}^{tot}} \else N$_{H}^{tot}$\fi}
\newcommand{\nhz}{\ifmmode{ N_{H}^z} \else N$_{H}^z$\fi}
\newcommand{\oi}{\ifmmode{\mathrm [O\,II]} \else [O\,II]\fi}
\newcommand{\oii}{\ifmmode{\mathrm [O\,II]} \else [O\,II]\fi}
\newcommand{\oiii}{\ifmmode{\mathrm [O\,III]} \else [O\,III]\fi}
\newcommand{\Loiii}{\ifmmode{\mathrm L_{O\,III}} \else $L_{\rm O\,III}$\fi}
\newcommand{\optebl}{\ifmmode L_{\rm 2500\,\AA} \else $~L_{\rm 2500\,\AA}$\fi}
\newcommand{\opteml}{\ifmmode l_{\mathrm{2500\,\AA}} \else $~l_{\mathrm{2500\,\AA}}$\fi}
\newcommand{\Teff}{\ifmmode T_{\mathrm{Eff}} \else $T_{\mathrm{Eff}}$\fi}
\newcommand{\xebl}{\ifmmode L_X \else $~L_X$\fi}
\newcommand{\xeml}{\ifmmode l_{\mathrm{2\,keV}} \else $~l_{\mathrm{2\,keV}}$\fi}

\def\geqsim{\lower.73ex\hbox{$\sim$}\llap{\raise.4ex\hbox{$>$}}$\,$}
\def\leqsim{\lower.73ex\hbox{$\sim$}\llap{\raise.4ex\hbox{$<$}}$\,$}

\newcommand{\umg}{\ifmmode{\mathrm{(}u-g\mathrm{)}} \else ($u-g$)\fi}
\newcommand{\gmr}{\ifmmode{\mathrm{(}g-r\mathrm{)}} \else ($g-r$)\fi}
\newcommand{\rmi}{\ifmmode{\mathrm{(}r-i\mathrm{)}} \else ($r-i$)\fi}
\newcommand{\gmi}{\ifmmode{\mathrm{(}g-i\mathrm{)}} \else ($g-i$)\fi}
\newcommand{\imz}{\ifmmode{\mathrm{(}i-z\mathrm{)}} \else ($i-z$)\fi}
\newcommand{\jmh}{\ifmmode{\mathrm{(}J-H\mathrm{)}} \else ($J-H$)\fi}
\newcommand{\hmk}{\ifmmode{\mathrm{(}H-K\mathrm{)}} \else ($H-K$)\fi}
\newcommand{\ctwo}{\ifmmode C_2 \else C$_2$\fi}

\received{\today}
\shorttitle{Changing Look Quasars in the SDSS-IV Time Domain Spectroscopic Survey}
\shortauthors{Green et al.}

\shorttitle{Changing Look Quasars from the TDSS}
\shortauthors{Green et al.}

\begin{document}

\title{The Time Domain Spectroscopic Survey: Changing-Look Quasar Candidates from Multi-Epoch Spectroscopy in SDSS-IV}

\correspondingauthor{Paul J. Green}
\email{pgreen@cfa.harvard.edu}

\author[0000-0002-8179-9445]{Paul J. Green}
\affiliation{Center for Astrophysics $\vert$ Harvard \& Smithsonian, 60 Garden Street, Cambridge, MA 02138, USA}

\author{Lina Pulgarin-Duque}
\affiliation{Center for Astrophysics $\vert$ Harvard \& Smithsonian, 60 Garden Street, Cambridge, MA 02138, USA}

\author{Scott F. Anderson}
\affiliation{Department of Astronomy, University of Washington, Box 351580, Seattle, WA 98195, USA}

\author[0000-0003-3422-2202]{Chelsea L. MacLeod}
\affiliation{BlackSky, 1505 Westlake Ave N \#600, Seattle, WA 98109, USA}

\author[0000-0002-3719-940X]{Michael Eracleous}
\affiliation{Department of Astronomy \& Astrophysics and Institute for Gravitation and the Cosmos, 525 Davey Laboratory, The Pennsylvania State University, University Park, PA 16802, USA}

\author[0000-0001-8665-5523]{John J. Ruan}
\affiliation{Department of Physics and Astronomy, Bishop's University, 2600 College St., Sherbrooke, QC J1M 1Z7, Canada}
 
\author[0000-0001-8557-2822]{Jessie Runnoe}
  \affiliation{Department of Physics and Astronomy, Vanderbilt University, Nashville, TN 37235, USA}
  
\author[0000-0002-3168-0139]{Matthew Graham}
  \affiliation{Cahill Center for Astronomy and Astrophysics, California Institute of Technology, 1216 E. California Boulevard, Pasadena, CA 91125, USA}

\author[0000-0002-9453-7735]{Benjamin R. Roulston}
  \affiliation{Center for Astrophysics $\vert$ Harvard \& Smithsonian, 60 Garden Street, Cambridge, MA 02138, USA}
  \affiliation{Department of Astronomy, Boston University, 725 Commonwealth Avenue, Boston, MA 02215, USA}

\author[0000-0001-7240-7449]{Donald P. Schneider}
  \affiliation{Department of Astronomy \& Astrophysics, 525 Davey Laboratory, The Pennsylvania State University, University Park, PA
  16802, USA} 
  \affiliation{Institute for Gravitation and the Cosmos, The Pennsylvania State University, University Park, PA 16802, USA}   


\author{Austin Ahlf} 	     	   
   \affiliation{Department of Astronomy, University of Washington, Box 351580, Seattle, WA 98195, USA}

\author[0000-0002-3601-133X]{Dmitry Bizyaev}
   \affiliation{Apache Point Observatory and New Mexico State University, P.O. Box 59, Sunset, NM, 88349, USA}
   \affiliation{Sternberg Astronomical Institute, Moscow State University, Moscow, Russia}

\author[0000-0002-8725-1069]{Joel R. Brownstein}
  \affiliation{University of Utah, Department of Physics and Astronomy, 115 S. 1400 E., Salt Lake City, UT 84112, USA}

\author{Sonia Joesephine del Casal} 
  \affiliation{Department of Astronomy, University of Washington, Box 351580, Seattle, WA 98195, USA}

\author[0000-0002-3696-8035]{Sierra A. Dodd}     	   
  \affiliation{Department of Astronomy, University of Washington, Box 351580, Seattle, WA 98195, USA}
  \affiliation{Department of Astronomy and Astrophysics, University of California, Santa Cruz, 1156 High Street, Santa Cruz, CA 95064 USA}

\author{Daniel Hoover}		   
  \affiliation{Department of Astronomy, University of Washington, Box 351580, Seattle, WA 98195, USA}

\author{Cayenne Matt}
  \affiliation{Department of Astronomy, University of Washington, Box 351580, Seattle, WA 98195, USA}
  \affiliation{Department of Astronomy, University of Michigan, 1085 S. University Ann Arbor, MI 48109, USA} 
 
\author{Andrea Merloni}
  \affiliation{Max-Planck-Institut für extraterrestrische Physik (MPE), Giessenbachstr. 85748 Garching, Germany}
  
\author[0000-0002-2835-2556]{Kaike Pan}
   \affiliation{Apache Point Observatory and New Mexico State University, P.O. Box 59, Sunset, NM, 88349, USA}
   
\author{Arnulfo Ramirez}    	 
  \affiliation{Department of Astronomy, University of Washington, Box 351580, Seattle, WA 98195, USA}

\author{Margaret Ridder}
  \affiliation{Department of Astronomy, University of Washington, Box 351580, Seattle, WA 98195, USA}
  \affiliation{Department of Physics, University of Alberta, Edmonton, AB T6G 2E1, Canada}




\begin{abstract}
Active galactic nuclei (AGN) can vary significantly in their
rest-frame optical/UV continuum emission, and with strong associated changes in
broad line emission, on much shorter timescales than predicted by
standard models of accretion disks around supermassive black holes.
Most such ``changing-look'' or "changing-state" AGN -- and at higher luminosities, 
changing-look quasars (CLQs) -- have been found via spectroscopic follow-up of known quasars
showing strong photometric variability.  The Time Domain Spectroscopic
Survey of SDSS-IV includes repeat spectroscopy of large numbers
of previously-known quasars, many selected irrespective of
  photometric variability, and with spectral epochs separated by
months to decades.  Our visual examination of these repeat spectra for
strong broad line variability yielded 61 newly-discovered CLQ
candidates.  We quantitatively compare spectral epochs to measure
changes in continuum and H$\beta$ broad line emission, finding 19 CLQs, of which 15 are
newly-recognized. The parent sample includes only broad line quasars,
so our study tends to find objects that have dimmed, i.e., turn-off
CLQs. However, we nevertheless find 4 turn-on CLQs that meet our
criteria,  albeit with broad lines in both dim and bright
states.  We study the response of H$\beta$ and Mg\,II emission
lines to continuum changes.  The Eddington ratios of CLQs are low,
and/or their H$\beta$ broad line width is large relative to the
overall quasar population. Repeat quasar spectroscopy in the upcoming
SDSS-V Black Hole Mapper program will reveal significant
numbers of CLQs, enhancing our understanding of the frequency
and duty-cycle of such strong variability, and the physics and
dynamics of the phenomenon.
\end{abstract}



\section{Introduction}
\label{s:intro}

Supermassive black holes in the cores of galaxies have garnered a special fascination since the first quasars were discovered as strong, variable radio sources (e.g., \citealt{sandage64,matthews63} with bright optical quasi-stellar (point source) counterparts \citep{oke63}, with optical spectra showing a cosmological redshift and emission lines broadened by 10,000\kms\, or more \citet{schmidt63}.  Since then, massive efforts to study supermassive black holes (SMBHs) have revealed that they span from ten thousand to ten billion solar masses, being found in the core of our own Milky Way, in nearby dwarf galaxies (e.g., \citealt{baldassare15,afanasiev18}), in the dominant galaxies of galaxy clusters (e.g., \citealt{mcConnell11}), and out to redshifts above seven, when the Universe was just 5\% of its current age (e.g., \citealt{Fan06, Banados18,Wang21}). SMBHs may be luminous and easily detected in active galactic nuclei (AGN) if they are accreting significantly, or they may be dim and quiescent, as in our own galaxy.  SMBHs grow primarily by gas accretion
or by BH-BH coalescence (e.g. \citealt{kauffmann00}), which should become detectable via gravitational waves in the near future with the pulsar timing arrays (e.g., \citealt{Ransom19,Arzoumanian20} and NASA's Laser Interferometer Space Antenna (LISA) mission \citep{Danzmann18}.

SMBH masses (\Mbh) in AGN observationally correlate tightly with a variety of their host galaxy properties, such as the bulge mass and luminosity (e.g., \citealt{kormendy95,mclure02,mcconnell13}), and the stellar velocity dispersion $\sigma$, \citep{Ferrarese00,gebhardt00,gultekin09,mcconnell13}. Copious research has assumed or analyzed the possibility of feedback, where the observed $\Msigma$ relation (i.e., the relationship between central BH mass and is the stellar velocity dispersion $\sigma_*$ in the bulge of a galaxy; \citealt{Ferrarese00,gebhardt00}) is mediated by powerful outflows from the accreting SMBHs that regulate galaxy growth  (e.g., \citealt{silk98,fabian12}).  However, the reality may be more complex e.g., the \Msigma\, relationship may simply be the result of stochastic averaging over galaxy merger histories \citep{Jahnke11}, and AGN outflows may enhance rather than quench star formation \citep{Zubovas17}.  There are complex, high-scatter relationships between AGN activity, outflows, feedback, and galaxy properties (e.g., \citealt{Fiore17}).  Powerful and increasingly complex cosmological  simulations such as IllustrisTNG \citep{pillepich18} and FIRE \citep{hopkins18} seek to tie together SMBH growth with cosmological structure formation. 

All this makes clear that the accretion rate of SMBHs i.e., quasar activity, is important to understand and characterize on all accessible timescales.  Indeed, variability of 10 -- 20\% observed on timescales of days to years is a fundamental characteristic of quasars that has often been used to identify samples with high efficiency \citep[e.g.,][]{Palanque_Delabrouille13}.  Larger amplitudes (a magnitude or more) of variability are also known, usually found on longer timescales and within large samples, but also sometimes serendipitously (e.g., \citealt{Luo2020,kynoch19}, and references therein).  On cosmic timescales, variability may be crudely characterized by the total average time, or ``duty cycle'' spent in active accretion \citep{Martini2003}. The present-day SMBH mass function basically depends on quasar lifetime and duty cycle.  We know that SMBH accretion activity varies significantly on timescales of millennia or longer.  Kiloparsec-scale jets and cavities driven by SMBH activity are seen in massive radio galaxies, and piercing the hot plasma of galaxy clusters (e.g., \citealt{birzan04}).  ``Double-double'' radio galaxies show jets that have stopped and restarted on such timescales \citep{mahatma19}.  Extended emission line regions lacking a present-day ionizing source (``Voorwerps"; \citealt{lintott09, keel12}) are also being discovered that testify directly to long periods of AGN quiescence (up to $\sim 10^5$yrs so far; \citealt{schawinski15}).  

We know that quasars are variable, and that emission lines respond to changes in the ionizing continuum, albeit with a less than linear relationship, known as the Baldwin effect (Beff), which is seen both in samples of quasars with single epoch spectra (the ensemble BEff or eBeff e.g., \citealt{baldwin78,green01}) and between spectral epochs of individual quasars (the intrinsic Beff or iBeff; \citealt{kinney90,goad04}).  The temporal lag in broad emission line (BEL) response to continuum changes continues to be used to measure SMBH masses via reverberation mapping (RM; \citealt{peterson93,shen16,grier17}).  On occasion, both the accretion luminosity (as represented by a power-law continuum component in the optical spectrum) and broad emission lines are seen to vary quite strongly, and even more rarely, both may dim below detectability. Extreme variability observed in - particularly dis/appearance of - the BELs  is called the ``changing look" phenomenon in AGN (CLAGN) or quasars (CLQs).\footnote{All quasars are AGN, but in common astronomical parlance, quasars have larger luminosities, usually above about $10^{45}$\,erg/s.  Such high luminosity AGN are rare and so normally found in the larger volumes of space encompassed at higher redshifts.  Given their large distances and high nuclear luminosity, the host galaxy is often difficult to detect; hence the most luminous AGN appear quasi-stellar, and are called quasars or quasi-stellar objects (QSOs).}
As the narrow emission lines (NELs), which originate from much larger regions (e.g.,, \citealt{Bennert02}), tend to remain steady relative to the BELs, AGN may change their classification, based on which emission lines continue to show broad components (e.g., \citealt{Osterbrock1981}) from Type~1 to Type~2, or by smaller increments e.g., Type~1.8 to 1.5.



CLQs have been intensively studied of late, because of both their mystery and their utility. Depending on the precise definition of a CLQ (e.g., continuum and broad \Hb\, luminosity and their change), nearly 70 have been found just in the last few years, mostly identified from multi-epoch optical photometry and follow-up spectroscopy of previously known (spectroscopically-identified) quasars \citep{LaMassa15,MacLeod16,Runnoe16,Ruan16,Gezari17,Yang18,MacLeod19,Frederick19,Ross20,Sheng20}.  The strong interest arises because for standard thin accretion disc models \citep{Shakura73}, large luminosity changes in the optical-emitting regions of an SMBH accretion disc due to overall accretion rate changes are expected to occur on the viscous timescales, corresponding to thousands of years or more \citep{kro99}, whereas CLQs have been seen to change significantly on timescales of years or even as little as about a month (though the latter usually at lower Seyfert-like luminosities e.g., \citealt{Trakhtenbrot19}). Therein lies the mystery.
If instead, the optical continuum-emitting regions are reprocessing radiation from the inner UV and X-ray-emitting regions, observed timescales for CLQ variability may be compatible with predictions, as discussed in \citet{LaMassa15} and references therein.   
Accretion discs supported by magnetic pressure may also allow powerful and rapid variability \citep{Dexter2019, Scepi2021}.  However, there may be theoretical caveats to the timescale problem, as described elsewhere (e.g., in \citealt{law12,law16}).  

The utility of CLQs arises because we may witness quasars essentially turning on and/or off.  This clearly demands revision of our understanding of the duty cycles and timescales of quasar activity, but also allows us to see clearly (without the contamination of a bright point source nucleus) what a quasar host galaxy looks like. For instance, though the $\Msigma$ relationship demands it, measuring the stellar velocity dispersion in a quasar host illuminated by a luminous active SMBH can be difficult, but becomes simple when the accretion activity ceases \citep{Dodd21, Charlton19}. A CLQ in a dim state also enables direct measurement of the host luminosity for accurate subtraction from the overall spectrum in brighter states (e.g., \citealt{bentz06}).

Tidal disruption events (TDEs) have been suggested as an explanation for intrinsic extreme variability.  Debris from a star whose orbit brings it too close to the SMBH is pulled into a stream, of which half forms (or augments) an accretion disk.  The resulting thermal radiation with $T_{eff}\sim (\LEdd/(4\pi\,R_t^2\sigma))^{1/4}$ (where $R_t$ is the tidal radius of the black hole) is expected to peak in the EUV or soft X-rays (0.25 - 2.5 $\times 10^5$K) but also irradiate circumnuclear gas resulting in enhanced optical emission.  The fall-back of debris is expected to rise for about a month, and then decline as $t^{-5/3}$ \citep{Rees1988,Ulmer1999}. The CLQs found to date have light curves that show a wide variety of shapes. On rare occasions, what appears to be a CLQ may be a TDE (e.g., see \citealt{mer15} analysis of the light curve of \citealt{LaMassa15}). However, most CLQs last too long ($\sim$years or decades) in the bright state for the emission to be primarily from a TDE (e.g., \citealt{Runnoe16,MacLeod19}) while others have a very different decay trend (e.g., \citealt{Ruan16,Gezari17}). Still, rare, potentially longer-lasting light curve patterns may emerge from unusual TDE systems involving e.g., giant stars, or binaries.

Extreme variability may also occur due to strong changes in intrinsic absorption that could block the continuum source and shade the
broad emission line region, but cloud crossing timescales ($t_{cross}$=$\,24\,M_8^{-1/2}\,L_{44}^{3/4}$\,years) are usually considered much too long.  While modelling of dust extinction of the optical/UV quasar power-law continuum  can sometimes reproduce the observed changes in the optical continuum emission, the observed BEL changes indicate a near-disappearance of the ionizing flux  \citep{LaMassa15,MacLeod16}. Obscuration of the continuum should also result in high linear polarization, which has not been found in dim state CLQs \citep{Hutsemekers19}.  \citet{Sheng17} find that mid-infrared (MIR) variability of 10 CLQs appears to follow variability in the optical band with a time lag consistent with dust reprocessing. Their results  argue that obscuration is not the cause of dimming in CLQs.  \citet{Yang18}
found the mid-IR WISE colors of CLQs are redder when brighter, indicating a strong hot dust contribution rather than reddening.  In several cases, the IR tracks the optical variability (e.g.,\,\citealt{Nagoshi21}), again suggesting that obscuration is not the cause of dimming in CLQs.

TDEs and variable obscuration may contribute to samples of CLQ candidates, but larger samples are key to understand the role and frequency of these phenomena relative to the predominant effects of accretion rate changes.  The characterization of strong changes in accretion rate leading to changes in the BLR also demand larger samples, to study the dependence of of the CLQ phenomenon on such fundamental parameters as luminosity, SMBH mass, Eddington ratio, and redshift.

For these reasons, the hunt for CLQs has intensified in the last few years, especially as large-area, multi-epoch surveys - both photometric and spectroscopic - are maturing.  In this paper, we describe a search for CLQs among objects with multi-epoch spectroscopy from a dedicated program targeting point-source variables for spectroscopy, the Time Domain Spectroscopy Survey (TDSS) of the Sloan Digital Sky Survey (SDSS).  


Throughout, we adopt a flat $\Lambda$ cold dark matter cosmology with $H_0 = 70\,$km s$^{-1}$\,Mpc$^{-1}$ and $\Omega_m = 0.30$.

\section{CLQ Candidate Selection}
\label{s:select}


Our parent sample consists of a subset of objects classified as quasars by the SDSS pipeline that appear in the SDSS Data Release 14 (DR14) quasar catalog \citep{paris18}, which included 526,356 quasars.  All the reduced and calibrated SDSS spectra analyzed herein were publicly released as part of SDSS DR14, which includes data through 2016 May 11.
The Sloan Digital Sky Survey (SDSS-IV) final quasar catalog 
(Data Release 16; \citealt{lyke20}) describes an additional 225,082 new quasars, and additional epochs obtained for previously-known SDSS quasars.  

The Time Domain Spectroscopy Survey (TDSS) is a subprogram of the 
{\emph extended Baryon Oscillation Spectroscopic Survey} (eBOSS; \citealt{dawson16}
within SDSS-IV \citep{blanton17}.  A pilot survey for the TDSS dubbed SEQUELS
actually started during SDSS-III \citep{Eisenstein11}, and is described in \citet{Ruan15}. 
The TDSS is the first major program intended to spectroscopically characterize generic variables, i.e.,
 without any explicit selection criteria based on color or light curve characteristics.  The TDSS targeted photometric objects having morphology consistent with a point source, and that were imaged both by SDSS and  Pan-STARRS1 (PS1; \citealt{kaiser10}), and showed significant variability within and/or between those two imaging surveys, as described by \citet{Morganson15}.  The bulk of the TDSS sought single-epoch spectroscopy (SES) of such point-source, photometric variables. Results up through SDSS Data Release 14 are described in \citet{Anderson22}.  However, the TDSS also included a smaller program for Few-Epoch Spectroscopy (FES), to examine the {\emph spectroscopic} variability of certain select samples of objects with existing SDSS spectroscopy, totalling about 20,000 spectra, as described in \citet{MacLeod18}.  These included hypervariable quasars, broad absorption line (BAL) quasars, quasars with possible double-peaked broad-line profiles, white dwarf/main sequence binaries, and dwarf carbon stars \citep{roulston19}.  The TDSS is also described on the web.\footnote{\url{ https://www.sdss.org/dr16/spectro/extragalactic-observing-programs/tdss/}}.

Another TDSS subprogram targeting objects with existing spectroscopy is the Repeat Quasar Spectroscopy program, described in detail in \citet{MacLeod18}. The RQS program explicitly targeted known quasars for repeat
spectroscopy so that quasar spectroscopic variability could be studied across a wide range of timescales.  The majority (about 70\%) of RQS targets are simply a bright magnitude-limited sample (RQS1; $i<19.1$), but quasars with $i<20.5$ and more than one spectrum in the archive were also prioritized. The RQS also includes fainter ($i<21$) subsets with and without detected photometric variability above a tuned threshold\footnote{The variability selection threshold is a chosen value of the reduced $\chi^2$ (assuming a constant mean magnitude) for the PS1 and SDSS $g$ and/or $r$ band light curves that is tuned in each region to achieve a target density of 10 - 15 TDSS spectroscopic targets per square degree.}:  RQS2, RQS2v, RQS3, and RQS3v.  Unlike the TDSS SES or FES subprograms, the RQS sample also includes quasars with extended morphologies in SDSS imaging. 


All TDSS DR14 spectra of quasars were reviewed in visual inspections for both quality assurance and potential science discoveries.  We asked student co-authors at the University of Washington to study example spectra of normal quasars, and to expect such spectra for the great majority of new TDSS quasars.  However, we asked the visual inspectors to be especially alert for new-epochs of TDSS spectra that did not look like those of normal quasars, including e.g., CLQs, quasars with odd emission line profiles, broad absorption lines, etc. Wherever an incoming  new-epoch TDSS quasar spectrum was deemed potentially  unusual by the inspector, that new-epoch spectrum was compared side by side with earlier epoch spectra contained in SDSS DR14.  We note that this process strongly favors finding turn-off CLQs (in their fading or dim state).  Furthermore, the method is both qualitative and incomplete, so that any statistical utility of the number of CLQs found here is at best limited to being a lower limit. However,
human visual inspection can detect small changes that might be missed by a computer algorithm, and also discount a variety of apparent changes that are clearly caused by artifacts.  There are also cases in between, where an apparently significant change may potentially be due to a problem e.g., with spectral fiber plug  placement in the plate, or glitches in the reduction.  In the majority of such cases, we have verified whether the observed spectral change is likely real by checking against the behavior of the quasar's photometric light curve at the spectral epoch(s) in question (e.g., see  Figure\,\ref{f:lcspecs}). However, several cases (see Table\,\ref{t:clqcans}), called for follow-up spectroscopy with different telescopes and instruments to confirm the reality of the apparent changes first seen between the SDSS spectral epochs. 

We restrict the TDSS CLQ candidate sample to redshift $z<0.9$, for which the \Hb/\oiii\, emission line complex remains below about 9500\AA\ in the observed frame, close to the red limit of early epoch SDSS spectra.  

For clarity and for statistical purposes, we define a bona fide CLQ to have a 3$\sigma$ change in the flux of the broad component of \Hb, as decribed in detail in \S\,\ref{s:clqdef}.  As can be seen in Table\,\ref{t:TDSStypes}, the largest share (23) of our CLQ candidates, of which 4 are bona fide CLQs, comes from the TDSS FES\_HYPQSO sample.\footnote{While TDSS FES\_HYPQSOs include both BAL QSOs and BL\,Lacs, we exclude those from our TDSS CLQ candidate sample.}  This makes perfect sense, since they were prioritized for TDSS spectroscopy from the 2\% most variable QSOs, outside the elliptical contours of a variability space defined both by variability within PS1 epochs, and between SDSS and PS1 epochs \citep{Morganson15,MacLeod18}.  
However, while large continuum changes enhanced the likelihood of their visual identification as CLQ candidates, we note (with a strong caveat for small-number statistics) that the HYPQSOs do not represent the largest fraction of confirmed CLQs.

The next largest contributing TDSS selection is the TDSS RQS1 program, yielding 18 CLQ candidates and 5 confirmed CLQs. The RQS1 designation arises from a simple magnitude cut, with no variability criterion, so the similar final number of CLQs to the HYPQSO-selected sample is probably just a result of the larger RQS1 parent sample size.  Indeed, given the small number of CLQs in Table\,\ref{t:TDSStypes}, we can only say with confidence that a random spectroscopic re-sampling of quasars on these timescales (or longer) will yield a useful number of new CLQs.

 As highlighted by \citet{Ruan16} and \citet{Anderson22}, variability selection results in a broader quasar color distribution than traditional color selection, such as used to select most SDSS quasars, where the primary goal is to distinguish quasars from stars in photometric color space (e.g., \citealt{Bovy12}), and the result is a bias against red quasars.  The larger fraction of red quasars uncovered in variability-selected samples can be attributed to more high redshift objects, and/or redder continuum emission due e.g., to their intrinsic continuum slope and/or internal dust reddening.  

This work primarily focuses on (1) repeat spectroscopy, for which the \emph{initial} target selection was most often via colors \citep{Richards02} and (2) redshifts where \Hb\, is visible, so our parent CLQ candidate sample has fairly typical initial (bright state) color distribution.  Dimmer states have often lost much of their bright UV/blue continuum and are rather more dominated by the host galaxy, so that the dim state population is indeed relatively red.  In some quasars with redshift $z<0.35$, strong broad \Ha\ may persist even in the dim state, adding further flux to the red filter bands.

\begin{deluxetable}{lccccc}
\tablecaption{CLQ Candidate Selection}
\tablewidth{1.0\textwidth}
\tablehead{\colhead{TDSS TargetType} & \colhead{N$_{\rm QSOs}$} &\colhead{N$_{Cand}$ } & \colhead{\%\,Cand} & \colhead{N$_{\rm CLQ}$} & \colhead{\%\,CLQ }  }
\startdata
A           &  17511 &  0  &  0.0  & 0  & 0    \\
B           &  44832 &  1  &  0.0  & 0  & 0    \\ 
FES\_DE     &  653   &  9  &  1.4  & 6  & 0.9  \\
FES\_HYPQSO &  1237  & 23  &  2.0  & 4  & 0.3  \\
FES\_MGII   &  64    &  1  &  1.6  & 1  & 1.5  \\
FES\_NQHISN &  744   &  2  &  0.3  & 2  & 0.3  \\
RQS1        &  10520 & 18  &  0.2  & 5  & 0.0  \\
RQS2        &  2243  &  1  &  0.0  & 1  & 0.0  \\
RQS2v       &  1111  &  2  &  0.1  & 0  & 0.0  \\
RQS3        &  1056  &  0  &  0    & 0  & 0    \\
RQS3v       &  1579  &  4  &  0.3  & 0  & 0    \\
Total       &  64039 &  61 &  0.1  & 19 & 0.03 \\
\enddata
\tablecomments{Target selection methods for our TDSS CLQs.  These abbreviations correspond to several different target  selection algorithms used for the TDSS program of SDSS-IV, and are described briefly in \S\,\ref{s:select}.  Some QSOs were selected by TDSS using more than one method. Small number statistics dominate the percentages listed, which are derived simply by dividing by $N_{QSOs}$.  See discussion in \S\,\ref{s:select}.} 
\end{deluxetable}
\label{t:TDSStypes}

\section{Spectroscopy and Analysis}\label{s:spec}

\subsection{Spectroscopic Data}
Table\,\ref{t:clqcans} lists all 61 CLQ candidates in our visually-selected DR14 quasar sample, their redshifts, spectral epochs and associated telescopes, along with best-fit modeled values for their continuum luminosities at 2800, 3240, and 5100\AA.  Emission line luminosities and their uncertainties are provided for \Hb\ and \MgII , where available. We also mark our designation of a single dim and bright state spectrum for each object, and the difference between their broad \Hb\ emission expressed in units of flux uncertainty $\sigma$, as described in more detail in \S\,\ref{s:clqdef}. 

For most epochs, we use spectra from the SDSS project, which are generally well flux-calibrated and corrected for telluric absorption.  For analysis of follow-up spectra from other telescopes described below, we correct for telluric absorption where needed by using a standard star observation at similar airmass and the empirical
method described in \citet{wade88} and \citet{oster90}.  In several cases, no such star was observed, nor any telluric correction performed, as is visible in some of the follow-up spectra. 

\subsubsection{SDSS/BOSS/eBOSS}

The SDSS spectra we use herein were obtained with the 2.5m Sloan telescope at Apache Point \citep{Gunn06}.  The oldest archival SDSS spectra we analyze here are from the original SDSS spectrograph, which hosted 640 3\arcsec\, fibers, plugged by hand for each designated field into aluminum plates that were then mounted at the focal plane, and which subtended 3\degsq\, on the sky. Resulting spectra have a resolution of $R\sim2000$, and wavelength coverage of about 3900 - 9100\AA.

The BOSS spectrographs were rebuilt from the original SDSS spectrographs \citep{Smee13} as part of the SDSSI/II, SDSS-III/BOSS, and SDSS-IV/eBOSS observation campaigns.  They host 1000 2-arcsec fibers in their plug plates, each covering 7\degsq\, of sky.  Resulting spectra have a resolution of $R\sim2000$, and wavelength coverage of about 3600 - 10400\AA.  The reduction pipelines for SDSS I/II and BOSS spectroscopy respectively, are described in \citet{Stoughton02} and \citep{Bolton12}. 


\subsubsection{MMT}\label{s:mmt}

For quasars fainter than mag $\sim19.5$ we obtained dozens of spectra using the 6.5M MMT on Mount Hopkins, Arizona.  The half dozen such spectra used in the current TDSS sample were obtained using the Blue Channel Spectrograph. Here, the 300 $\ell\, {\rm mm}^{-1}$ grating was used with a clear filter and $2\times$ binning in the spatial direction, yielding a resolution of $R\sim1300$. Typically, 
we used a central wavelength near 6560\AA, with resulting wavelength coverage
from 3900 - 9240\AA. To reduce MMT data, we used the {\tt pydis} software adapted for use with 
Blue Channel data.\footnote{{\tt http://jradavenport.github.io/2015/04/01/spectra.html}}

Depending on spectrograph availability and our allotted time, we also used 
BinoSpec \citep{fabricant19}, with the 270$\ell\, {\rm mm}^{-1}$  grating blazed at 6650\AA.
We used the median of three exposures to mitigate the effects of cosmic rays, and BinoSpec data were reduced via the pipeline described in \citet{chilingarian19}.  The resulting wavelength coverage is about 3820 - 9200\AA, at a resolution of $\sim1400$.

\subsubsection{Magellan}\label{s:magellan}
For some faint objects near the celestial equator, we 
used the Magellan Clay 6.5m telescope with the Low Dispersion Survey
Spectrograph 3 (LDSS3)-C spectrograph.  We used the VPH-All grism (covering
4250--10000\AA) with the standard $1\farcs0 \times 4\farcm0$ center
long slit mask.  Reductions were carried out both using standard IRAF techniques, and
the {\tt pydis} software adapted for use with LDSS3 data.

\subsubsection{William Herschel Telescope (WHT)}
\label{s:wht}

Some follow-up spectra in our sample were obtained
with the 4.2m William Herschel Telescope (WHT) in La Palma. 
As described in \citet{MacLeod19}, observations were performed using the Intermediate 
dispersion Spectrograph and Imaging System (ISIS). The 5300 dichroic was used
along with the R158B and R300B gratings in the red and blue arms,
respectively, along with the GG495 order sorting filter in the red
arm. With a slit width of 1\farcs{0}, we achieved spectral resolution of $R\sim$1500 at 5200\,\AA\, in the blue and $R\sim$1000 at 7200\,\AA\, in the red, and nominal total
coverage of $\sim 3100\textup{--}10600$\,\AA . 
WHT data were
reduced using custom \textsc{pyraf} scripts and standard techniques.


\begin{longrotatetable}
\begin{deluxetable}{lcccccccccccccccc}
\label{t:clqcans}
\tabletypesize{\scriptsize}
\centerwidetable
\tablecaption{TDSS CLQ candidates}
\tablehead{ \colhead{SDSSJID } & \colhead{ $z$} & \colhead{Spec} &  \colhead{Facility} & \colhead{$N_{\sigma}$} & \colhead{State} & \colhead{$L_{3240}$} & \colhead{$\sigma_{L3240}$} & \colhead{$L_{5100}$} & \colhead{$\sigma_{L5100}$} & \colhead{$L_{\Hb}$} & \colhead{$\sigma_{L\Hb}$ } & \colhead{$L_{2800}$} & \colhead{$\sigma_{L2800\AA}$} & \colhead{$L$(\MgII)} & \colhead{$\sigma_{L\MgII}$ }  & \colhead{Notes}  \\
\cline{7-16}
\colhead{Selection} & \colhead{ } &  \colhead{  MJD } & \colhead{ } & \colhead{broad $\Hb$} & \colhead{(1) }  &  \multicolumn{8}{c}{ ($10^{42}$\,erg$\,s^{-1}$)} & \colhead{ }}              
\startdata
002311.06+003517.53 & 0.422 & 51816 & SDSS   & 1.26    &    & 286.4  & 9.2   &  339.1  &  9.2   & 3.60  & 0.12 & 448.02  & 11.12  & 15.66  & 0.49   & CLQ,M16c,H19 \\
MgII                &       & 55480 & BOSS   & 4.30    & B  & 688.3  & 5.9   &  512.0  &  5.9   & 7.07  & 0.13 & 800.40  & 12.44  & 20.57  & 0.44   &  \\
                    &       & 56979 & eBOSS  & \ldots  & D  & 109.1  & 2.1   &  120.1  &  3.5   & 1.91  & 0.06 & 142.77  & 2.67   & 9.10   & 0.10   & \ldots \\
                    &       & 57597 & Mag/L3 & 1.69    &    &  86.3  & 3.7   &   91.9  &  5.7   & 2.84  & 0.16 & 83.50   & 3.59   & 11.81  & 0.96   & \ldots \\
                    &       & 58037 & MMT/BC & 1.36    &    & 127.1  & 4.2   &  105.8  &  6.0   & 2.24  & 0.09 & 109.71  & 3.47   & 7.50   & 0.22   & \ldots \\
                    &       & 58069 & eBOSS  & 0.97    &    & 155.1  & 3.6   &  188.9  &  3.9   & 2.06  & 0.08 & 144.87  & 1.32   & 4.67   & 0.09   & \ldots \\
002548.50-003351.38 & 0.569 & 52262 & SDSS   & 1.06    & B  & 144.2  & 9.1   &  203.9  & 14.3   & 3.52  & 0.29 & 230.26  & 13.29  & 4.55   & 1.03   & \ldots \\
RQS3v               &       & 58048 & BOSS   & \ldots  & D  &  42.7  & 6.1   &   57.3  & 12.4   & 0.82  & 0.27 & 58.63   & 2.36   & 1.27   & 0.19   & \ldots \\
003841.53+000226.73 & 0.650 & 52261 & SDSS   & \ldots  & D  & 230.3  & 9.0   &  242.2  & 16.3   & 3.32  & 0.28 & 341.28  & 4.92   & 9.03   & 0.35   & \ldots \\
RQS2v               &       & 55182 & BOSS   & 1.70    & B  & 361.8  & 8.2   &  404.5  & 12.0   & 5.91  & 0.26 & 473.01  & 4.62   & 8.41   & 0.15   & \ldots \\
                    &       & 58051 & eBOSS  & 0.87    &    & 160.3  & 9.4   &  213.9  & 17.1   & 1.61  & 0.32 & 231.04  & 4.85   & 6.73   & 0.30   & \ldots \\
004645.99+002729.60 & 0.467 & 52199 & SDSS   & 1.66    & B  & 174.7  & 6.4   &  189.6  &  9.6   & 4.35  & 0.12 & 167.35  & 6.07   & 5.50   & 0.27   & R18 \\
RQS1                &       & 58080 & BOSS   & \ldots  & D  & 144.0  & 8.8   &  172.1  & 15.3   & 3.99  & 0.15 & 144.67  & 7.87   & 3.29   & 0.16   & \ldots \\
005813.68+001409.09 & 0.679 & 52520 & SDSS   & 0.73    &    & 175.2  &16.7   &  142.6  & 23.2   & 0.88  & 0.55 & 147.55  & 5.76   & 7.54   & 0.43   & R18 \\
RQS3v               &       & 57360 & eBOSS  & 0.96    & B  &  75.6  & 8.8   &   36.9  &  7.0   & 0.16  & 0.00 & 63.02   & 0.67   & 1.24   & 0.11   & \ldots \\
                    &       & 58097 & eBOSS  & \ldots  & D  &  67.3  & 8.2   &   46.2  &  9.5   & 0.00  & 0.05 & 71.05   & 1.36   & 1.58   & 0.50   & \ldots \\
010929.23+014917.05 & 0.779 & 57282 & eBOSS  & 0.60    & B  & 211.9  & 8.2   &  307.2  & 23.7   & 8.86  & 0.44 & 188.02  & 8.28   & 5.50   & 0.22   & \ldots \\
B                   &       & 58112 & eBOSS  & \ldots  & D  & 120.2  & 6.0   &  204.0  & 14.3   & 4.99  & 0.64 & 101.56  & 5.36   & 6.02   & 0.43   & \ldots \\
012256.19-000252.68 & 0.341 & 51817 & SDSS   & \ldots  & D  &  35.1  & 2.0   &   39.3  &  3.5   & 0.32  & 0.08 & \ldots  & \ldots & \ldots & \ldots & \ldots \\
RQS1                &       & 58113 & eBOSS  & 2.31    & B  & 112.1  & 1.6   &  139.4  &  3.2   & 1.03  & 0.04 & \ldots  & \ldots & \ldots & \ldots & \ldots \\
013015.10+002557.19 & 0.336 & 51820 & SDSS   & 2.05    & B  & 139.3  & 2.2   &  174.9  &  4.3   & 0.76  & 0.07 & \ldots  & \ldots & \ldots & \ldots & \ldots \\
RQS1                &       & 55483 & BoSS   & 1.46    &    & 105.8  & 0.9   &  151.9  &  1.6   & 0.49  & 0.04 & 180.93  & 1.77   & 2.51   & 0.21   & \ldots \\
                    &       & 58113 & eBOSS  & \ldots  & D  &  76.7  & 1.5   &  104.2  &  2.6   & 0.30  & 0.05 & 123.30  & 2.75   & 2.62   & 0.27   & \ldots \\
013435.90+022839.85 & 0.177 & 56899 & eBOSS  & \ldots  & D  &  21.4  & 0.1   &   28.0  &  0.4   & 0.07  & 0.01 & \ldots  & \ldots & \ldots & \ldots & \ldots \\
RQS1                &       & 58124 & eBOSS  & 1.40    &    &  26.6  & 0.7   &   41.1  &  1.4   & 0.41  & 0.03 & \ldots  & \ldots & \ldots & \ldots & \ldots \\
                    &       & 58699 & Mag/L3 & 2.81    & B  &  22.6  & 0.4   &   34.1  &  0.8   & 0.28  & 0.01 & \ldots  & \ldots & \ldots & \ldots & \ldots \\
015653.16-004623.28 & 0.615 & 52199 & SDSS   & 1.29    &    & 215.9  &11.4   &  269.8  & 21.0   & 3.05  & 0.26 & 317.74  & 16.01  & 7.26   & 0.39   & \ldots \\
HYP                 &       & 55182 & BOSS   & 2.28    & B  & 343.4  & 8.3   &  380.4  & 11.8   & 5.90  & 0.17 & 365.83  & 6.01   & 9.94   & 0.22   & \ldots \\
                    &       & 56960 & eBOSS  & 2.10    &    & 415.8  & 8.2   &  355.4  &  7.4   & 5.44  & 0.17 & 434.49  & 8.13   & 8.99   & 0.19   & \ldots \\
                    &       & 58080 & eBOSS  & \ldots  & D  & 180.2  & 9.4   &  281.6  & 22.7   & 2.17  & 0.26 & 381.31  & 14.11  & 6.17   & 0.21   & \ldots \\
015726.00-000602.15 & 0.802 & 55182 & BOSS   & 0.75    &    & 264.9  &13.0   &  253.9  & 24.2   & 1.84  & 0.77 & 424.14  & 11.45  & 5.32   & 0.19   & \ldots \\
RQS3v               &       & 55449 & BOSS   & 1.10    & B  & 333.0  &16.1   &  342.4  & 22.5   & 2.07  & 0.49 & 354.61  & 12.35  & 6.47   & 0.27   & \ldots \\
                    &       & 58103 & eBOSS  & \ldots  & D  &  86.4  &15.1   &   73.8  & 23.7   & 2.72  & 0.98 & 126.51  & 13.59  & 4.59   & 0.33   & \ldots \\
020222.00+010557.11 & 0.502 & 51871 & SDSS   & 1.43    & B  & 409.2  & 5.5   &  354.8  &  5.9   & 5.19  & 0.26 & 428.48  & 5.62   & 6.34   & 0.31   & R18 \\
RQS1                &       & 58079 & eBOSS  & \ldots  & D  &  91.3  & 4.5   &   89.1  &  7.0   & 1.20  & 0.12 & 117.93  & 2.28   & 5.57   & 0.17   & \ldots \\
020514.77-045639.74 & 0.363 & 55944 & BOSS   & 3.75    & B  &  92.2  & 1.8   &  111.2  &  2.7   & 1.01  & 0.03 & 77.37   & 1.75   & 2.14   & 0.16   & CLQ \\
RQS1                &       & 56660 & BOSS   & 2.09    &    &  37.6  & 1.1   &   50.5  &  2.1   & 0.53  & 0.02 & 40.69   & 0.64   & 2.50   & 0.11   & \ldots \\
                    &       & 57723 & eBOSS  & \ldots  & D  &  18.2  & 1.7   &   22.6  &  2.8   & 0.26  & 0.07 & 17.53   & 0.20   & 0.67   & 0.10   & \ldots \\
                    &       & 58699 & Mag/L3 & 1.50    &    &  28.4  & 0.9   &   31.7  &  1.2   & 0.68  & 0.05 & 0.77    & 0.01   & 6.71   & 0.79   & \ldots \\
021259.59-003029.43 & 0.395 & 51816 & SDSS   & 5.91    & B  & 833.5  &25.1   &  627.2  & 27.4   & 13.43 & 0.15 & 904.86  & 11.05  & 6.45   & 0.26   & CLQ \\
HYP,DE              &       & 56979 & eBOSS  & \ldots  & D  & 174.0  & 2.4   &  216.3  &  3.2   & 8.25  & 0.06 & 142.49  & 1.53   & 4.73   & 0.22   & \ldots \\
                    &       & 58097 & eBOSS  & 1.29    &    & 176.1  & 4.3   &  208.7  &  6.0   & 6.84  & 0.10 & 166.99  & 2.56   & 6.51   & 0.15   & \ldots \\
021359.08-025352.81 & 0.168 & 57336 & eBOSS  & 2.13    & B  &  17.4  & 0.1   &   24.3  &  0.3   & 0.14  & 0.00 & \ldots  & \ldots & \ldots & \ldots & \ldots \\
RQS1                &       & 58070 & BOSS   & \ldots  & D  &  10.2  & 0.1   &   13.2  &  0.3   & 0.04  & 0.01 & \ldots  & \ldots & \ldots & \ldots & \ldots \\
021359.79+004226.81 & 0.182 & 51816 & SDSS   & 6.54    & B & 232.6  & 0.9   &  203.3  &  1.1   & 3.97  & 0.04 & \ldots  & \ldots & \ldots & \ldots & CLQ,R18 \\
NQHISN              &       & 57043 & eBOSS  & 9.43    &    &  89.9  & 0.5   &  105.5  &  0.7   & 2.67  & 0.07 & \ldots  & \ldots & \ldots & \ldots & \ldots \\
                    &       & 58097 & eBOSS  & \ldots  & D  &  39.1  & 1.2   &   57.4  &  2.7   & 1.14  & 0.02 & \ldots  & \ldots & \ldots & \ldots & \ldots \\
022930.91-000845.37 & 0.609 & 52200 & SDSS   & 1.05    & B  & 521.5  &18.9   &  473.3  & 22.6   & 10.83 & 0.50 & 538.12  & 19.17  & 18.37  & 0.36   & R18 \\
DE                  &       & 58028 & eBOSS  & 0.81    &    & 115.5  & 2.6   &  232.1  &  8.6   & 5.32  & 0.21 & 93.17   & 2.26   & 5.16   & 0.15   & \ldots \\
                    &       & 58072 & eBOSS  & \ldots  & D  & 141.8  & 1.1   &  250.2  &  6.5   & 7.32  & 0.38 & 119.57  & 1.12   & 6.01   & 0.12   & \ldots \\
023030.38-004513.94 & 0.504 & 51820 & SDSS   & 1.52    &    & 162.3  & 6.4   &  169.5  &  8.4   & 3.45  & 0.19 & 104.25  & 3.22   & 1.15   & 0.33   & \ldots \\
HYP                 &       & 52200 & SDSS   & 2.25    & B  & 182.0  & 7.9   &  204.6  & 11.6   & 3.88  & 0.21 & 117.07  & 4.64   & 1.10   & 0.23   & \ldots \\
                    &       & 56238 & BOSS   & 1.21    &    &  68.6  & 3.8   &   66.5  &  5.4   & 1.24  & 0.12 & 34.53   & 1.93   & 1.66   & 0.07   & \ldots \\
                    &       & 56267 & BOSS   & 0.61    &    &  79.6  & 2.6   &   74.1  &  4.0   & 1.92  & 0.12 & 31.31   & 1.37   & 1.53   & 0.09   & \ldots \\
                    &       & 56577 & BOSS   & 1.02    &    &  76.0  & 2.3   &   66.0  &  3.1   & 1.30  & 0.09 & 35.97   & 1.72   & 1.16   & 0.09   & \ldots \\
                    &       & 57039 & eBOSS  & 0.50    &    &  65.6  & 2.8   &   60.5  &  4.2   & 0.99  & 0.20 & 22.73   & 1.89   & 1.71   & 0.10   & \ldots \\
                    &       & 58096 & eBOSS  & \ldots  & D  &  23.2  & 1.8   &   56.6  &  7.6   & 1.42  & 0.19 & 18.03   & 1.57   & 1.01   & 0.12   & \ldots \\
023214.86-024845.45 & 0.785 & 55557 & BOSS   & \ldots  & D  & 204.5  & 6.3   &  293.4  & 13.3   & 10.29 & 0.97 & 182.50  & 6.12   & 13.35  & 0.26   & \ldots \\
RQS2v,HYP           &       & 57723 & eBOSS  & 1.85    &    & 708.7  & 6.6   &  988.5  & 12.8   & 23.97 & 2.36 & 636.18  & 6.53   & 13.31  & 1.82   & \ldots \\
                    &       & 57727 & eBOSS  & 2.19    & B  & 822.6  &12.0   & 1021.0  & 15.1   & 29.47 & 0.68 & 767.44  & 8.71   & 14.13  & 1.44   & \ldots \\
024508.67+003710.68 & 0.299 & 51871 & SDSS   & 4.21    & B  &  83.4  & 2.2   &  102.5  &  3.1   & 0.72  & 0.03 & \ldots  & \ldots & \ldots & \ldots & CLQ \\
RQS1                &       & 58081 & eBOSS  & \ldots  & D  &  16.1  & 1.5   &   17.6  &  2.0   & 0.00  & 0.06 & 25.21   & 1.38   & 0.90   & 0.16   & \ldots \\
024932.01+002248.35 & 0.345 & 52175 & SDSS   & 4.99    & B  & 148.1  & 2.4   &  134.8  &  2.7   & 1.76  & 0.04 & \ldots  & \ldots & \ldots & \ldots & CLQ \\
HYP                 &       & 57041 & eBOSS  & 1.66    &    &  28.2  & 0.9   &   32.7  &  1.5   & 0.44  & 0.03 & 21.16   & 0.92   & 2.03   & 0.09   & \ldots \\
                    &       & 58081 & eBOSS  & \ldots  & D  &  21.2  & 1.5   &   23.7  &  2.0   & 0.35  & 0.04 & 13.06   & 0.75   & 0.92   & 0.14   & \ldots \\
085704.02+261818.14 & 0.711 & 53381 & SDSS   & 1.65    & B  & 906.1  &28.1   &  982.1  & 46.0   & 14.57 & 0.59 & 1550.68 & 29.76  & 30.64  & 0.57   & \ldots \\
RQS1                &       & 58133 & eBOSS  & \ldots  & D  & 367.5  &15.4   &  401.3  & 19.0   & 3.99  & 0.27 & 514.32  & 6.45   & 18.72  & 0.24   & \ldots \\
090258.39+252915.15 & 0.264 & 53401 & SDSS   & 1.21    & B  &  76.2  & 1.3   &   94.9  &  2.0   & 0.67  & 0.05 & \ldots  & \ldots & \ldots & \ldots & \ldots \\
RQS1                &       & 58133 & eBOSS  & \ldots  & D  &  32.6  & 0.7   &   40.1  &  1.0   & 0.47  & 0.05 & \ldots  & \ldots & \ldots & \ldots & \ldots \\
090339.98+241257.6  & 0.605 & 53401 & None   & \ldots  & B  & 328.4  &21.8   &  242.8  & 35.8   & 9.63  & 0.46 & 426.31  & 6.39   & 7.12   & 0.53   & M19n \\
HYP                 &       & 58085 & None   & 1.38    & D  &  98.3  &14.9   &   80.7  & 25.1   & 2.46  & 0.62 & 113.54  & 2.73   & 3.38   & 0.16   & \ldots \\
091234.00+262828.32 & 0.551 & 53415 & SDSS   & 1.68    & B  & 374.7  &15.0   &  246.0  & 16.0   & 6.88  & 0.41 & 429.08  & 18.41  & 12.14  & 0.39   & M19n \\
HYP                 &       & 58131 & eBOSS  & \ldots  & D  & 212.0  & 6.8   &  244.8  & 12.5   & 4.05  & 0.41 & 227.33  & 7.15   & 6.40   & 0.41   & \ldots \\
091407.16+243633.08 & 0.681 & 56009 & BOSS   & 1.48    & B  &  98.6  & 7.3   &  123.7  & 12.7   & 3.98  & 0.54 & 91.76   & 7.34   & 6.74   & 0.23   & \ldots \\
HYP                 &       & 58137 & eBOSS  & \ldots  & D  &  44.4  & 3.3   &   80.5  &  8.3   & 1.52  & 0.34 & 36.66   & 3.28   & 2.87   & 0.17   & \ldots \\
091634.67+195952.70 & 0.771 & 56017 & BOSS   & \ldots  & D  & 317.1  & 9.3   &  347.4  & 14.2   & 5.89  & 0.58 & 366.26  & 7.74   & 7.39   & 0.58   & \ldots \\
HYP                 &       & 57817 & eBOSS  & 0.78    & B  & 485.1  &12.0   &  565.1  & 20.6   & 9.67  & 0.54 & 705.69  & 6.61   & 6.53   & 1.04   & \ldots \\
093148.13+234837.08 & 0.891 & 56246 & BOSS   & 2.25    & B  & 582.7  &52.1   &  473.2  & 40.9   & 14.09 & 0.76 & 623.33  & 50.84  & 23.51  & 1.13   & \ldots \\
HYP                 &       & 57814 & eBOSS  & \ldots  & D  & 275.7  & 2.6   &  205.3  &  4.0   & 8.29  & 0.31 & 303.14  & 2.86   & 10.66  & 0.40   & \ldots \\
094231.68+233613.44 & 0.795 & 53735 & SDSS   & 1.30    & B  &2075.5  &32.2   & 1673.7  & 37.1   & 21.55 & 1.37 & 2224.29 & 35.19  & 41.54  & 1.14   & \ldots \\
RQS1                &       & 58132 & eBOSS  & \ldots  & D  & 527.5  &10.7   &  423.9  & 13.0   & 41.14 & 1.33 & 603.75  & 6.20   & 27.77  & 0.49   & \ldots \\
100220.18+450927.30 & 0.401 & 52376 & SDSS   & 2.26    & B  & 167.0  & 4.7   &  204.0  &  8.3   & 2.51  & 0.11 & 196.10  & 6.50   & 6.09   & 0.32   & VIc,M16c,M19n,H19 \\
DE                  &       & 56683 & BOSS   & \ldots  & D  &  48.8  & 2.5   &   55.8  &  3.5   & 0.53  & 0.06 & 59.89   & 6.51   & 4.40   & 0.17   & \ldots \\
100302.62+193251.28 & 0.467 & 53762 & SDSS   & \ldots  & D  & 131.0  & 5.1   &  147.7  & 10.1   & 4.02  & 0.13 & 123.74  & 5.54   & 14.60  & 0.15   & CLQ,TurnOn,M19r,S17 \\
HYP                 &       & 57817 & eBOSS  & 4.31    & B  & 296.0  & 5.1   &  271.2  &  7.0   & 4.56  & 0.15 & 315.89  & 19.32  & 12.91  & 0.16   & \ldots \\
                    &       & 58183 & MMT/BC & 1.91    &    & 358.9  & 9.3   &  173.2  &  6.7   & 3.62  & 0.34 & 206.35  & 11.50  & 7.31   & 0.19   & \ldots \\
101322.43+214021.74 & 0.640 & 56274 & BOSS   & 1.05    & B  & 152.3  & 4.7   &  159.3  &  6.5   & 2.05  & 0.42 & 150.33  & 5.01   & 4.27   & 0.12   & \ldots \\
HYP                 &       & 57815 & eBOSS  & \ldots  & D  &  72.2  & 8.1   &  125.1  & 17.8   & 2.07  & 0.39 & 60.58   & 6.35   & 3.83   & 0.29   & \ldots \\
102844.36+230419.98 & 0.304 & 53763 & SDSS   & 1.34    & B  & 141.6  & 1.6   &  161.3  &  2.4   & 1.77  & 0.05 & \ldots  & \ldots & \ldots & \ldots & \ldots \\
RQS1                &       & 58136 & eBOSS  & \ldots  & D  &  57.3  & 0.8   &   59.6  &  1.3   & 0.50  & 0.04 & 32.69   & 0.45   & 2.08   & 0.17   & \ldots \\
102913.80+271853.66 & 0.615 & 53794 & SDSS   & 1.72    & B  & 277.1  &15.1   &  189.8  & 11.8   & 5.49  & 0.41 & 313.07  & 16.35  & 7.65   & 0.81   & \ldots \\
HYP                 &       & 58143 & eBOSS  & \ldots  & D  &  28.5  & 4.5   &   31.8  &  7.7   & 0.46  & 0.19 & 21.56   & 1.92   & 1.97   & 0.15   & \ldots \\
                    &       & 58456 & eBOSS  & 0.67    &    &  45.3  & 5.5   &   54.7  &  8.7   & 0.34  & 0.34 & 45.99   & 1.91   & 2.46   & 0.18   & \ldots \\
103624.70+302411.96 & 0.452 & 56358 & BOSS   & 1.14    & B  &  62.1  & 2.0   &   72.1  &  2.8   & 0.40  & 0.05 & 58.44   & 2.34   & 1.84   & 0.12   & \ldots \\
HYP                 &       & 58155 & eBOSS  & \ldots  & D  &  30.0  & 2.1   &   34.6  &  3.8   & 0.13  & 0.07 & 28.98   & 1.73   & 1.08   & 0.21   & \ldots \\
                    &       & 58522 & eBOSS  & 0.79    &    &  44.8  & 2.7   &   52.5  &  4.4   & 0.11  & 0.06 & 47.36   & 4.30   & 2.22   & 0.20   & \ldots \\
105058.42+241351.18 & 0.270 & 54068 & SDSS   & 5.01    & B  &  82.6  & 1.4   &  110.2  &  2.8   & 0.99  & 0.03 & \ldots  & \ldots & \ldots & \ldots & CLQ \\
RQS1                &       & 54086 & SDSS   & 4.67    &    &  81.2  & 1.2   &  104.9  &  1.7   & 1.01  & 0.02 & \ldots  & \ldots & \ldots & \ldots & \ldots \\
                    &       & 58138 & eBOSS  & \ldots  & D  &  27.5  & 1.9   &   33.9  &  3.2   & 0.28  & 0.04 & \ldots  & \ldots & \ldots & \ldots & \ldots \\
105325.40+302419.34 & 0.249 & 53463 & SDSS   & 3.47    & B  &  71.6  & 1.3   &   90.7  &  2.2   & 1.42  & 0.05 & \ldots  & \ldots & \ldots & \ldots & CLQ,M19n \\
DE                  &       & 58141 & BOSS   & \ldots  & D  &  37.5  & 0.4   &   39.2  &  0.6   & 0.23  & 0.03 & \ldots  & \ldots & \ldots & \ldots & Dbl Pk \\
                    &       & 58514 & BOSS   & 1.09    &    &  45.1  & 0.4   &   52.0  &  0.7   & 0.52  & 0.02 & \ldots  & \ldots & \ldots & \ldots & \ldots \\
105513.88+242553.69 & 0.496 & 53793 & SDSS   & 5.93    & B  & 158.7  & 8.5   &  135.6  &  8.6   & 2.57  & 0.11 & 167.06  & 8.70   & 7.17   & 1.12   & CLQ,M19c,R18 \\
HYP                 &       & 58138 & eBOSS  & \ldots  & D  &  29.2  & 3.3   &   29.9  &  5.4   & 0.54  & 0.24 & 10.63   & 2.18   & 1.89   & 0.10   & \ldots \\
111329.68+531338.78 & 0.239 & 52649 & SDSS   & 4.00    & B  &  54.3  & 1.3   &   64.7  &  1.8   & 0.98  & 0.02 & \ldots  & \ldots & \ldots & \ldots & CLQ,M19c,R18 \\
DE                  &       & 57374 & eBOSS  & \ldots  & D  &  14.1  & 0.6   &   15.8  &  0.9   & 0.32  & 0.03 & \ldots  & \ldots & \ldots & \ldots & \ldots \\
                    &       & 57425 & WHT    & 0.69    &    &  0.46  & 0.0   &    2.5  &  0.1   & 0.71  & 0.02 & \ldots  & \ldots & \ldots & \ldots & \ldots \\
113651.66+445016.48 & 0.116 & 53083 & SDSS   & \ldots  & D  &  21.0  & 0.6   &   32.0  &  1.2   & 0.17  & 0.01 & \ldots  & \ldots & \ldots & \ldots & CLQ,TurnOn \\
NQHISN              &       & 57732 & eBOSS  & 6.41    & B  &  51.3  & 0.6   &   82.6  &  1.4   & 0.62  & 0.01 & \ldots  & \ldots & \ldots & \ldots & \ldots \\
113706.93+481943.68 & 0.220 & 53054 & SDSS   & 4.70    & B  &  36.3  & 2.1   &   74.9  &  4.5   & 1.71  & 0.05 & \ldots  & \ldots & \ldots & \ldots & CLQ \\
DE                  &       & 56412 & BOSS   & 1.92    &    &  58.9  & 0.5   &   65.8  &  0.8   & 0.88  & 0.04 & \ldots  & \ldots & \ldots & \ldots & \ldots \\
                    &       & 57162 & BOSS   & \ldots  & D  &  24.8  & 0.3   &   30.1  &  0.6   & 0.23  & 0.02 & \ldots  & \ldots & \ldots & \ldots & \ldots \\
115039.32+363258.43 & 0.340 & 53436 & SDSS   & 2.94    & B  &  55.8  & 2.3   &   61.9  &  3.8   & 0.34  & 0.04 & \ldots  & \ldots & \ldots & \ldots & VIc,Y18c \\
HYP                 &       & 57422 & eBOSS  & \ldots  & D  &  17.4  & 1.5   &   22.7  &  2.5   & 0.14  & 0.07 & 20.90   & 0.37   & 1.42   & 0.11   & \ldots \\
121831.61+581636.85 & 0.448 & 56420 & BOSS   & 2.76    & B  &  66.7  & 4.2   &  107.3  &  9.8   & 0.97  & 0.10 & 57.24   & 4.32   & 2.64   & 0.24   & \ldots \\
HYP                 &       & 58197 & eBOSS  & \ldots  & D  &  63.4  & 2.5   &   69.3  &  3.6   & 0.13  & 0.05 & 34.18   & 1.38   & 2.46   & 0.25   & \ldots \\
123431.08+515629.27 & 0.296 & 52379 & SDSS   & 2.46    & B  &  66.9  & 2.5   &   81.6  &  3.5   & 0.79  & 0.06 & \ldots  & \ldots & \ldots & \ldots & \ldots \\
DE                  &       & 58198 & eBOSS  & \ldots  & D  &  28.5  & 1.1   &   32.3  &  1.9   & 0.22  & 0.05 & 49.11   & 0.35   & 2.78   & 0.29   & \ldots \\
124523.21+584047.06 & 0.611 & 54570 & SQLS   & 2.12    & B  & 188.8  & 7.7   &  195.9  &  9.9   & 1.60  & 0.12 & 187.01  & 7.58   & 4.89   & 0.31   & M19n \\
HYP                 &       & 58191 & eBOSS  & \ldots  & D  &  39.5  & 4.1   &   38.9  &  7.3   & 0.13  & 0.09 & 39.61   & 1.94   & 2.39   & 0.20   & \ldots \\
124614.85+582624.10 & 0.611 & 52765 & SDSS   & 2.24    & B  & 428.7  & 8.1   &  281.6  &  9.3   & 6.27  & 0.38 & 491.91  & 9.68   & 9.69   & 1.05   & \ldots \\
HYP                 &       & 58193 & eBOSS  & \ldots  & D  &  91.1  & 7.0   &   97.6  & 12.3   & 1.43  & 0.30 & 36.67   & 3.56   & 4.35   & 0.35   & \ldots \\
124946.13+365204.21 & 0.595 & 53772 & SDSS   & 2.74    & B  & 386.4  &13.9   &  414.5  & 20.7   & 6.17  & 0.34 & 752.55  & 17.09  & 13.83  & 0.41   & VIc \\
HYP                 &       & 57781 & eBOSS  & \ldots  & D  & 130.8  & 5.7   &  140.7  &  8.7   & 1.05  & 0.19 & 235.81  & 2.86   & 7.60   & 0.19   & \ldots \\
131834.43+531641.04 & 0.894 & 56415 & BOSS   & 1.22    & B  & 709.1  &22.7   &  595.7  & 26.7   & 15.78 & 1.56 & 750.27  & 28.29  & 25.48  & 0.58   & \ldots \\
HYP                 &       & 57431 & eBOSS  & \ldots  & D  & 170.4  &17.6   &  179.2  & 33.5   & 5.31  & 0.85 & 178.57  & 8.30   & 15.51  & 0.45   & \ldots \\
135415.54+515925.77 & 0.320 & 52465 & SDSS   & 4.83    & B  & 204.9  & 2.7   &  146.2  &  2.7   & 1.99  & 0.06 & \ldots  & \ldots & \ldots & \ldots & CLQ \\
HYP                 &       & 57427 & eBOSS  & \ldots  & D  &  17.2  & 1.0   &   33.0  &  2.3   & 0.93  & 0.06 & 14.32   & 0.61   & 1.10   & 0.19   & \ldots \\
143040.58+364903.90 & 0.570 & 53089 & SDSS   & \ldots  & D  & 121.6  & 9.4   &  136.5  & 15.3   & 0.81  & 0.17 & 183.34  & 4.50   & 5.35   & 0.67   & \ldots \\
HYP                 &       & 58242 & eBOSS  & 2.55    & B  & 222.7  & 5.8   &  285.4  &  8.8   & 3.35  & 0.22 & 394.93  & 3.80   & 13.01  & 0.41   & \ldots \\
143455.30+572345.10 & 0.174 & 52346 & SDSS   & 4.87    & B  &  95.2  & 0.4   &  132.0  &  0.9   & 1.90  & 0.04 & \ldots  & \ldots & \ldots & \ldots & CLQ,M19c \\
DE                  &       & 57426 & WHT    & 0.71    &    &  22.8  & 1.6   &   12.9  &  1.6   & 0.56  & 0.03 & \ldots  & \ldots & \ldots & \ldots & \ldots \\
                    &       & 57870 & eBOSS  & \ldots  & D  &  11.6  & 0.1   &   15.4  &  0.3   & 0.04  & 0.01 & \ldots  & \ldots & \ldots & \ldots & \ldots \\
163620.38+475838.36 & 0.236 & 52144 & SDSS   & 3.81    & B  &  66.1  & 0.7   &   84.4  &  1.3   & 0.54  & 0.02 & \ldots  & \ldots & \ldots & \ldots & CLQ \\
DE                  &       & 58257 & eBOSS  & \ldots  & D  &  28.6  & 0.3   &   34.2  &  0.5   & 0.09  & 0.02 & \ldots  & \ldots & \ldots & \ldots & \ldots \\
163934.03+464314.23 & 0.731 & 56190 & BOSS   & \ldots  & D  & 119.7  &14.2   &  154.9  & 27.2   & 1.15  & 0.36 & 198.37  & 3.71   & 8.80   & 0.26   & \ldots \\
HYP                 &       & 58257 & eBOSS  & 1.38    & B  & 308.0  & 9.1   &  298.1  & 14.4   & 4.34  & 0.24 & 370.74  & 3.10   & 13.43  & 0.20   & \ldots \\
164829.25+410405.55 & 0.852 & 52054 & SDSS   & 1.07    & B  &1349.4  &51.0   & 1046.5  & 54.4   & 8.89  & 1.91 & 1413.84 & 63.52  & 25.67  & 1.40   & \ldots \\
HYP                 &       & 58016 & eBOSS  & \ldots  & D  & 154.0  &30.6   &  254.5  & 88.8   & 11.70 & 1.29 & 183.40  & 17.70  & 12.79  & 0.72   & \ldots \\
211838.12+005640.72 & 0.385 & 52443 & SDSS   & 2.10    & B  & 137.9  & 3.7   &  147.1  &  4.2   & 1.27  & 0.06 & 187.66  & 3.05   & 3.34   & 0.15   & \ldots \\
RQS1                &       & 57691 & eBOSS  & 0.95    &    &  73.5  & 2.0   &   75.1  &  2.6   & 0.42  & 0.05 & 87.02   & 1.47   & 3.85   & 0.11   & \ldots \\
                    &       & 57938 & eBOSS  & \ldots  & D  &  79.8  & 2.1   &   82.3  &  2.7   & 0.65  & 0.06 & 100.29  & 0.87   & 2.64   & 0.13   & \ldots \\
215130.35+003522.26 & 0.340 & 55479 & BOSS   & \ldots  & D  &  41.5  & 0.7   &   60.7  &  1.2   & 0.26  & 0.02 & 103.01  & 1.28   & 3.77   & 0.14   & \ldots \\
RQS3v               &       & 58035 & eBOSS  & 1.17    &    &  21.2  & 2.3   &   33.7  &  4.3   & 0.61  & 0.06 & 47.66   & 0.61   & 0.05   & 0.05   & \ldots \\
                    &       & 58699 & Mag/L3 & 1.61    & B  &  66.0  & 0.5   &   50.9  &  0.6   & 0.56  & 0.02 & 121.37  & 1.50   & 6.89   & 2.43   & \ldots \\
222132.41-010928.70 & 0.288 & 52140 & SDSS   & 1.44   & D &  66.4  & 1.4   &   76.7  &  2.2   & 0.76  & 0.06 & \ldots  & \ldots & \ldots & \ldots & \ldots \\
RQS1,FES\_DE        &       & 58013 & eBOSS  & \ldots & B & 101.6  & 0.3   &  133.6  &  0.5   & 0.65  & 0.01 & 47.08   & 1.38   & 2.96   & 9.73   & \ldots \\
                    &       & 58038 & eBOSS  & 1.09    &    & 103.7  & 0.7   &  143.5  &  1.3   & 1.18  & 0.04 & 151.65  & 1.01   & 0.31   & 0.15   & \ldots \\
                    &       & 58673 & MMT/BS & 1.62    &    &  65.7  & 2.7   &   82.5  &  4.0   & 1.58  & 0.10 & \ldots  & \ldots & \ldots & \ldots & \ldots \\
224113.54-012108.84 & 0.058 & 55824 & BOSS   & \ldots  & D  &   2.9  & 0.0   &    9.0  &  0.1   & 0.04  & 0.00 & \ldots  & \ldots & \ldots & \ldots & CLQ,TurnOn,DblPk \\
RQS1                &       & 57724 & eBOSS  & 10.33   &    &  10.5  & 0.3   &   15.8  &  0.7   & 0.21  & 0.00 & \ldots  & \ldots & \ldots & \ldots &  \\
                    &       & 58043 & eBOSS  & 12.24   &    &  19.2  & 0.4   &   20.3  &  0.1   & 0.23  & 0.00 & \ldots  & \ldots & \ldots & \ldots &  \\
                    &       & 58367 & MMT/BC & 14.09   & B  &   5.6  & 0.0   &   11.0  &  0.2   & 0.22  & 0.00 & \ldots  & \ldots & \ldots & \ldots & \ldots  \\
                    &       & 58699 & Mag/L3 & 9.13    &    &   6.0  & 0.0   &    8.0  &  0.1   & 0.13  & 0.00 & \ldots  & \ldots & \ldots & \ldots & \ldots  \\
230614.18-010024.45 & 0.267 & 51811 & SDSS   & \ldots  & D  &  19.2  & 2.6   &   26.7  &  4.3   & 0.85  & 0.24 & \ldots  & \ldots & \ldots & \ldots & \ldots \\
RQS1                &       & 57684 & eBOSS  & 1.49    &    &  43.1  & 0.9   &   56.5  &  1.6   & 0.67  & 0.02 & \ldots  & \ldots & \ldots & \ldots & \ldots \\
                    &       & 58022 & eBOSS  & 1.64    & B  &  53.0  & 1.4   &   71.2  &  2.2   & 0.62  & 0.03 & \ldots  & \ldots & \ldots & \ldots & \ldots \\
231625.39-002225.50 & 0.297 & 51816 & SDSS   & 3.16    & B  &  81.6  & 1.8   &   97.5  &  2.5   & 0.64  & 0.04 & \ldots  & \ldots & \ldots & \ldots & CLQ,R18 \\
RQS1                &       & 57715 & eBOSS  & 1.20    &    &  64.5  & 0.6   &   73.9  &  1.1   & 0.57  & 0.04 & 35.87   & 0.56   & 1.17   & 0.12   & \ldots \\
                    &       & 58013 & eBOSS  & \ldots  & D  &  49.4  & 0.6   &   61.2  &  1.1   & 0.32  & 0.03 & 40.33   & 2.05   & 3.10   & 0.22   & \ldots \\
234623.42+010918.11 & 0.509 & 52524 & SDSS   & \ldots  & D  & 139.7  & 6.3   &  138.1  &  8.5   & 1.03  & 0.15 & 67.81   & 4.10   & 4.79   & 0.31   & CLQ,TurnOn, R18,H20 \\
RQS2,RQS2v          &       & 57595 & Mag/L3 & 1.01    &    & 125.4  &10.9   &  115.8  & 13.8   & 2.76  & 0.44 & 93.69   & 6.34   & 7.99   & 2.06   &  \\
                    &       & 57684 & eBOSS  & 3.39    & B  & 346.2  & 5.0   &  325.8  & 30.1   & 5.92  & 0.09 & 420.30  & 15.42  & 12.27  & 0.14   & \ldots \\
                    &       & 57989 & MMT/BC & 1.62    &    & 230.3  & 1.2   &  172.2  &  0.7   & 4.21  & 0.06 & 232.99  & 0.37   & 9.06   & 0.07   & \ldots \\
                    &       & 58081 & eBOSS  & 2.48    &    & 306.1  &26.5   &  321.2  &  3.3   & 6.03  & 0.15 & 371.14  & 2.52   & 12.51  & 0.23   & \ldots \\
                    &       & 58367 & MMT/BC & 1.69    &    & 241.2  & 1.6   &  213.2  &  1.7   & 5.79  & 0.10 & 242.25  & 1.28   & 7.63   & 0.08   & \ldots \\
                    &       & 58669 & MMT/BC & 2.23    &    & 177.5  & 2.0   &  223.7  &  3.2   & 9.03  & 0.10 & 159.55  & 1.80   & 9.19   & 0.15   & \ldots
\enddata
\tablecomments{ The full list of 169 quasar/epochs is available in the electronic version. (1) Designated state B=Bright, D=Dim. Facility: Mag/L3=  Magellan$\sim$6.5m Clay/LDSS3.  MMT/BC = MMT$\sim$6.5m/BlueChannel.   MMT/BS =  MMT$\sim$6.5m/BinoSpec.  WHT = WHT$\sim$4.2m/ISIS. CLQ listed in Notes for a quasar when $\Nsigma >=3$ between the designated bright and dim spectral epochs.. TurnOn notes that when a CLQ brightened with time. VIc is listed for $\Nsigma >=2$ quasars where strong CLQ nature seems evident upon visual inspection (see Figure\,\ref{f:lcspecs2}). References cited on first line of each quasar. M16c = \citet{MacLeod16} CLQ based there on $\Nsigma >3$. \citep{MacLeod19} M19c = CLQ based on $\Nsigma >3$; M19v = CLQ based on VI only; M19n = candidate, not analyzed; M19r = rejected. H19 = noted by \citet{Hutsemekers19} as low ($<1\% $) spectro-polarization. R18 = noted by \citet{Rumbaugh18} as an Extremely Variable Quasar (EVQ; $\Delta\,g>1$\,mag). Y18 = CLQ noted by \citet{Yang18}. H20 = \citet{Homan20}.   } 
\end{deluxetable}
\end{longrotatetable}

\subsection{Visual Inspection and Rescaling}
\label{s:rescale}

After compiling and overplotting the multi-epoch spectra of each QSO in our initial sample, we visually examined all the spectra.  For a given QSO, we limited spectral model fitting 
to the wavelengths spanned by the spectral epoch with the smallest wavelength coverage.
(For example the first SDSS ``Legacy" spectrograph typically covered as red as about 9100\,\AA, whereas the BOSS spectrograph was sensitive to beyond 1\,\micron.)
This ensures a more fair comparison of changes in luminosity because the best-fit power-law continuum model can vary simply by inclusion of different spectral regions.   

By selection, some of the spectral epochs for each QSO look quite different, and may have significantly different continuum levels or shapes.  In some cases, the apparent differences may not be intrinsic, but rather due to technical issues with the observations (such as a poorly-seated spectroscopic fiber) or data reduction (e.g., an incorrect flux calibration). One way to check the normalization (but not the continuum slope) is by measuring the \oiii\, line emission at 4959 and 5007\,\AA, adjacent to \Hb. In general, the \oiii\, emission line luminosity should not be intrinsically variable and the total \oiii\, emission is not expected to vary significantly over the timespan of our observing project\footnote{Variations are known, at least in lower luminosity AGN (e.g., \citealt{Peterson2013}), but especially for higher luminosity QSOs, that variability is expected to be much smaller than what may occur due to changes in seeing and/or slit or fiber placement.}. The \oiii\, emission line luminosity originates either from the quasar narrow line region or from circumnuclear regions of high star formation rate within the spectral aperture. Star formation regions change on timescales exceeding Myr, and the extent of the narrow line region is known to extend over several kiloparsecs \citep{Bennert02}. 
For each QSO, we therefore perform full PyQSOFit modeling (see \S\,\ref{s:decomp}) of each available spectral epoch to isolate and measure the \oiii\, luminosity.  We then compare the derived \Loiii\, values across all available epochs to determine whether a spectrum warrants rescaling. For a given QSO, we first derive the mean and standard deviation of \Loiii\, from all spectra, along with the deviation from the mean of each epoch.  For QSOs with more than two spectra, we iteratively determine which may be outliers, and recalculate the cleaned mean and standard deviation without them.  If a spectrum meets two criteria, that (1) its log\,\Loiii\, deviates by at least 0.04 from the cleaned mean (about 10\% in linear luminosity), and (2) that deviation is at least twice the cleaned standard deviation - we renormalize it by a simple multiplicative factor so that its \Loiii\, matches the mean value.  Such rescaling was used on only eight spectra (of five quasars), most of which were follow-up spectra of our own, although one was an SDSS spectrum with a suspect calibration, which may have been caused by a fiber misplaced within its plug hole on the plate, or an unusual kink or blockage in the spectral pathway.  In general, the SDSS flux calibrations are excellent, since they employ detailed and consistent methods incorporating dozens of flux standards across each night \citep{Bolton12}.
 
\subsection{Light Curves and Spectra}\label{s:lcs}
Ideally, strong spectral changes should be confirmed by photometric variability at similar epochs in the same direction (brightening or dimming). This is especially important if the changes are unusual, extreme, or potentially attributable to technical mishaps in the spectroscopic observations or reduction as described above.  We therefore sought complementary multi-epoch and multi-band photometry from public databases. 

\begin{figure}
\centering
\plotone{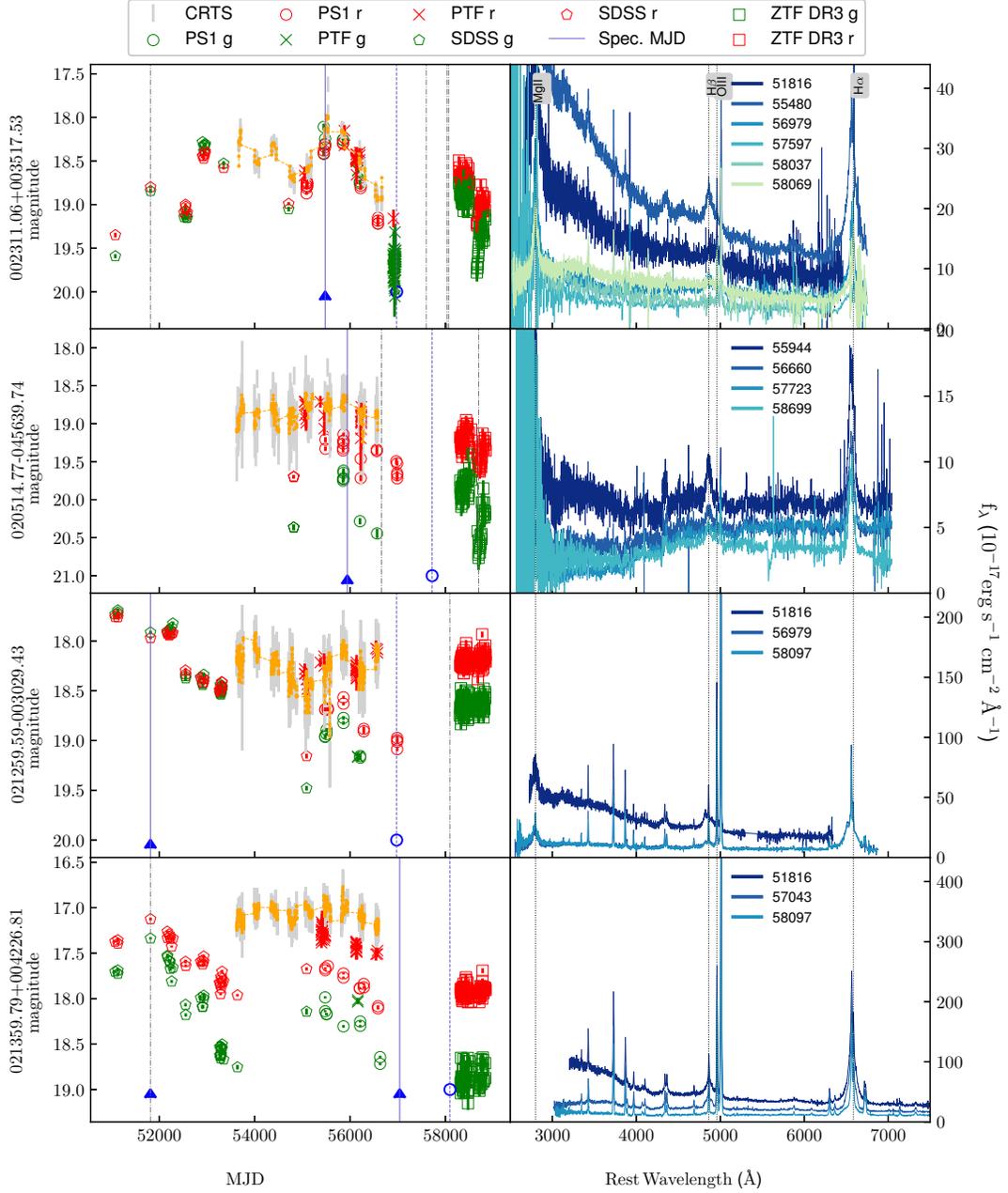}
\vspace{-1.5cm}
\caption{Light curves and spectra for confirmed TDSS CLQs.
\emph{Left:}  As shown in the key, CRTS magnitudes are represented by gray error bars centered on the CRTS ($V$-band) magnitude.  Five  day CRTS averages are orange points.  For all other surveys, green (red) points represent $g$ ($r$) band magnitudes. The earliest photometry here, from SDSS, is shown with open pentagons. PS1 magnitudes are shown as open circles, PTF as crosses, ZTF DR3 as open boxes. The spectral epoch we choose for the bright state is marked with a solid vertical blue line, and the dim state with a dashed blue line. Other spectral epochs are shown with dot-dashed grey lines. Any spectral epoch with a blue upward arrow indicates \Nsigma$>3$ relative to the dim state, which has its epoch marked by an open blue circle.
\emph{Right:}. Spectra of all epochs for every TDSS CLQ, shown in the rest frame.  Prominent lines such as \MgII, \Hb, \oiii, and \Ha\ are labeled in the topmost panel, with corresponding vertical dashed lines crossing the spectra for every quasar.  For comparison to the light curves at left, the spectral epoch in MJD for each spectrum is provided with a short solid line of the corresponding color. The complete figure set of light curves and spectra for all 19 of our CLQ candidates is available (presented as 5 images) in the online journal.  For seven of the spectral plots, we chose a maximum flux value below the peak of [OIII]\,5007\AA\ for clarity.}
\label{f:lcspecs}
\end{figure}

 We obtained $g$ and $r$ band photometry from the following sources: PanSTARRS1 (PS1;\citealt{kaiser10}), Palomar Transient Factory (PTF;\citealt{law09}), Zwicky Transient Facility (ZTF;\citealt{bellm19}) Data Release~3 and the Sloan Digital Sky Survey (SDSS;\citealt{York00}). We also obtained photometry from the Catalina Surveys (CRTS; \citealt{drake09}), which provides a $V$-band calibrated magnitude. PTF and ZTF data were obtained through the NASA/IPAC Infrared Science Archive\footnote{\url{https://irsa.ipac.caltech.edu/}; \citealt{https://doi.org/10.26131/irsa156}}. PS1 data were obtained through the PanStarrs catalog search via MAST hosted at the  STScI\footnote{\url{https://catalogs.mast.stsci.edu/panstarrs/}}. CRTS data was obtained through the Caltech Survey Data Release~2 online query tool\footnote{\url{http://nesssi.cacr.caltech.edu/DataRelease/}}. SDSS photometry was obtained through the CasJobs Skyserver\footnote{\url{https://skyserver.sdss.org/casjobs/}}. All searches were done using a cone search within 2\,arcsec of the coordinates of the individual source. Figure\,\ref{f:lcspecs} shows light curves and spectra for all of the confirmed CLQs in our TDSS sample, ordered by right ascension.  The first four are printed in  Figure\,\ref{f:lcspecs}, while all 19 CLQs appear in the associated online figureset.

It is worth noting the 
overall characteristics of this spectroscopically-selected sample, as well as 
some interesting features in individual CLQs, as evident from inspection of Figure\,\ref{f:lcspecs}.  Quasars are typically bluer when brighter, sometimes interpreted to be the result of an increased accretion rate resulting in heating of the accretion disk, with consequent blue/UV upturn. However, 
the Balmer recombination (free-bound) continuum and the high-order Balmer lines blend together and make a pseudo-continuum in the blue. These two components of the continuum vary with \Ha\ and \Hb\ without any lag and can create the bluer-when-brighter effect even when only \Ha\ and \Hb\ are observed to vary.  

All the CLQ spectra displayed in Figure\,\ref{f:lcspecs} are bluer when brighter; there are no examples of spectra that are brighter but redder. Prime examples include J002311.06+003517.53 and J135415.54+515925.77 which show extreme spectral changes in the blue, with corresponding photometric changes spanning more than a magnitude in $g$-band. The strength of broad \Hb\ emission seems to correlate directly with brightness. This is demonstrated later in \S\,\ref{s:belvar}.  

Since the spectra are shown in rest-frame, it is easy to note in general that little change in the spectral continuum shape occurs redder than about 5000\AA, the \Hb/\oiii\, region. This is especially obvious, for instance, in the spectra of J113651.66+445016.48. Except for the \Ha/[NII] line complex, the bulk of emission at these longer rest wavelengths originates in the host galaxy itself. 

In this paper, we do not perform any detailed analysis of photometric variability. However, it is instructive to scan the light curves and spectra shown in Figure\,\ref{f:lcspecs}.  We discuss these CLQs {\sout in detail}  individually in the Appendix.  However, in \S\,\ref{s:diversity}, we also discuss in general the diverse behavior observed even in our modest sampling of CLQ variability. 

\subsection{Spectral Decomposition}
\label{s:decomp}

\begin{figure}[h!]
\label{f:024932model}
\centerline{
\includegraphics[angle=90,scale=.3]{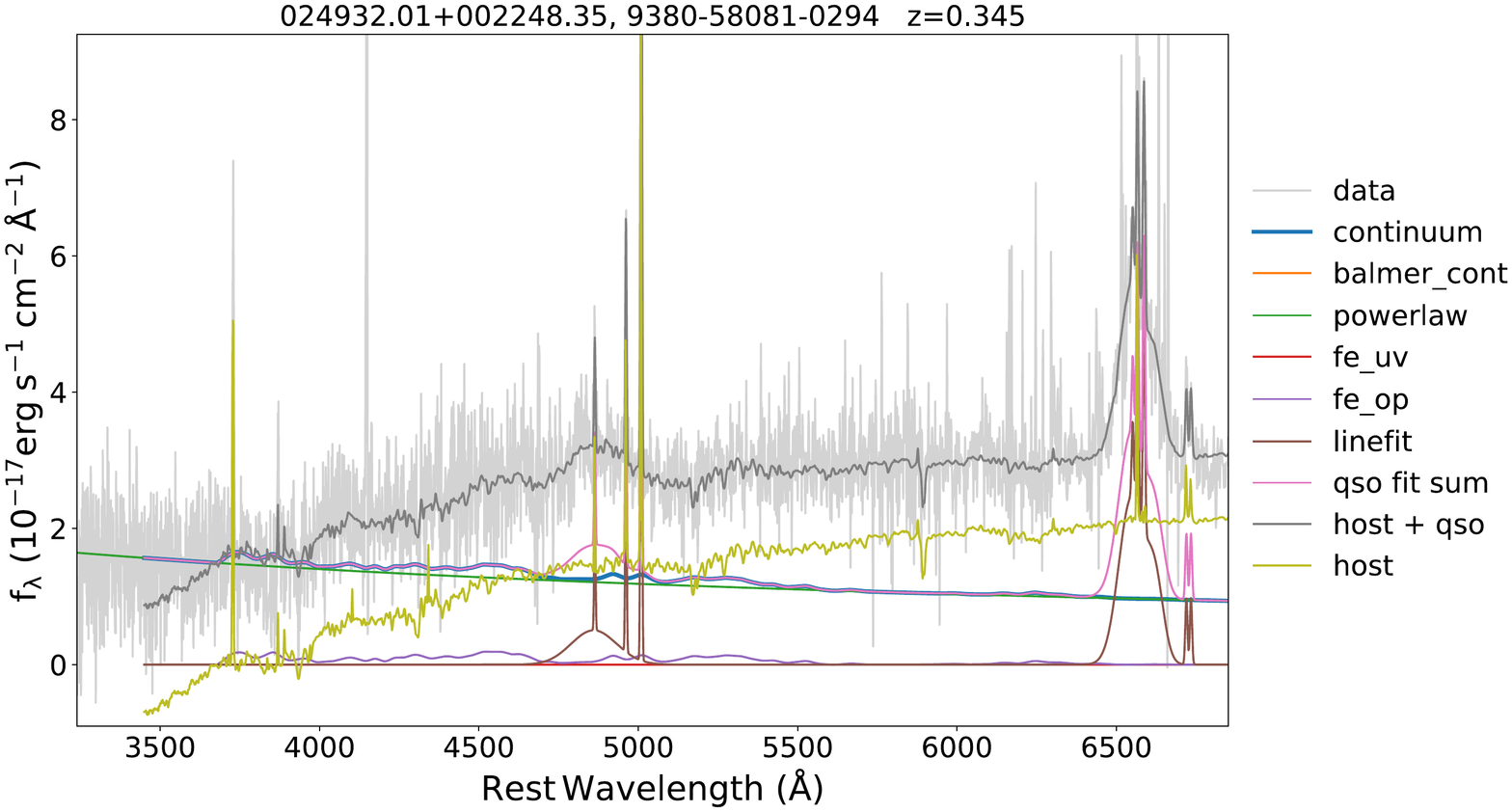}
\includegraphics[angle=90,scale=.3]{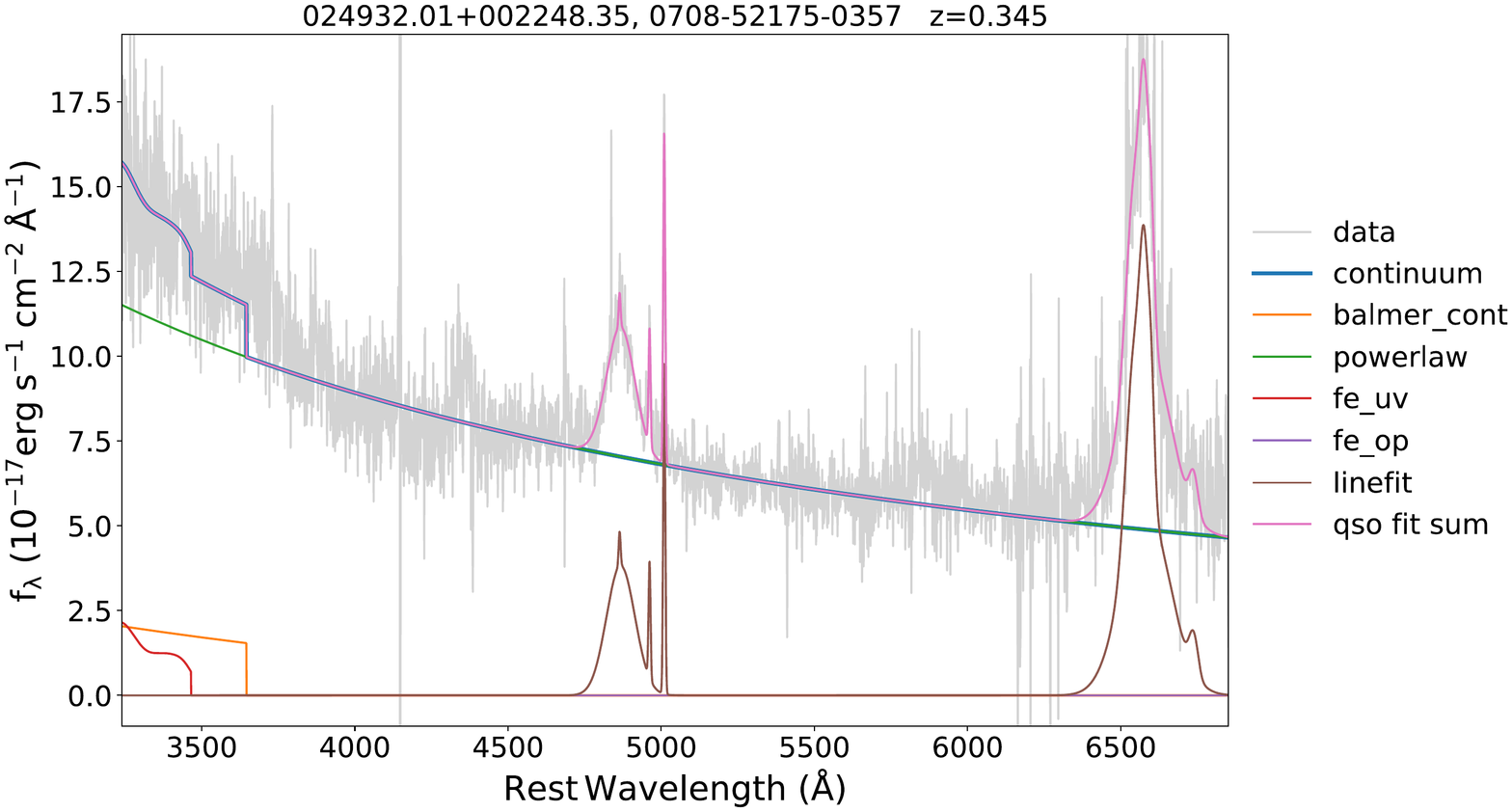}
}
\vspace{-1cm}
\caption{\footnotesize{
Spectral decomposition using PyQSOFit for the dim and bright state spectra of SDSS\,J024932.01+002248.3.
\emph{Left:} Spectral decomposition using PyQSOFit for the dim state spectrum MJD 58081. The light grey line shows the original spectrum.  Fitted spectral model components may include  line fits for \Hb\, \oiii\, H$\alpha+$N[II] (brown), a power-law continuum (green), Balmer continuum (orange), optical Fe\,II (violet) and UV Fe\,II (red; only below 3500\AA). The blue {\tt continuum} line shows the sum of the best-fit full continuum including the power-law, Balmer continuum, and Fe\,II components.   The blue {\tt qso fit sum} includes the continuum plus the host fit.
Broad \Hb\, emission is visible, but quite weak compared to the bright state.
\emph{Right:} PyQSOFit spectral decomposition for the bright state spectrum MJD 52175. For this spectrum the host galaxy component is not detectable.  The power-law continuum luminosity is greatly increased.  Broad \Hb\, emission is stronger by 4.99$\sigma$ compared to the dim state.
}}
\end{figure}

We use PyQSOFit\footnote{PyQSOFit is a Python adaptation of the IDL code QSOFit, referenced in \citet{Shen19}} for spectral decomposition \citep{2018ascl.soft09008G} of all of our SDSS spectra as well as for the follow-up spectra we obtained on other telescopes. We correct the spectra to the rest frame and correct for Galactic extinction using  the extinction curve of \citet{Cardelli89} and dust map of \citep{SFD98}.  We then perform a host galaxy decomposition using galaxy eigenspectra from \citep{Yip_2004a} as well as quasar eigenspectra from \citep{Yip_2004b}.  If more than half of the pixels from the resulting host galaxy fit are negative, then the host galaxy and quasar eigenspectral fits are not applied.  We then fit the power law, UV/Optical FeII and Balmer continuum models. The optical Fe\,II emission template spans 3686 $-$ 7484\AA, from \citet{Boroson92}, while the UV Fe\,II template spans 1000--3500\AA, adopted from  \citet{vestergaard01}, \citet{Tsuzuki06}, and  \citet{Salviander07}. PyQSOFit fits these empirical Fe\,II templates using a normalization, broadening, and wavelength shift. Next we perform emission line fits, using Gaussian profiles as described in \citet{Shen2019}. 
Depending on redshift and spectral coverage, we fit the following emission lines: \Ha$\lambda$6564.6 broad and narrow, [NII]$\lambda$6549,6585, [SII]$\lambda$6718,6732, \Hb$\lambda$4863 broad and narrow, \oiii$\lambda$5007,4959, \MgII$\lambda$2800 broad and narrow, CIII]$\lambda$1908, CIV$\lambda$1549 broad and narrow, Ly$\alpha\lambda$1215 broad and narrow.
We run all of these fits using Monte Carlo simulation based on the actual observed spectral error array, which in turn yields an error array for all our decomposition fits. An example spectral decomposition is shown in Figure\,\ref{f:024932model}.


The host galaxy fits used in PyQSOFit are limited to rest-frame wavelengths between 3450 -- 8000\AA.  Due to this limitation, to fit the \MgII\, line complex, we also run PyQSOFit separately on all our spectra and epochs with host decomposition off. 

We do not fit a polynomial continuum, as we found it often competed strongly with the power law continuum, yielding unreasonable fits for both continuum components.  At first, we experimented with fitting a host galaxy model only to the dim state spectrum, since it should be of highest contrast there, and easiest to fit without contamination from the quasar continuum.  However, we found that applying the best-fit dim state host model to all epochs often resulted in poor overall fits, likely due to differing combinations of seeing and spectral aperture for different epochs.  For this reason, we fit a host galaxy component to every spectral epoch separately.

\figsetstart
\figsetnum{1}
\figsettitle{Light curves and spectra for 19 CLQs from the TDSS}

\figsetend

\subsection{Defining and Identifying Bona Fide CLQs}
\label{s:clqdef}

For purposes of comparison between quasars, spectral epochs, or different studies of quasar samples, it is crucial to have a common definition of what we mean by the CLQ phenomenon. For instance, it is insufficient to merely say that ``broad \Hb\, disappeared", since either a visual impression or a measurement can be strongly affected by S/N. 

After running PyQSOFit on all of our spectra, we use the fits for the full continuum model (power law, the UV and optical Fe component, and Balmer continuum models) along with the galaxy host model if applicable.  
We first subtract the full continuum model and the masked\footnote{The masked host fit excludes strong quasar emission line regions, as described in the PyQSOFit code.} galaxy host fits from each observed spectrum to arrive at a quasar line flux spectrum.  For each such spectrum, we rebin both the flux spectrum and corresponding variance spectrum to 2\AA/pix so that different epoch spectra can be directly compared.  

\begin{figure}[h!]
\label{f:024392DimBright}
\centerline{
\includegraphics[angle=90,scale=.3]{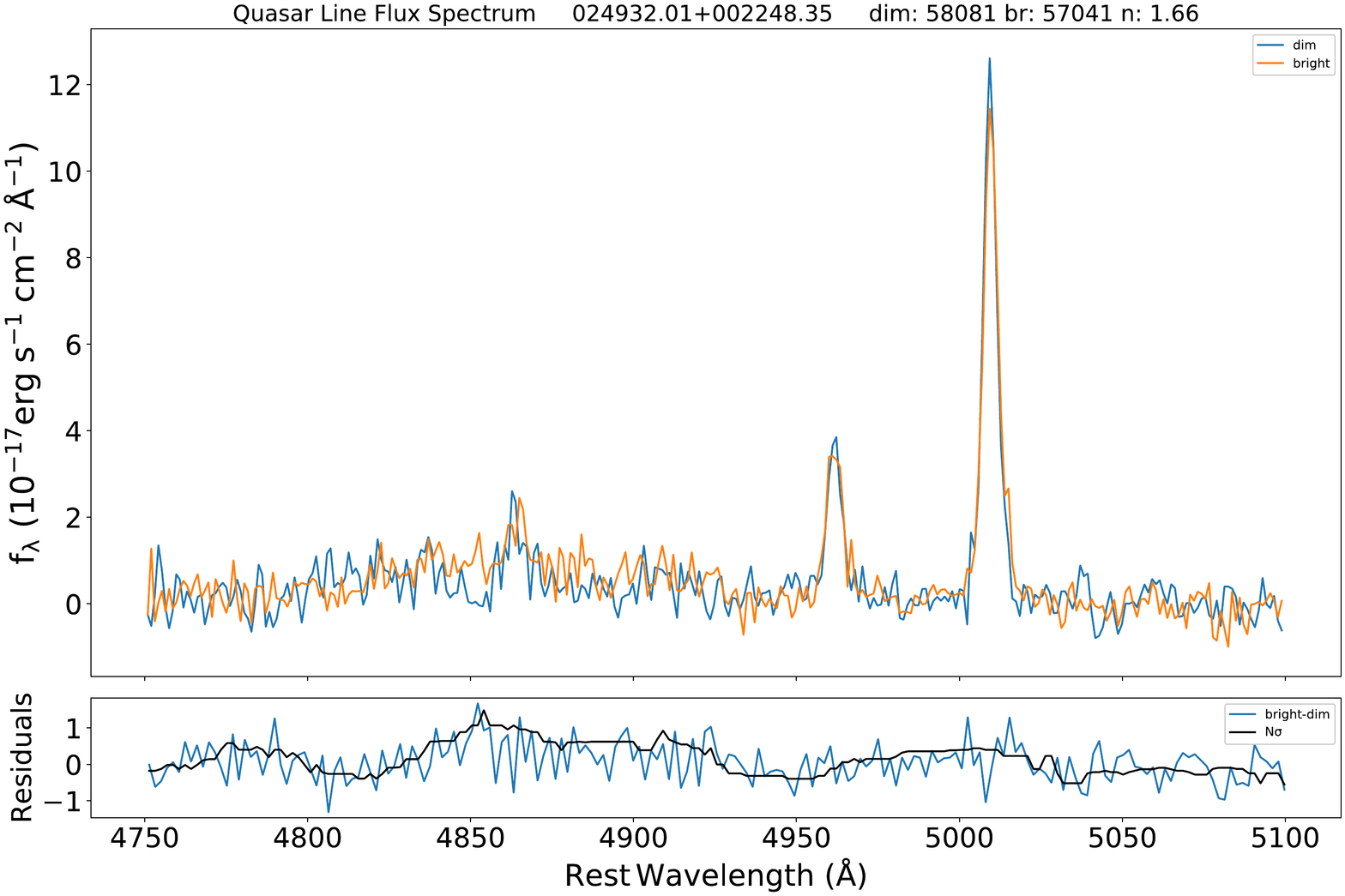}
\includegraphics[angle=90,scale=.3]{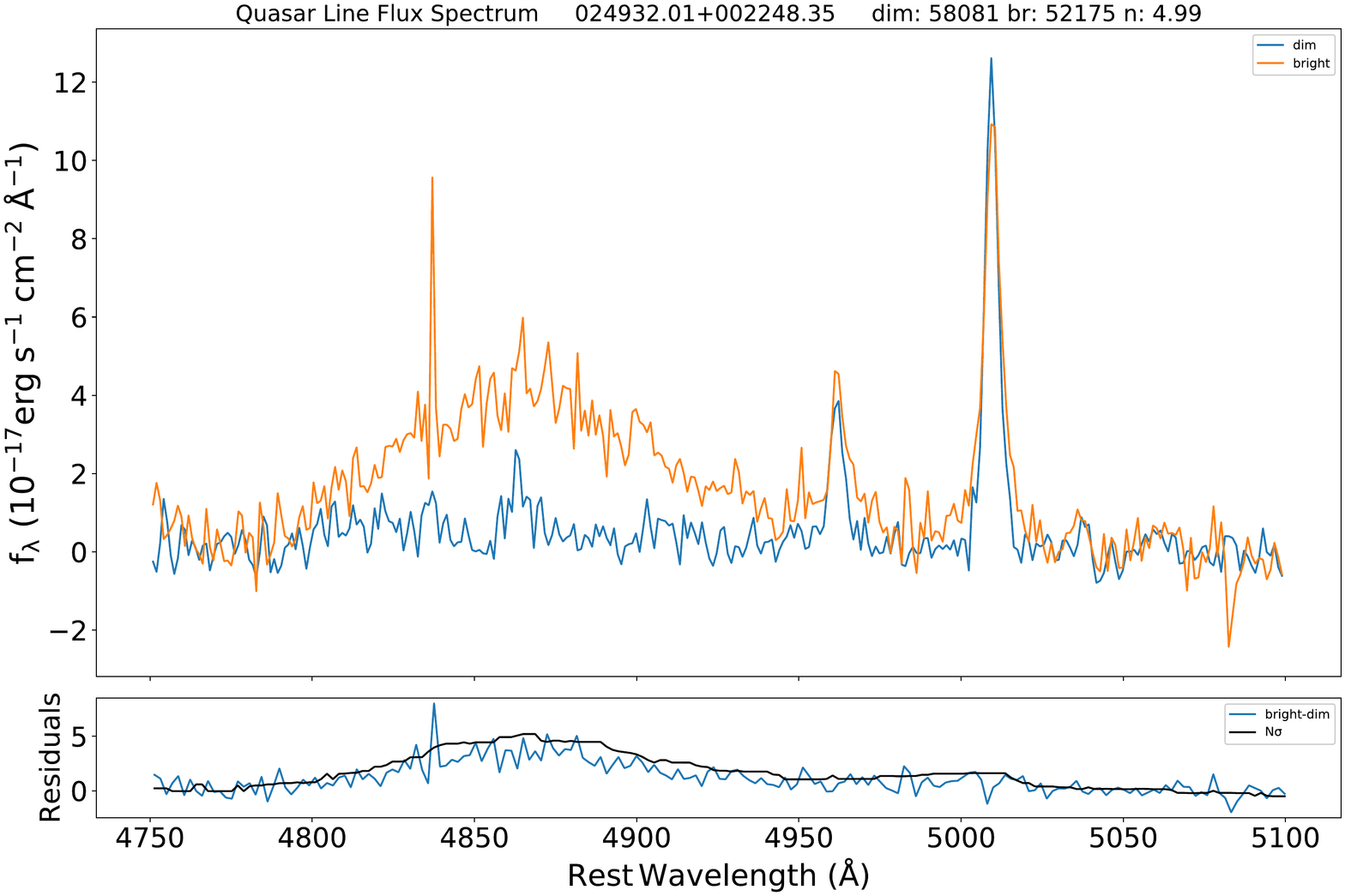}}
\caption{\footnotesize{Spectra of the CLQ SDSS\,J024932.01+002248.3 in the \Hb\, region, shown in the rest frame, with the best-fit continuum subtracted from all epochs.  \emph{Left:} Overplot of two dim spectral epochs, MJD 57041 (orange line), over our declared dim state spectrum at MJD 58081 (blue line).  The difference spectrum is shown in the lower panel (blue line), along with its smoothed version in units of the uncertainty $\sigma$ (black line).  The \Nsigma\, measurement we derive of 1.66 reflects the small change seen.
 \emph{Right:} Overplot of the brightest spectral epoch, at MJD 52175 (orange line), over the same dim state spectrum at MJD 58081 (blue line).  In accord with the strong evident change in the strength of the broad \Hb\, line, the \Nsigma\, measurement we derive here of 4.99 (the maximum value of the smoothed black line in the lower panel) affords a CLQ designation according to our criteria. }}
\end{figure}

In the column marked ``State" of Table\,\ref{t:clqcans}, we denote our choice of dim and bright spectral epoch for each quasar by D and B, respectively. The dim state is chosen as that with the lowest 3240\AA\, continuum luminosity $L_{3240}$, and marked with D in the State column.  For all other spectra, we run our \Nsigma\, calculation on this quasar line flux array,  at every pixel across the \Hb\, region from 4750 -- 4940\AA, as follows: 
\begin{equation} \label{eq1}
N_\sigma = (f_{bright} - f_{dim})/\sqrt{\sigma_{bright}^2 + \sigma_{dim}^2}
\end{equation}
Here, $f$ is the flux in \flamcgs\ in the pixel, whereas $\sigma$ is the spectral variance, including as usual the propagated uncertainties from statistical and instrumental noise and sky background subtraction. The \Nsigma\, array is then smoothed with a median filter using a kernel size normally of 16 pixels (32 divided by the sampling rate of the spectrum in \AA/pixel). We then subtract the $N_\sigma(4750)$ value from this heavily smoothed array, and thereby find the maximum relative value of \Nsigma, which is tabulated in Table\,\ref{t:clqcans}. This is the same method used in \citet{MacLeod19}, as inspired by a similar usage in \citet{FilizAk12}.  An example of normalized dim and bright state spectra overplotted can be seen in Figure\,\ref{f:024392DimBright}.

We generally choose for a CLQ determination the bright state spectral epoch with the largest value of \Nsigma. This is also the bright state we use wherever a comparison is made between just two epochs for each quasar, such as when plotting \Hb\, flux changes versus luminosity in Figure\,\ref{f:delta_hb}. Our selected bright state epochs are marked B in the State column of  Table\,\ref{t:clqcans}.  In very few cases for quasars with more than two spectral epochs, rather than using \Nsigma\, to select the bright spectral epoch, we instead choose the spectral epoch with the highest $L_{3240}$, if  visual inspection of the \Nsigma\, spectral overplots or the light curve trends present compelling evidence for that choice. For example, for J021359.79+004226.81, we chose MJD 51816 because while its \Nsigma\ value of 6.54 is not as high as that of 9.43 for MJD 57043, the enhanced blue spectral continuum is much stronger at MJD 51816.

\begin{figure}[h!]
\label{f:015629}
\centerline{
\includegraphics[angle=0,scale=.22]{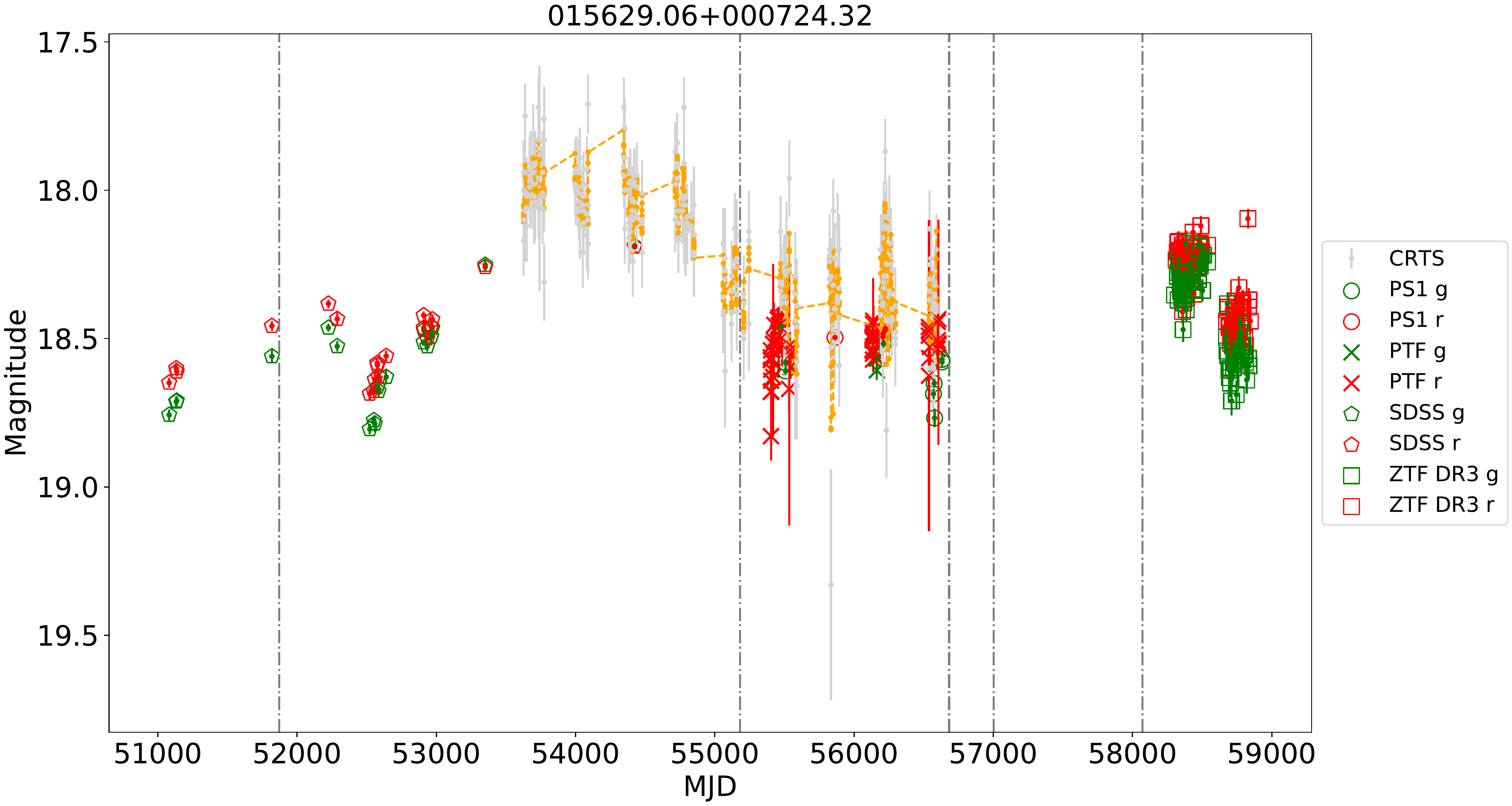}
\includegraphics[angle=90,scale=.32,trim=3.8cm 0 0 0cm]{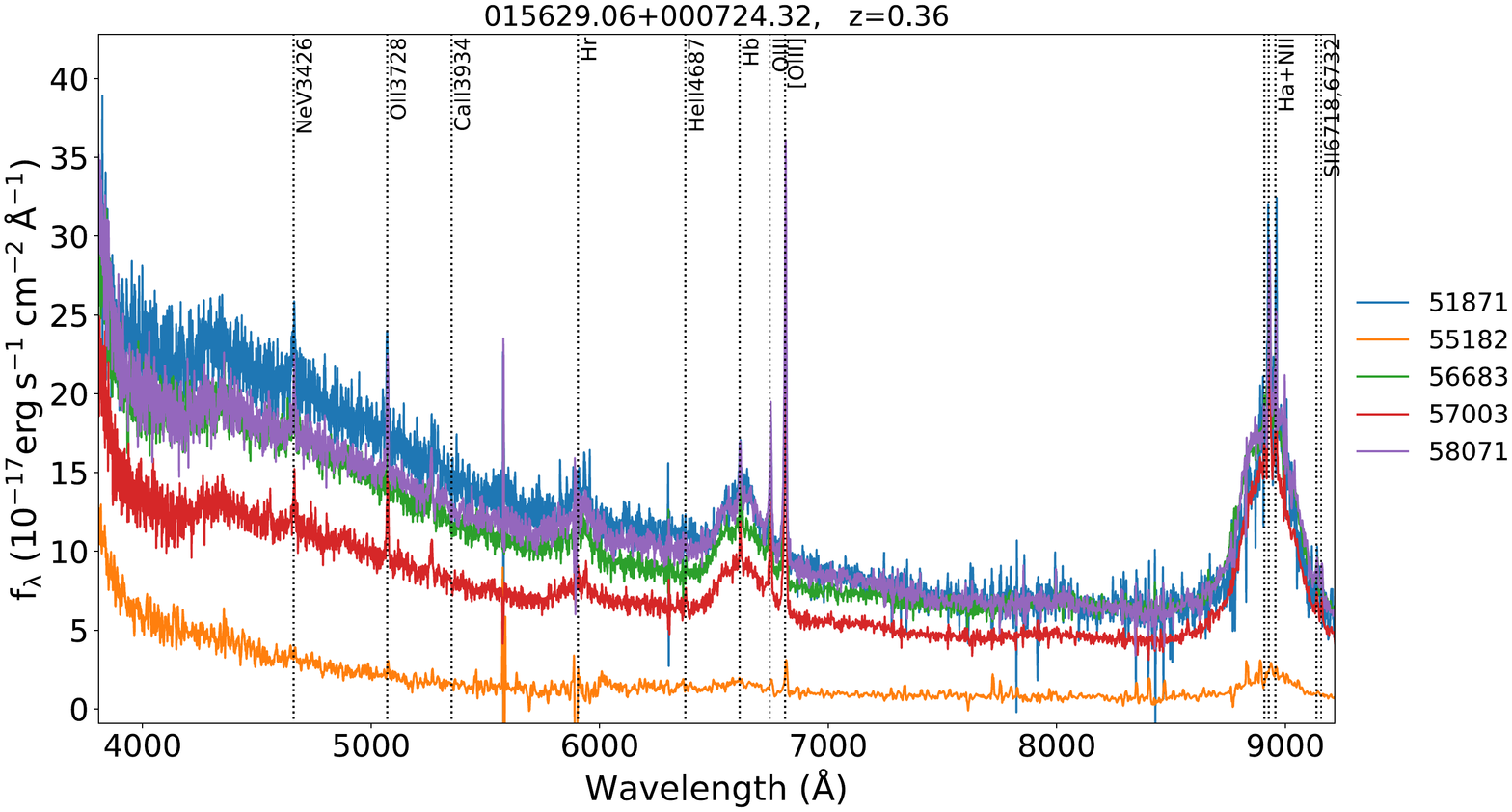}}
\vspace{0.35cm}
\caption{\footnotesize{A potentially spurious CLQ candidate, SDSS\,J015629.06+000724.32 was chosen by visual inspection as a potential CLQ primarily because of the low flux and lack of broad line emission seen in spectral epoch 55182. {\em LEFT:} Photometry and spectral epochs (shown as vertical lines) for the CLQ candidate SDSS\,J015629.
{\em RIGHT:}  The faint spectrum from MJD 55182 does show some narrow emission lines, and a blue upturn similar in shape to other spectral epochs.  However, the factor of 2 -- 3 drop in flux is not reflected in the light curve.  The faint spectrum thus probably represents a rare case of a misplaced spectral fiber.  Later epochs looked brighter and more similar in flux, so there was no point in dedicated spectroscopic followup, as performed for some other objects with suspect spectral epochs.  Therefore we've excluded this object from further analysis, but present it here simply as a warning to other researchers studying large samples of objects with repeat spectroscopy.
 }}
\end{figure}

Note that we do not use identical criteria to \citet{MacLeod19}, because they require {\em a priori} that $|\Delta g|>$1 mag and $|\Delta r|>$0.5 mag), and then verified CLQ status from pursuant followup spectroscopy.  In comparison, our candidate sample begins with multi-epoch spectroscopy, and is, when necessary, verified by multi-epoch photometry.

Our use of a significance threshold as a criterion for determining CLQ status, while well-defined and efficient, is not ideal in that it is dependent on the spectroscopic data quality rather than on intrinsic changes to the quasar spectra over time.  We suggest a potential intrinsic definition later in \S\,\ref{s:future}.

\begin{figure}[ht]
\centerline{
\includegraphics[scale=0.8]{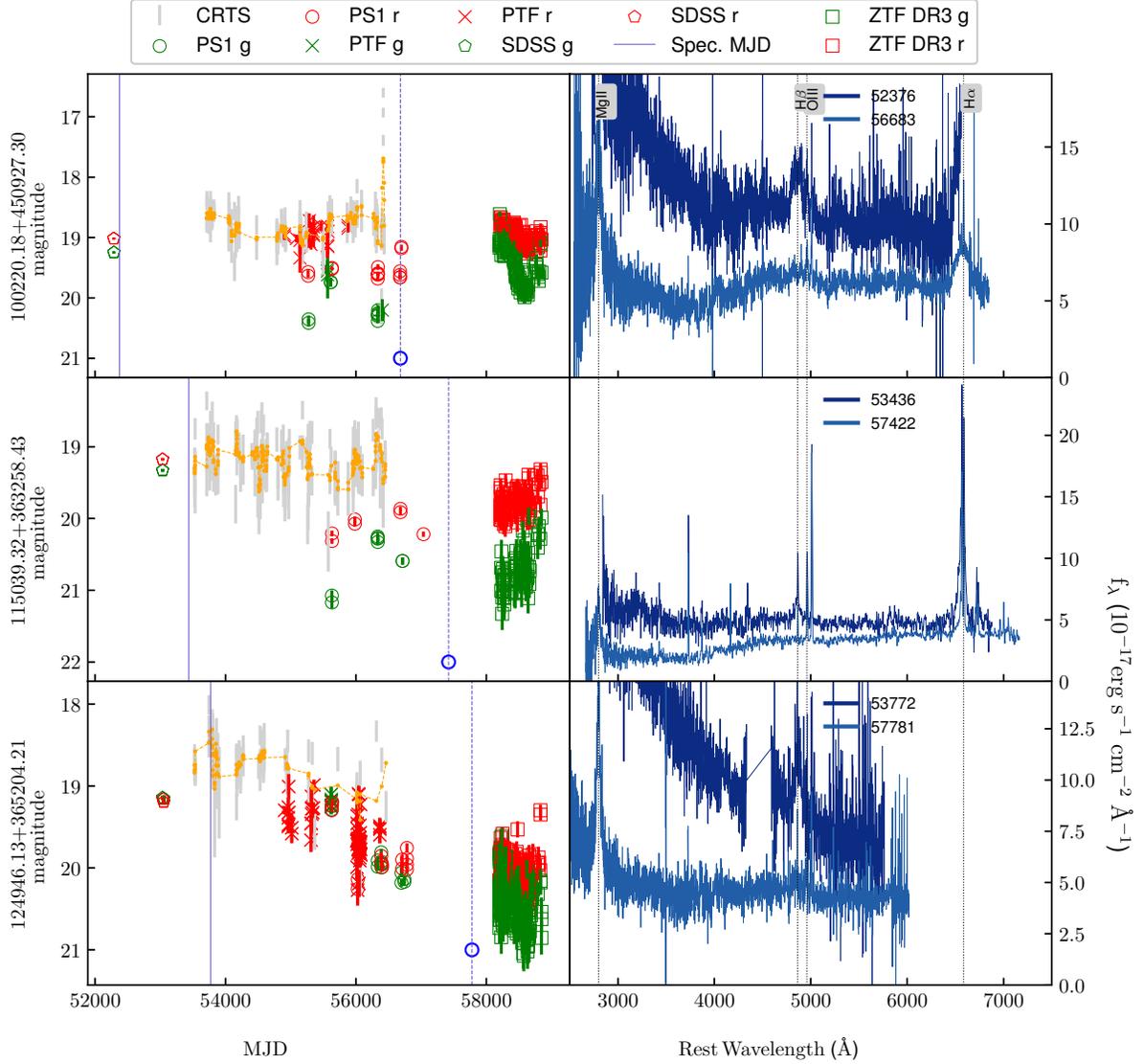}}
\vspace{-5.5cm}
\caption{Light curves and spectra for less significant TDSS CLQs. For these three CLQ candidates, $2<\Nsigma<3$.  While rather convincing visually, their lower quality spectra
prevent a firm CLQ classification in our adopted scheme.  Symbols are as explained in they key at the top of the figure, or in the caption of Figure\,\ref{f:lcspecs}.
}
\label{f:lcspecs2}
\end{figure}


\section{Results} \label{s:results}

In Table\,\ref{t:clqcans}, we summarize data and measurements of our visually-identified CLQ candidate sample.  Each quasar has at least two spectral epochs.  Under the SDSS ID of each quasar, we note the TDSS subprogram under which criteria it was targeted for repeat spectroscopy (see also Table\,\ref{t:TDSStypes}).  For each spectral epoch, we list the facility used to obtain the spectrum, the modified Julian day (MJD), and luminosities and their uncertainties derived from our model fits for 3240\AA\ continuum, broad \Hb, 2800\AA\ continuum, and \MgII\ emission, as available, in units of 10$^{42}$erg\,s$^{-1}$.  The \Nsigma\, value we derive, as described above in \S\,\ref{s:clqdef} is listed for every epoch but that of the dim state.

In the final (Notes) column of Table\,\ref{t:clqcans}, we confirm CLQ status for the bright state spectrum with the largest \Nsigma\, value above three, such that the CLQ designation appears at most once in the table for any given quasar. We find 19 CLQs in this TDSS sample, of which four were previously noted as CLQs in \citet{MacLeod19} (J1055+2425, J1113+5313, J1434+5723) or \citet{MacLeod16} (J0023+0035).  In most cases, we have additional spectral epochs available.

The light curves of the 19 CLQs span as much as 20 years, and show strong diversity in the character of variability.
We sometimes find significant changes in brightness on surprisingly short timescales; J002311.06+003517.53 dims by $\Delta g\sim 1.5$\,mag within $\sim 1000$\,days. J105513.88+242553.69 shows instead a slow, steady dimming over 15 years by $\Delta g\sim 2$\,mag. J143455.30+572345.10 dims by as much, on a similar timescale, but with sparser photometric coverage. 

By contrast, some quasars show relatively minor shifts in photometry, yet still show changes in $L_{3240}$ and \Nsigma\ that are significant, such as J024508.67+003710.68, which may have dimmed only about $\Delta g\sim 0.3$\,mag.  However, its light  curve, and those of many CLQ candidates, do not always sample epochs close to the spectroscopy.  Sometimes, large, rapid changes are seen in photometry, but without nearby spectral epochs; J111329.68+531338.78 re-brightens in recent epochs by $\Delta g\sim 1.5$\,mag within just a year. The variability of individual CLQs is discussed in detail in the Appendix.

Most of the CLQs that we identify are ``turn-off" CLQs, i.e., they satisfy our criteria by dimming with time. This is a consequence of our selection process, since our parent sample is a quasar catalog requiring identification of broad emission lines. We nevertheless find that four of the 19 CLQs are ``turn-on" CLQs, identified as such in the Notes column of Table\,\ref{t:clqcans}: J100302.62+193251.28, J113651.66+445016.48 (also noted in \citealt{MacLeod19}), J224113.54-012108.84, and J234623.42+010918.11.  The latter is particularly striking, showing a brightening of $\Delta g\sim1.5$\,mag early in the light curve.  In such instances, the early epoch pipeline classification as quasar may be due to broad emission lines other than \Hb, but in most cases, the continuum and broad \Hb\, emission were simply weaker, and have increased significantly during our monitoring.

There are three quasars that distinctly appear by visual inspection to be CLQs based on both spectra and light  curves, but we measure only $2\leq \Nsigma <3$. We mark these in the Notes column of Table\,\ref{t:clqcans} as VIc. Two of them were previously published as bona fide CLQs - J1002+4509 \citep{MacLeod16} and J1150+3632 \citet{Yang18}. These would likely have passed our CLQ criteria, had the spectral S/N been higher.  We show the light curves and spectra for these objects in Figure\,\ref{f:lcspecs2}.  

\begin{figure}[!ht]
\label{f:zmags}
\centerline{
\includegraphics[scale=.4]{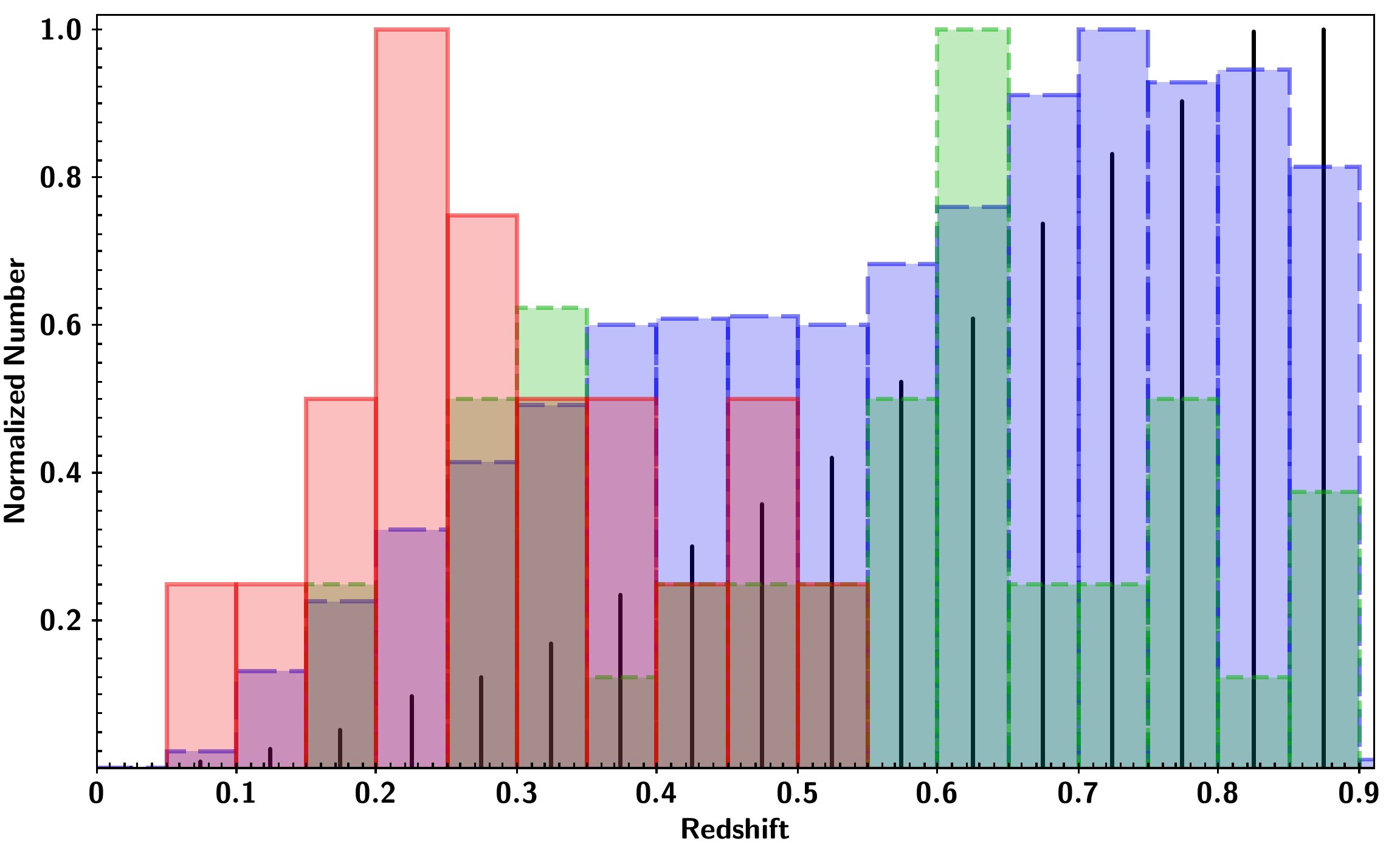}
\includegraphics[scale=.4]{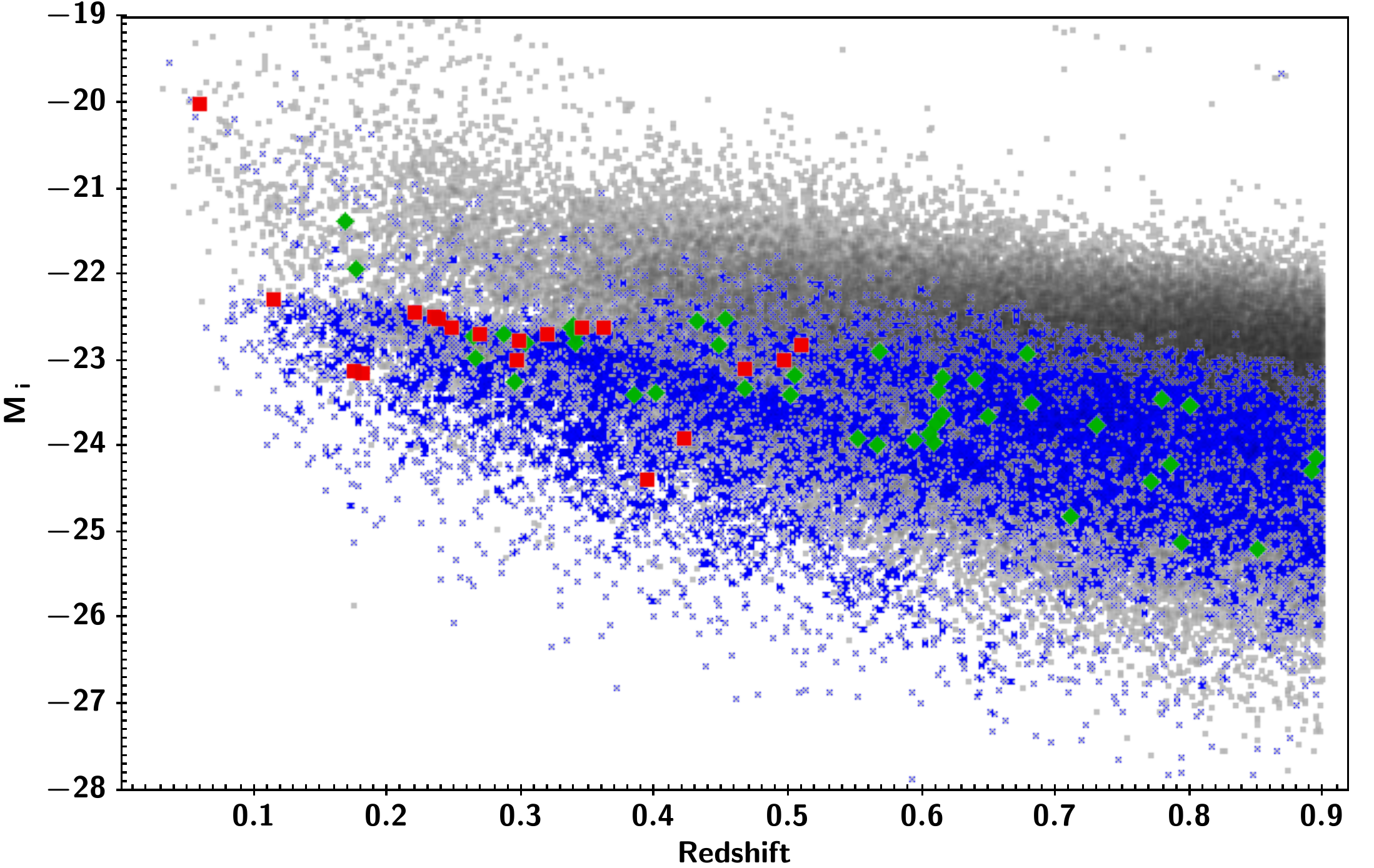}}
\centerline{
\includegraphics[scale=.4]{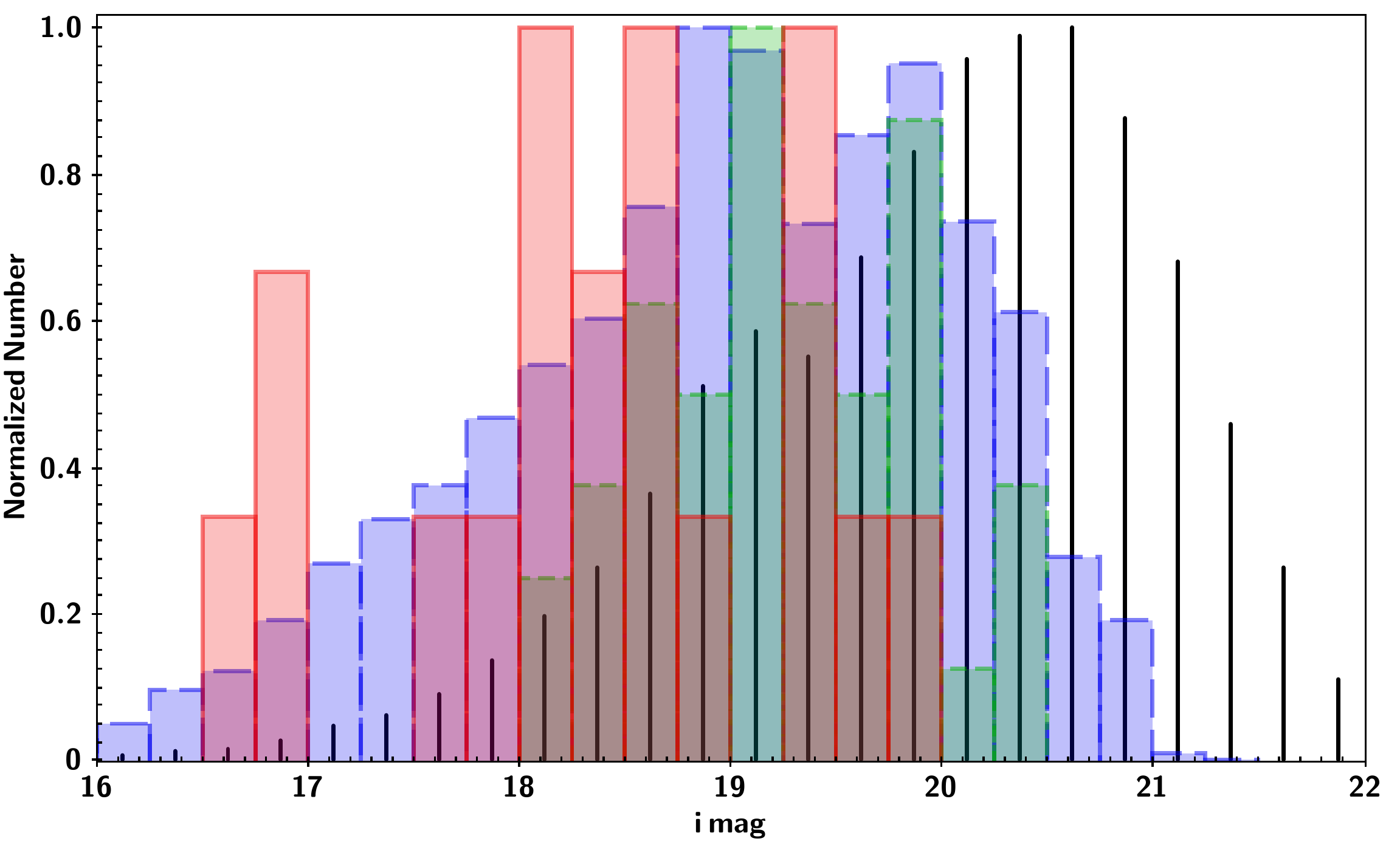}
\includegraphics[scale=.4]{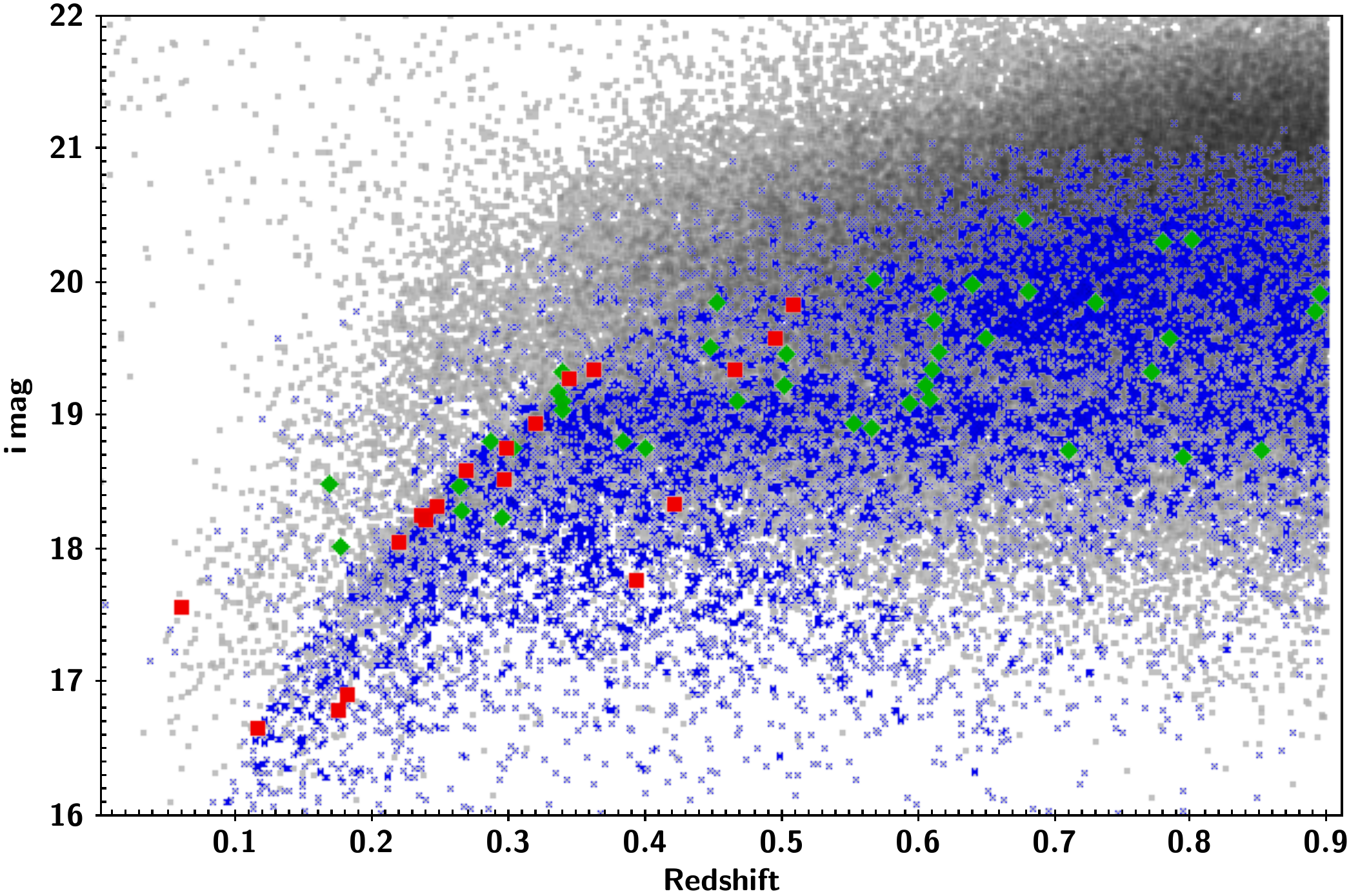}
}
\caption{\footnotesize{{\em TOP LEFT:} Redshift histogram for the full SDSS DR16 quasar sample below $z=0.9$ (about 105,000 QSOs, black spikes), 11,483 TDSS FES/RQS targets (blue bars, long-dashed lines), 61 CLQ candidates (green bars, short-dashed lines), and 19 CLQs (red bars, solid lines) with bins of width 0.05.  For clarity, each histogram is normalized so that it peaks at one. There are no bona fide CLQs in our sample beyond $z=0.5$.  {\em BOTTOM LEFT:} Histogram of $i$ band magnitudes.  {\em TOP RIGHT:} Absolute $i$ band magnitude $M_i$ versus redshift for the full DR16 sample (small grey small points), TDSS FES/RQS targets (small blue crosses), CLQ candidates (green diamonds) and CLQs (red squares). {\em BOTTOM RIGHT:} SDSS $i$ band magnitude versus redshift.  
}}
\end{figure}

While our candidate CLQs extend to redshift 0.9, the CLQs we identify are all at $z<0.5$.  Figure\,\ref{f:zmags} shows histograms of the full DR16 quasar sample between $0<z<0.9$ (black spikes), our TDSS FES$+$RQS targets (blue), CLQ candidates (green), and CLQs (red).  For clarity, each histogram is normalized so that it peaks at one.  We also plot the absolute and apparent $i$-band magnitudes against redshift for the same samples.

Despite small number statistics, these plots indicate that CLQs are likely from a typical quasar population, and even when compared to our TDSS FES$+$RQS target sample with its brighter magnitude limits and/or variability criteria, tend to be found at lower redshifts and brighter magnitudes, likely due to our S/N criterion. The $i$ band magnitudes shown in Figure\,\ref{f:zmags}  are taken from SDSS, so are typically from an earlier, brighter phase. The $i$ band, with effective wavelength 7628\AA\, \citep{fukugita96} samples rest-frame wavelengths near 5900\AA\, (5330, 4737, 4487\AA) for the median redshifts of the CLQs (0.3; candidates 0.43; FES$+$RQS 0.61, full DR16 sample 0.7).  This $i$ band is a reasonable choice, because these rest-frame wavelengths are normally redward of $\sim 4000$\AA, where the quasar continuum changes strongly between dim and bright states for quasars.

\subsection{Luminosity Changes, Timescales and Flux Ratios}

\begin{figure}[h!]
\label{f:dMJDdLHb}
\centerline{
\includegraphics[angle=90,scale=.33]{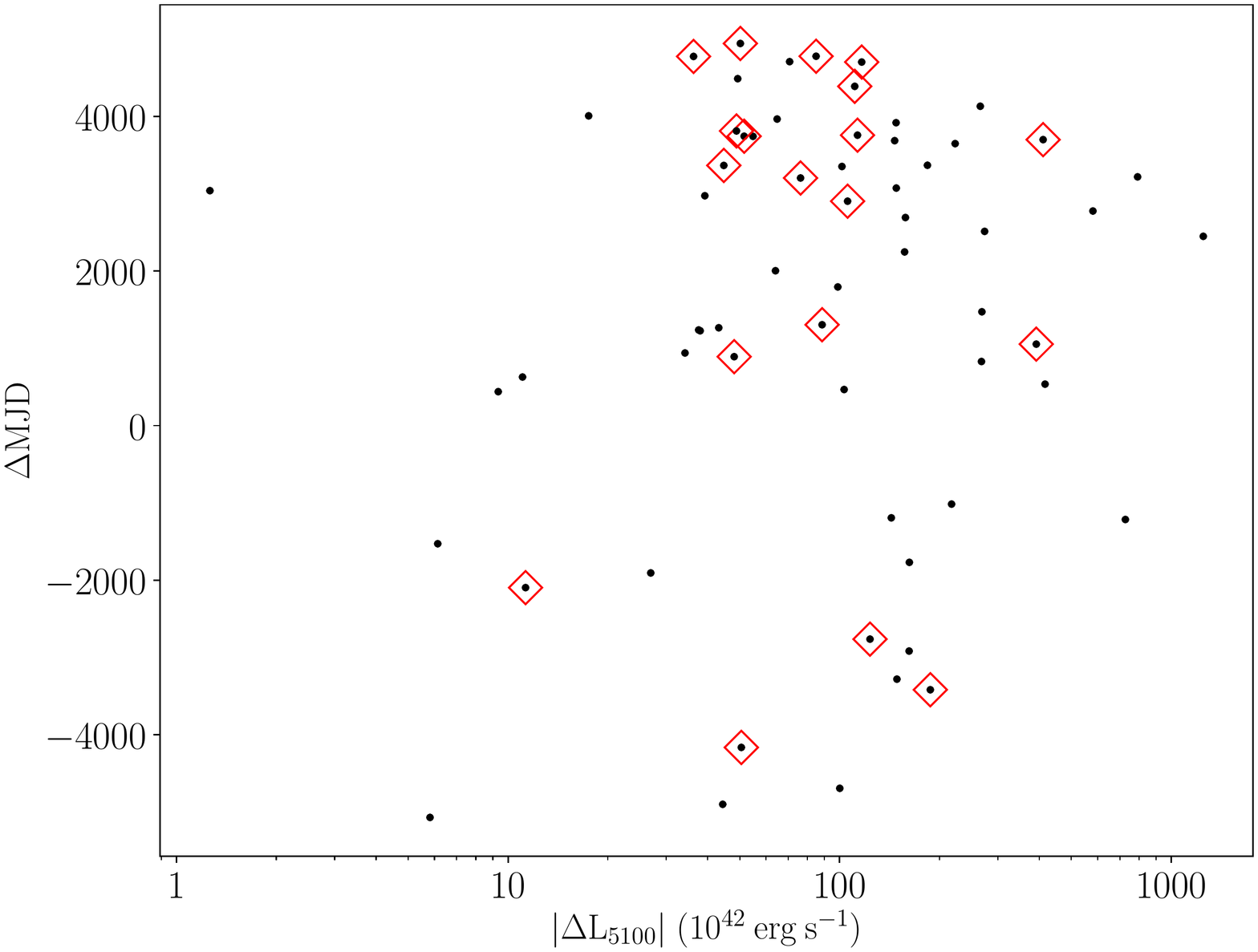}
\includegraphics[angle=90,scale=.33]{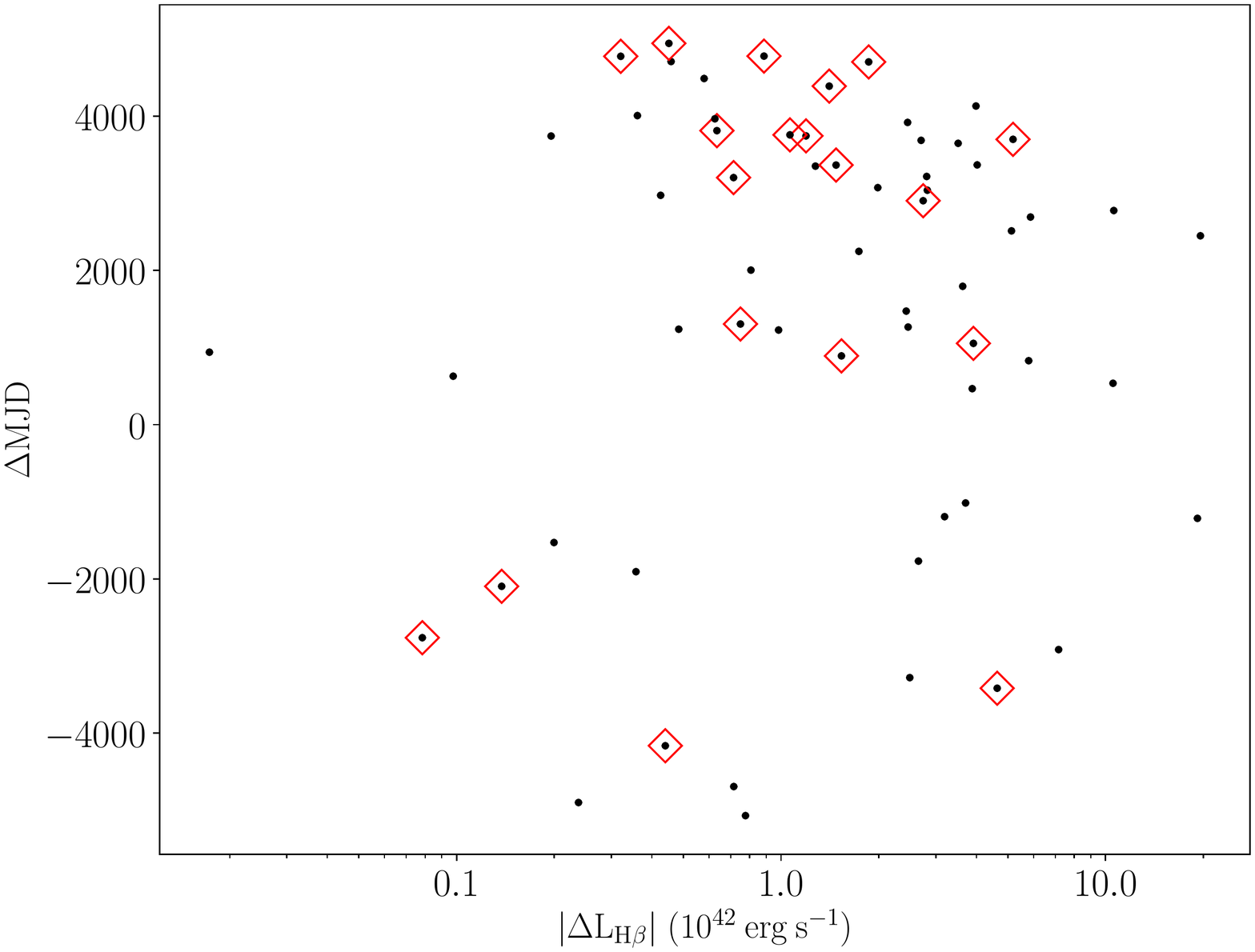}}
\caption{\footnotesize{Time in the rest-frame between spectral epochs $\Delta$MJD/(1+$z$) versus luminosity change for our full CLQ candidate sample (black dots).   $\Delta$MJD is calculated here as dim minus bright epoch, so is positive for ``turn-off" objects that dim with time.  As our parent sample consists of SDSS broad-line quasars, the majority of both CLQ candidates and bona fide (\Nsigma\,$>3$) CLQs (marked with open red diamonds) have $\Delta$MJD$>0$. The luminosity changes  are derived as the difference between the selected bright state spectrum having the largest measured 3240\AA\, continuum flux, and the dim state spectrum, each indicated in Table\,\ref{t:clqcans}. The available time spans between epochs reach up to a dozen years.  \emph{Left:} Timespan vs. the change in the luminosity at 5100\AA, in units of 10$^{42}\lcgs$.  More CLQs are found with larger $\Delta$MJDs. However, no clear trend is seen overall in luminosity change vs. timespan probed.  \emph{Right:} Timespan vs. the change in the luminosity in the broad \Hb\, emission, which is of order a hundred times lower than the continuum emission changes. Again, no strong trends are seen.  }}
\end{figure}

Figure\,\ref{f:dMJDdLHb} shows the rest-frame
time interval between spectral epochs $\Delta$MJD versus luminosity change for our full CLQ candidate sample, both for 5100\AA\, power-law continuum and broad \Hb. 
The only clear trend is that a larger fraction of candidates are bona fide CLQs at longer epoch separations, for turn-off CLQs ($\Delta$MJD$>0$). Statistics are inadequate in our sample to judge such trends in turn-on CLQs.  Interestingly, it is not readily apparent that larger multi-year time spans lead to larger luminosity changes in our sample. \citet{Luo2020} found that even among EVQs (i.e., photometrically-selected to have $\Delta g>1$mag), only a small fraction ($<4\%$) showed monotonic variability over a $\sim$16\ year time span\footnote{Most (57\%) showed complex behavior, while $\sim$40\% showed a single broad peak or dip in the light curve.}

For quasars in general, stochastic variability, with a structure function described by an asymptotic long-timescale rms variability SF$_{\infty}\sim$0.2\,mag and a rest-frame damping timescale $\tau\sim200$\, days provides a reasonable description of quasar variability \citep{MacLeod2010}.  EVQs are better represented by SF$_{\infty}\sim$0.4\,mag and timescale $\tau\sim600$\, days, but the relative fractions of longer-term variability trends (monotonic vs. single peak/dip vs. complex) are not well-reproduced by simple stochastic variability models \citep{Luo2020} . 

Our results indicate that large multi-epoch spectroscopic samples (or large time domain photometric samples with prompt spectroscopic follow-up of strong variables) are likely to find a small fraction of quasars to be CLQs, but that CLQs may be found spanning a wide range of timescales sampled.  Whether epoch separations less than a few years will reveal substantial CLQ behavior is not clear, since our sample is sparsely populated in that range (see Figure\,\ref{f:dMJDdLHb}).  However, the larger reverberation mapping program planned for SDSS-V could detect and track perhaps 2000 quasars over more than about 150 epochs to detect a larger number of short-timescale CLQs, such as SDSS\,J141324+530527, reported by \citet{Dexter2019}.
The medium tier of the All Quasar Multi-Epoch Spectroscopy
(AQMES) survey of SDSS-V will obtain hundreds more spectra with rest-frame intervals of months (see S\,\ref{s:future}).

\subsection{Broad Line Variability}
\label{s:belvar}

\begin{figure}[h!]
\label{f:delta_hb}
\centering
\includegraphics[angle=90,scale=.7]{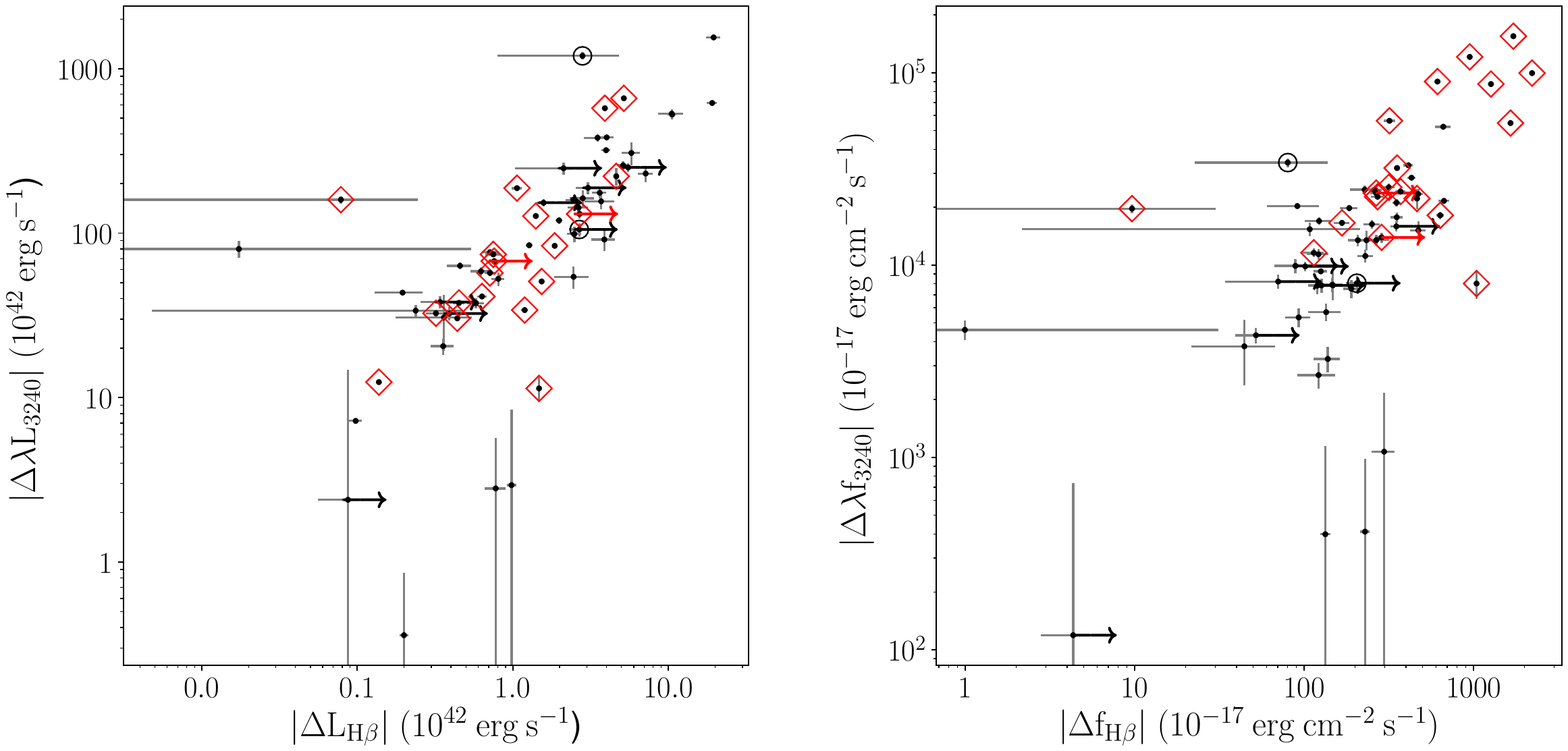}
\vspace{-3cm}
\caption{
The change in power-law luminosity (left) or flux (right) at 3240\AA, 
plotted against the change in broad \Hb\, luminosity or flux, for our full CLQ candidate sample (black dots). Values are derived between the selected dim state spectrum and the bright state spectrum having the largest measured $f_{\rm 3240}$ continuum flux. Horizontal arrows indicate lower limits on CLQs, corresponding to upper limits on the amount of dim-state broad H$\beta$ flux. Open red diamonds mark those quasars designated as CLQs herein due to measurements of \Nsigma$>3$.  Values are derived as the ratio between the selected bright state spectrum having the largest measured 3240\AA\, continuum flux, and the dim state spectrum, as indicated in Table\,\ref{t:clqcans}.  The relationship is strong and clear in both panels, suggesting a direct relationship between the ionizing flux and broad \Hb\, emission. 
}
\end{figure}

In Figure\,\ref{f:delta_hb}, we plot broad \Hb\ line strength against 3240\AA\ continuum strength, showing a strong correlation in both luminosity and flux space. The correlation is not surprising, since \Hb\ is the optical emission line most used for reverberation mapping studies. The success of RM confirms the strong influence of photoionization in the BLR.  
%
The ionizing source is assumed to be small (relative to the BLR) and quasi-isotropic, \Hb\ is indeed seen to react in many quasars to $L_{5100}$, with a delay from about 1 -- 100 days.  Thus in the great majority of cases here, the broad \Hb\ line flux has had time to react to continuum changes between the bright and dim states measured.

The \MgII\ emission line plays an important role in quasar studies, as it is used increasingly for RM estimates of \Mbh\, (e.g., \citealt{shen16}), but also for virial (single-epoch) estimates when \Hb\, becomes inaccessible from the ground ($z>1$; e.g., \citealt{Vestergaard2009,wang09,shen11}).

During our visual inspection of these multi-epoch spectra of strongly variable quasars, we noted that the variation in the strength of the \MgII\, emission line is generally much lower than that of \Hb.  For instance, the \MgII\ line may remain relatively strong even in the dim state, after the continuum and broad \Hb\ emission have significantly faded. The relatively weak response of \MgII\ to continuum changes has been noted repeatedly in the literature (e.g., \citealt{cla91,Kokubo2014,cac15,Zhu2017}). 

\citet{Homan20} presented a large study of \MgII\ variability in extremely variable ($|\Delta g|>1$) quasars, and found considerable complexity in different quasars' \MgII\ response to continuum changes, both for line strength and profile. The response of \MgII\ to continuum variations was different not only between quasars, but even between epochs in individual quasars. 

In an RM study, \citet{Sun2015} found that the \MgII\ line tends to vary less than \Hb\ , with a broader response function, and that no clear \MgII\ radius-luminosity relation may  exist for \MgII. This could be because while the Balmer lines are recombination dominated,  at BLR densities, \MgII\ is predominantly collisionally excited.  Furthermore, the \MgII\ BLR may not necessarily end where the photoionization conditions are no longer optimal, but rather at the dust sublimation radius \citet{Guo2020}.

\begin{figure}[h!]
\label{f:delta_mgii}
\centering
\includegraphics[angle=90,scale=0.7]{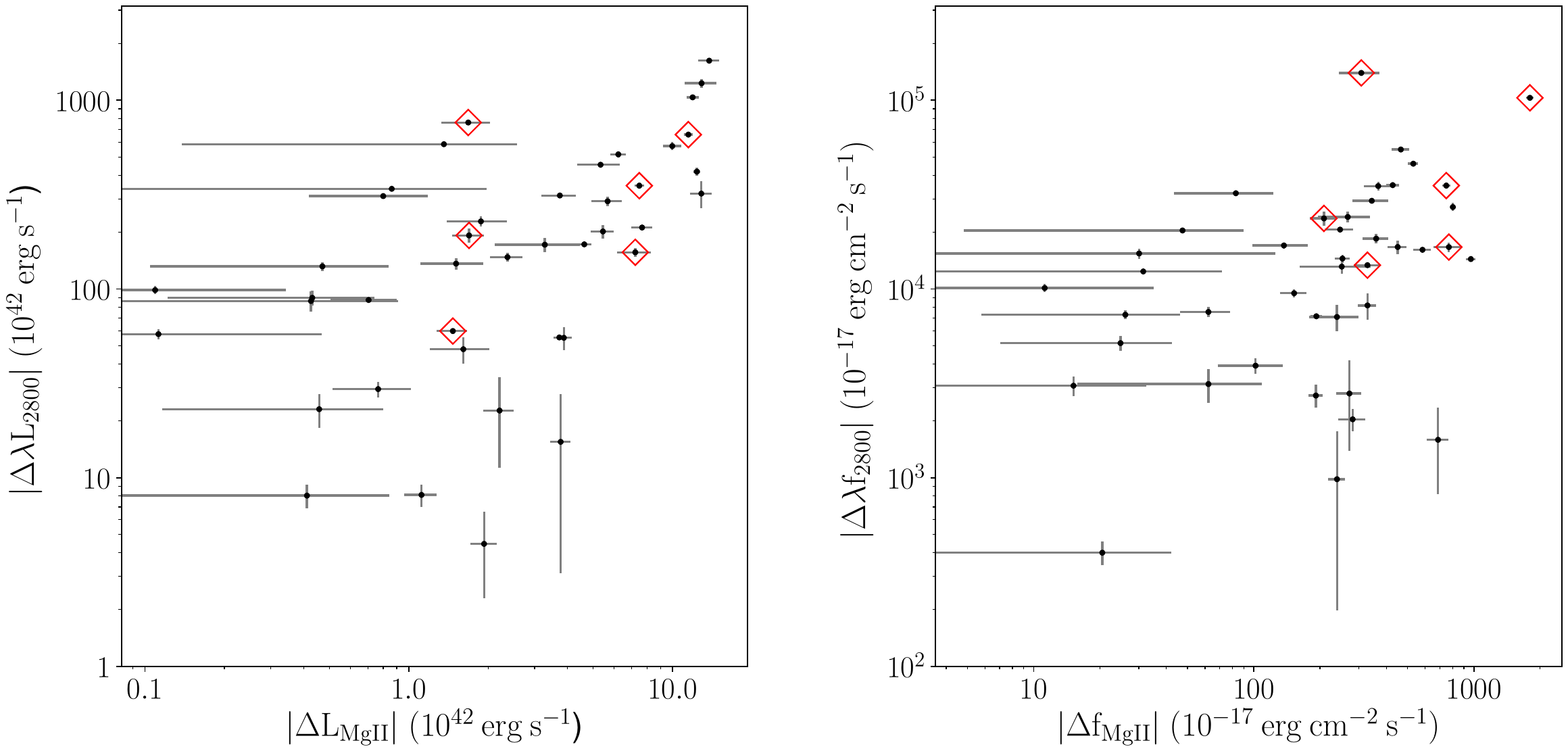}
\vspace{-3cm}
\caption{
The change in power-law luminosity (left) or flux (right) at 2800\AA, 
plotted against the change in broad \MgII\, luminosity or flux, for our full CLQ candidate sample (black dots), wherever $z>0.35$. Values are derived between the selected dim state spectrum and the bright state spectrum having the largest measured $f_{\rm 2800}$ continuum flux.  Open red diamonds mark those quasars designated as CLQs herein due to measurements of \Nsigma$>3$.   The relationship between the ionizing flux and broad \MgII\, emission is significantly weaker than for \Hb\ in Fig~\ref{f:delta_hb}, or perhaps non-existent.}
\end{figure}

Interestingly, \citep{Roig14} searched among some 250,000 luminous galaxies with SDSS/BOSS spectra in the redshift range $0.35<z<1.1$ and found 293 examples ($\sim0.1\%$) with strong, broad \MgII\ lines, but lacking a blue continuum or any broad Balmer line emission. If not for their broad \MgII\ emission, these spectra would normally be classified as Seyfert~2 or LINER galaxies.  If the \MgII\ BLR is larger than the \Hb\ BLR, these could be dim-state CLAGN where the \MgII\ region remains illuminated.  Indeed, \MgII\, changing look AGN have subsequently been discovered by \citet{Guo2019}.


 Significant broad \Ha\ emission sometimes remains after \Hb\ disappears in many CLQs, as evident in some of the $z<0.4$ CLQs shown in Fig\,\ref{f:lcspecs} and also in e.g., \citet{LaMassa15,Yang18,Sheng20}.
By contrast, in the luminous SDSS/BOSS sample of
galaxies with broad \MgII\ emission \citet{Roig14}, there is little evidence for broad Balmer line emission.
They stacked 162 broad \MgII\ galaxies with $0.35 < z < 0.57$
(covering both \MgII\ and \Ha) and detected only narrow Balmer line components (250 km/s for \Ha\ and 150 km/s for \Hb). 
This implies that the primary broad line emitting region size increases from \Hb\ to \Ha, to \MgII.  \citet{grier17} studied \Ha\ and \Hb\ lags in the SDSS RM project and found the \Ha\ lags to be $\sim$40\% longer than for \Hb\ (see their Figure 10). \citet{Homayouni2020} measured \MgII\ lags and also found they exceed \Hb\ lags by a similar amount.  

 \citet{yang20} investigated the variability of the \MgII\, line in a sample of extremely variable quasars (EVQs), finding that, in contrast to the Balmer lines, the FWHM of broad \MgII\, does not react strongly to continuum changes (Figure\,\ref{f:breathing}).  In Figure\,\ref{f:delta_mgii} we plot the change in \MgII\, against continuum, in both luminosity and flux; evidently, the trend is indeed  weak.

\begin{figure}[h!]
\centering
\includegraphics[angle=90,scale=.3]{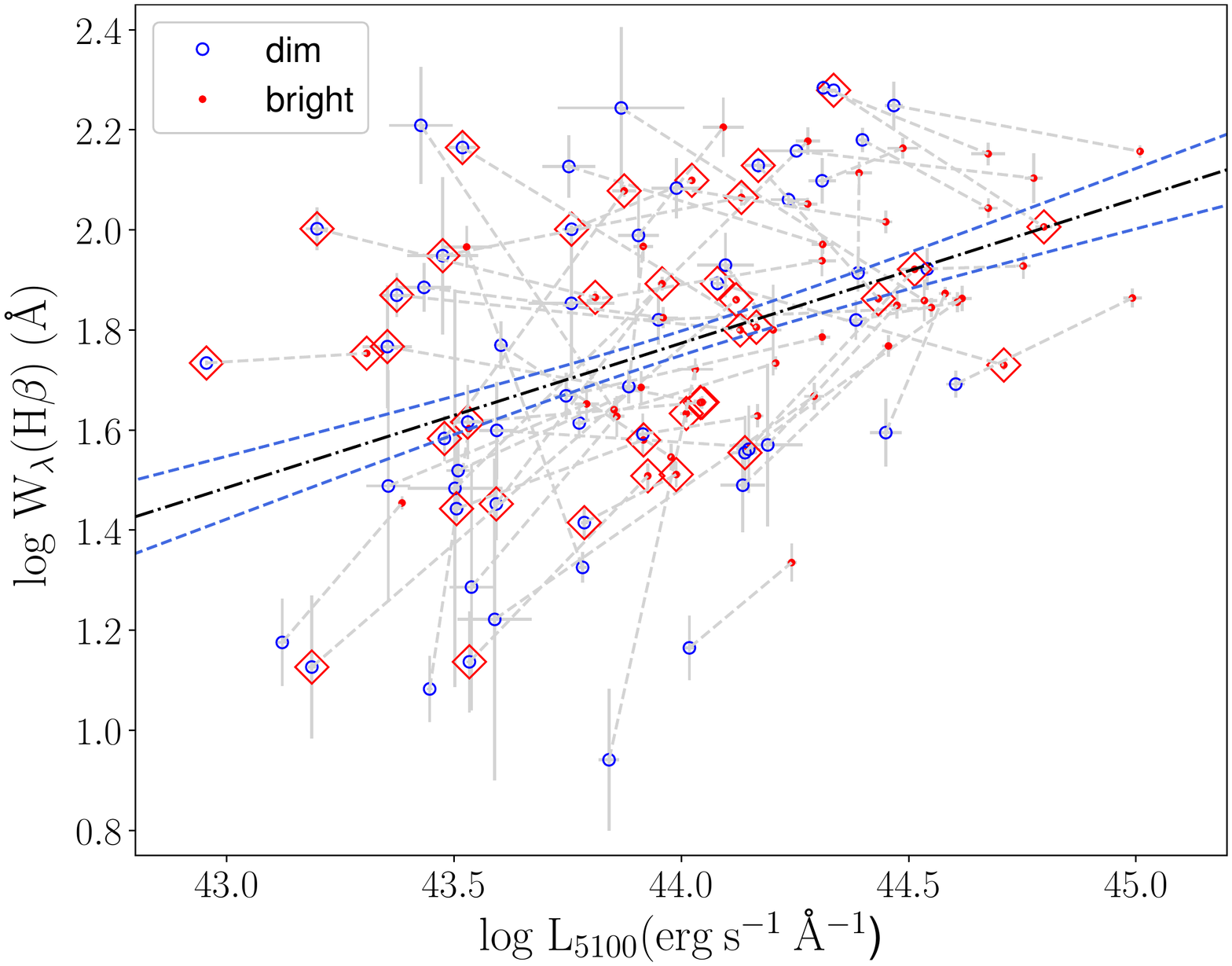}
\includegraphics[angle=90,scale=.3]{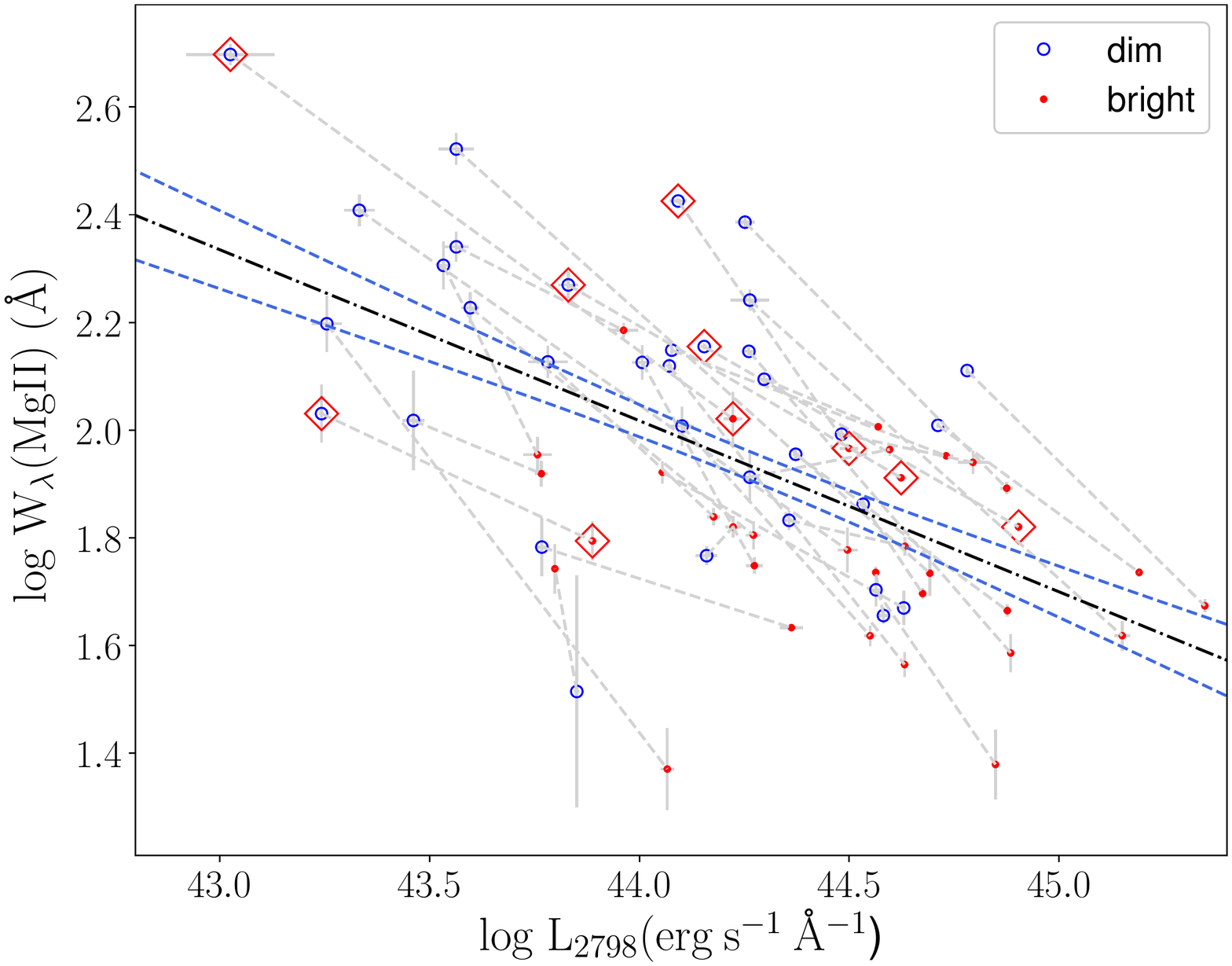}
\caption{
{\em LEFT:}
Equivalent width of the broad \Hb\, emission line vs. luminosity for all analyzed CLQ candidates, i.e., Baldwin effect (Beff) plots.  Both axes use the
best-fit power-law continuum value at rest 5100\AA\, for each quasar and spectral epoch.  Dim states are plotted with open blue circles, and bright states with filled red dots, with a dashed line connecting the two states for each QSO. The CLQs, defined as having broad H$\beta$ change of at
least $N_{\sigma}({\rm H}\beta)>3$, have points over-plotted as red diamonds.  
The global trend for \Hb\ represented by all the points is the ensemble Baldwin Effect, with a bivariate slope of $0.29\pm 0.06$, shown as a black dot-dash line. The light-grey dashed lines between epochs represents the intrinsic Baldwin Effect; the average slope of those lines is much steeper: $2.05\pm 0.07$. 
{\em RIGHT:} Equivalent width of the broad \MgII\, emission line vs. luminosity for all analyzed CLQ candidates, using the best-fit power-law continuum value at 2798\AA\, for each quasar and spectral epoch. 
The global BEff for \MgII\ has a bivariate slope of $-0.32\pm 0.05$, whereas the average iBEff slope of those lines is $-0.8283\pm 0.0008$.  
\label{f:baldwin}}
\end{figure}
\begin{figure}[h!]
\centering
\includegraphics[angle=90,scale=.3]{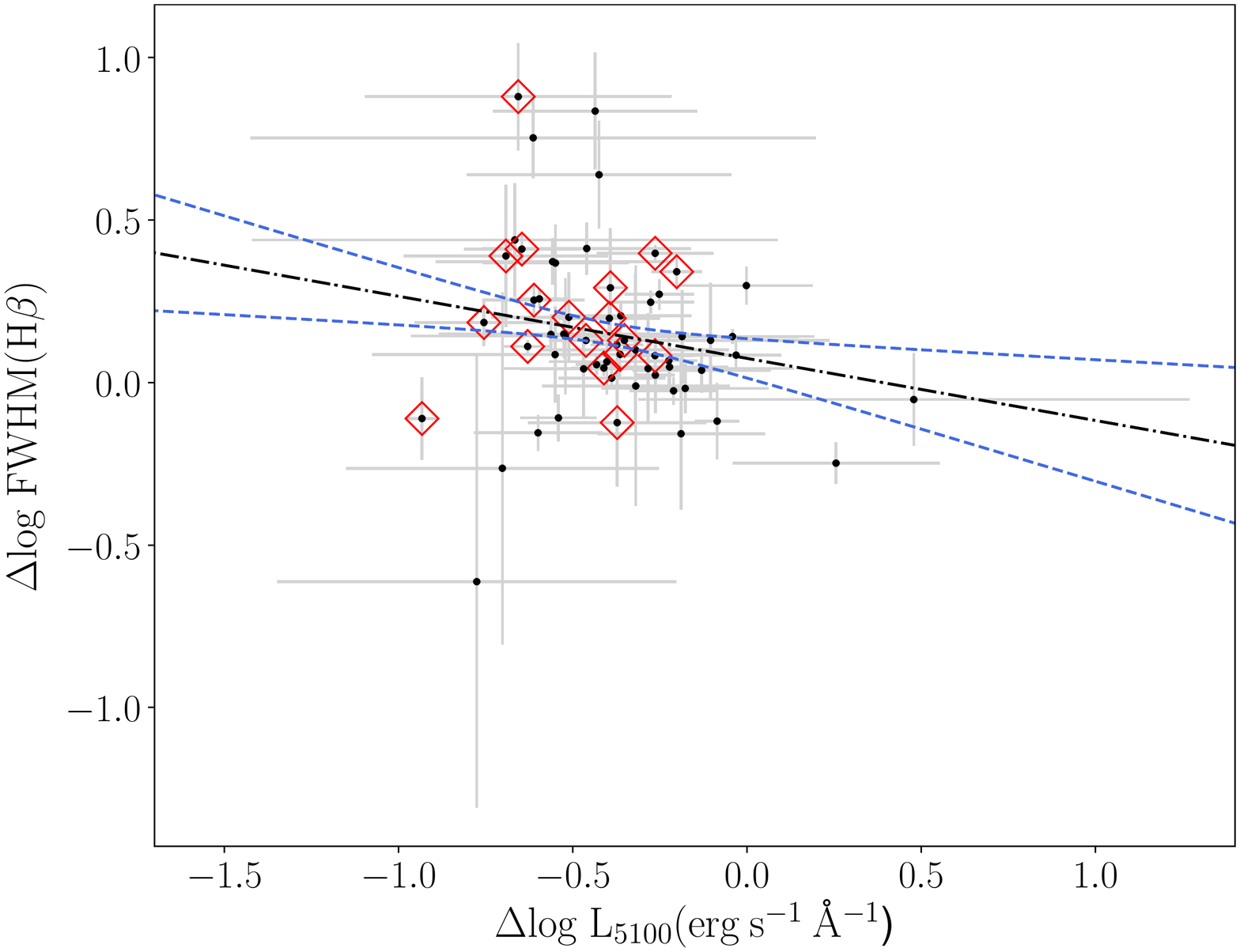}
\includegraphics[angle=90,scale=.3]{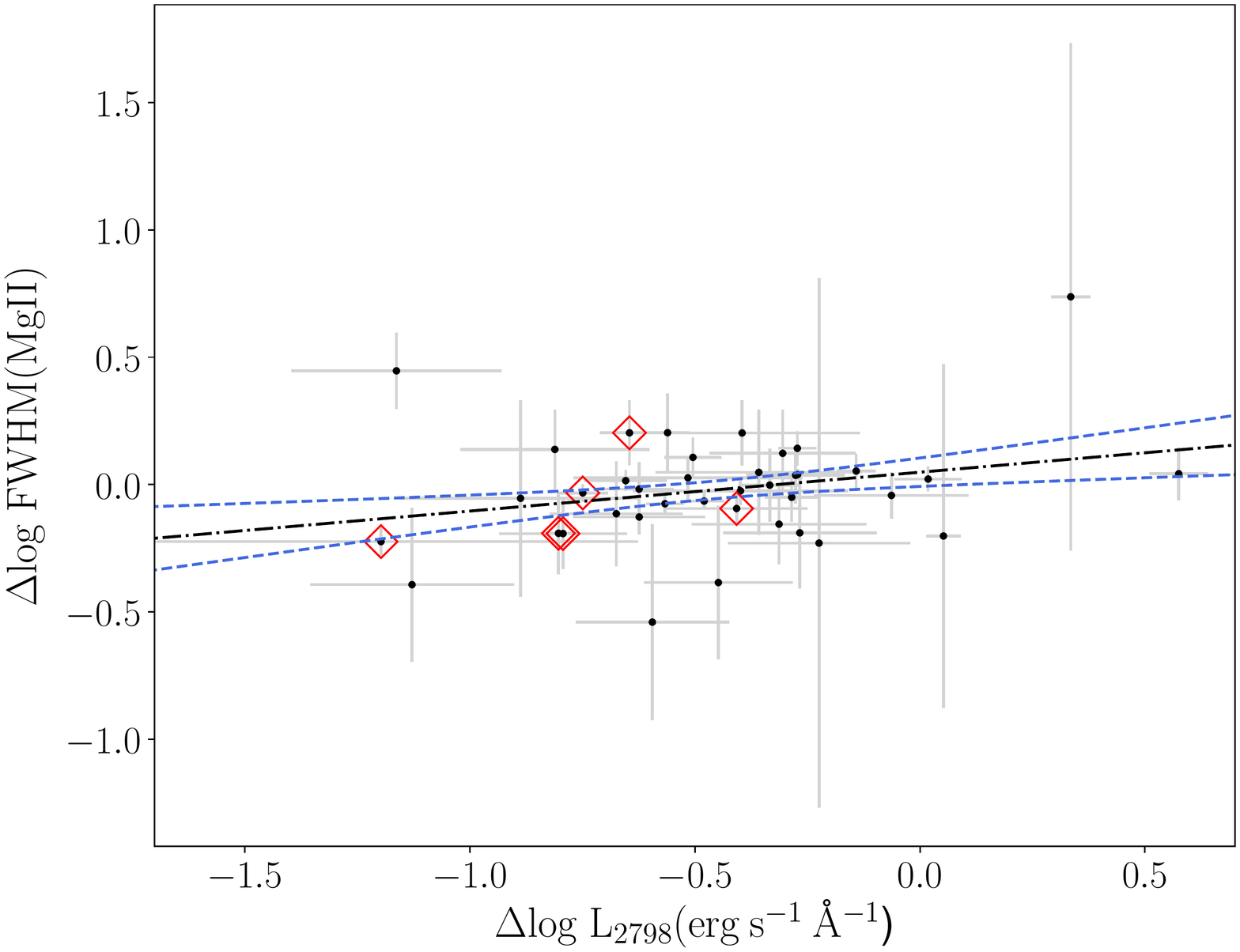}
\caption{
{\em LEFT:}
Change in FWHM of the broad \Hb\, emission line vs. luminosity change for all analyzed CLQ candidates, calculated between the bright and dim states designated in Table\,\ref{t:clqcans}.  Both axes use the
best-fit power-law continuum value at rest 5100\AA\, for each quasar and spectral epoch.  The CLQs, defined as having broad H$\beta$ change of at least $N_{\sigma}({\rm H}\beta)>3$, have points over-plotted as red diamonds.  
The trend for \Hb\ has a best-fit bivariate slope of $-0.19\pm 0.13$, shown as a black dot-dash line, with blue dashed lines indicating the 1-$\sigma$ confidence interval on the fit.
{\em RIGHT:} Same quantities for broad \MgII\, emission line and best-fit power-law continuum value at 2798\AA\.\, The trend for \MgII\ has a best-fit bivariate slope of $0.15\pm 0.09$, shown as a black dot-dash line, with blue dashed lines indicating the 1-$\sigma$ confidence interval on the fit.
\label{f:breathing}}
\end{figure}

In Figure\,\ref{f:baldwin}, we show the best-fit broad line equivalent width W$_{\lambda}$ vs. the underlying continuum luminosity for both \Hb\ and \MgII .\footnote{Model fits in the \MgII, region do not include a host galaxy, and are performed with PyQSOFit, but separately from the \Hb\ region fits.}   Treating every spectrum as if it were an individual QSO, there is a weak, possibly positive trend in the overall \Hb\ ensemble Baldwin effect, which has been noted before for Balmer lines in QSO samples \citep{Rakic2017, Greene2005}.  
The linear equation in this sample of spectra for the \Hb\ ensemble Baldwin Effect is 
${\rm log}\,W_{\lambda}(\Hb)= 0.29\pm 0.06\,{\rm log}(L_{5100}) - 10.9 \pm 2.5$.

This is expected, since broad \Hb\ emission reacts (albeit with a lag) directly and almost linearly to continuum changes.  The \Hb\ plot shows extreme scatter, largely because the primary selection for our CLQ candidate sample is to have extreme variability in continuum and broad \Hb\ emission. 
Expected sources of eBeff scatter in quasar samples can likely be attributed to differences between quasars in their inclination angles, BLR and absorber geometries, SMBH masses and accretion rates.  In contrast, for the intrinsic Baldwin effect (iBeff), only the last factor changes substantially in any given QSO.  Yet here, the \Hb\ intrinsic slopes, shown in Figure\,\ref{f:baldwin} by dashed lines between our designated dim and bright states (see Table\,\ref{t:clqcans}), seem to show at best a weak trend, and many of the lines are nearly vertical.  This is again likely due to our \emph{a priori} selection of extremely variable \Hb\ strength.
Indeed, the average slope of the \Hb\ iBEff is much steeper, at  $2.05 \pm 0.07$.

The line best fitting the ensemble \MgII\ Baldwin effect in our full candidate CLQ sample is: 
${\rm log}\,W_{\lambda}(\MgII)= -0.32\pm 0.05 {\rm log}(L_{2798}) + 16.0 \pm 2.4$.
In contrast, the average slope best fitting the intrinsic Baldwin effect is much steeper, at $-0.828 \pm 0.001$.
Steeper slopes for the iBeff compared to the eBeff are well-known, and indicate a subdued response to continuum changes.  This is especially true, and clearly visible in Figure\,\ref{f:baldwin} for \MgII.

\section{Black Hole Masses and Eddington Ratios}
\label{s:redd}

\citet{MacLeod19} and \citet{Rumbaugh18} have argued  that the EVQ and (even more so) CLQ populations have lower Eddington ratios $\Lbol/\LEdd$ than do normal quasars.  However, since their CLQ samples were selected starting from criteria of photometric variability, it is interesting to compare here with our primarily\footnote{Table\,\ref{t:clqcans} shows that about a third of our candidates were FES-HYPQSO targets, which are selected for strong photometric variability.} spectroscopically-selected sample.

\begin{figure}[h!]
\centering
\includegraphics[angle=90,scale=0.5]{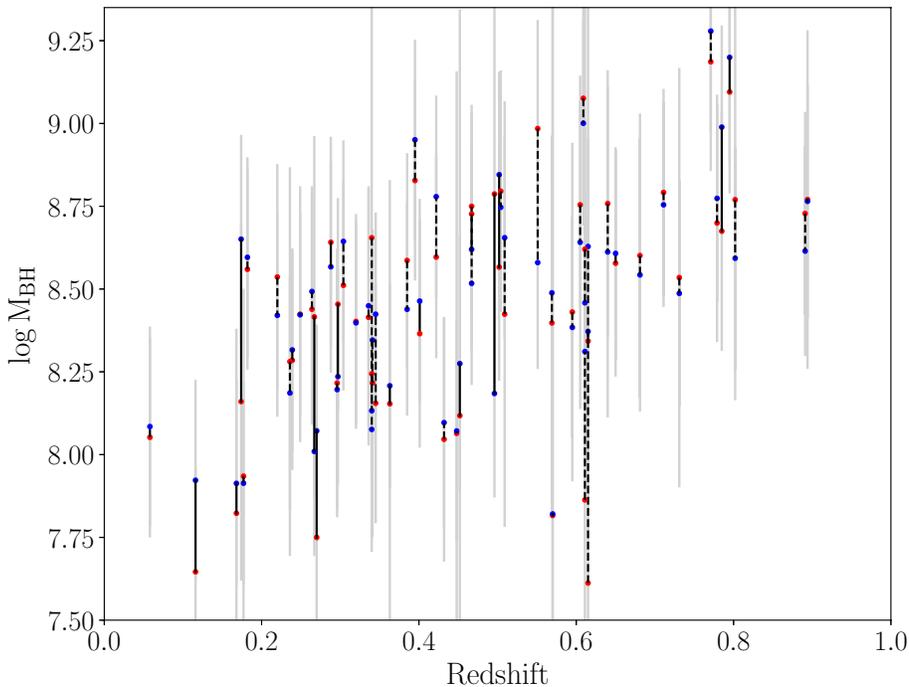}
\caption{Black hole mass estimates using the \Hb\ emission line vs. redshift for our CLQ candidate sample.  For each QSO, the bright epoch (red dot) is connected to the dim epoch (blue dot) at the QSO redshift.  Bona fide CLQs are connected by solid black lines, others by dashed lines. FWHM(\Hb) tends to decrease as luminosity increases (the 'breathing' effect; see Figure\,\ref{f:breathing}) in such a way that
the \Mbh\ estimate does not change significantly; propagated uncertainties on individual \Mbh\, estimates are shown as transparent grey bars, and normally exceed the shift in best-fit values between epochs.  The apparent correlation between log\Mbh\ and redshift is well-known, and primarily due to selection effects in the parent QSO sample; the SDSS magnitude limit enforces a luminosity-redshift correlation, and despite a range of accretion rates, the luminosity-\Mbh\ correlation thereby creates a secondary \Mbh - redshift correlation.
\label{f:mbh_z}}
\end{figure}

The Eddington luminosity is $\LEdd=1.26\times 10^{38}\Mbh$ in erg/s, with \Mbh\ in solar units.  
For our TDSS sample, we derive \Mbh\, separately from our own PyQSOFit model fits to broad \Hb\ in all spectra.  Assuming that the BLR is virialized, we use 
$$\Mbh = f_{BLR} \frac{R_{BLR}\,\Delta V^2}{G} $$
where $ f_{BLR}$ is a virial factor to
characterize the kinematics, geometry, and inclination of the BLR
clouds, $R_{BLR}$ is the "characteristic radius" of the BLR, 
$\Delta V$ is the velocity of the BLR clouds, usually traced
by the FWHM or the line dispersion
$\sigma_{\Hb}$ of the broad \Hb\ line, and $G$ is the gravitational constant.  Specifically, we use FWHM from the
PyQSOFit output {\texttt Hb\_whole\_br\_fwhm}. 
Errors in single-epoch virial estimates are larger than for the more robust reverberation mapping measurements \citep{peterson93,shen16,grier17}, but can be reduced by taking account of the mass accretion rate, as in \citet{Yu2020}.  Their best fit for the virial factor,
which incorporates the affects of BLR geometry, orientation and kinematics, is 
$$ {\rm log}\,f_{BLR} = -1.1\, {\rm log\,[FWHM}(\Hb)/2000] + 0.48$$.
For the characteristic radius of the BLR, we use their best fit
   $$\rm{log}\,R_{BLR} = 0.42 \,\rm{log}\,L_{44} - 0.28 R_{Fe} + 1.53$$
where $L_{44}$ is our $L_{5100}$ continuum fit values from Table\,\ref{t:clqcans} in units of $10^{44}$\,erg/s, 
$R_{Fe}$ is the ratio of iron to \Hb\ flux (taken from PyQSOFit outputs {\texttt Hb\_whole\_br\_area} and {\texttt fe\_op}),  
and $R_{BLR}$ is in light-days.  The resulting \Mbh\ estimates are shown plotted against redshift in Figure\,\ref{f:mbh_z} for the designated bright and dim state for each QSO in our sample  (red and blue points, respectively), with those values connected by a dashed line.  The apparent overall correlation  between log\Mbh\ and redshift is due to a combination of the SDSS flux limit (which due to volume sampled and the quasar luminosity function induces a correlation between luminosity and redshift) and the known correlation between quasar luminosity and \Mbh.  
The sometimes significant differences in mass estimates for a single quasar glaringly highlight the uncertainties in single-epoch virial mass estimates, expected to be particularly severe in a CLQ sample defined by large changes in both luminosity and broad \Hb\ line emission.  However, the epochal differences in virial mass estimates do not generally exceed the individual uncertainties.

\begin{figure}[h!]
\centering
\includegraphics[scale=0.9]{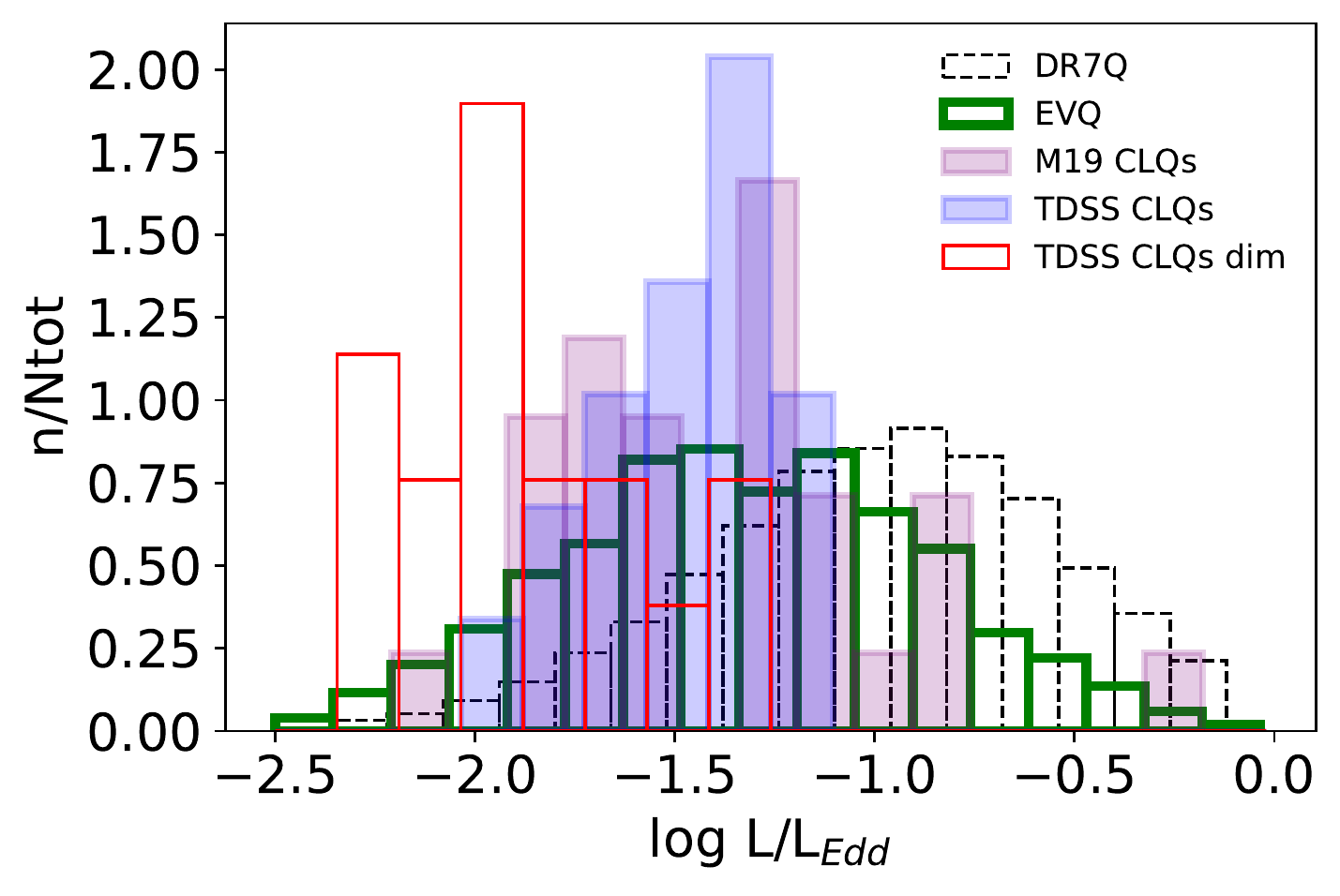}
\includegraphics[scale=0.9]{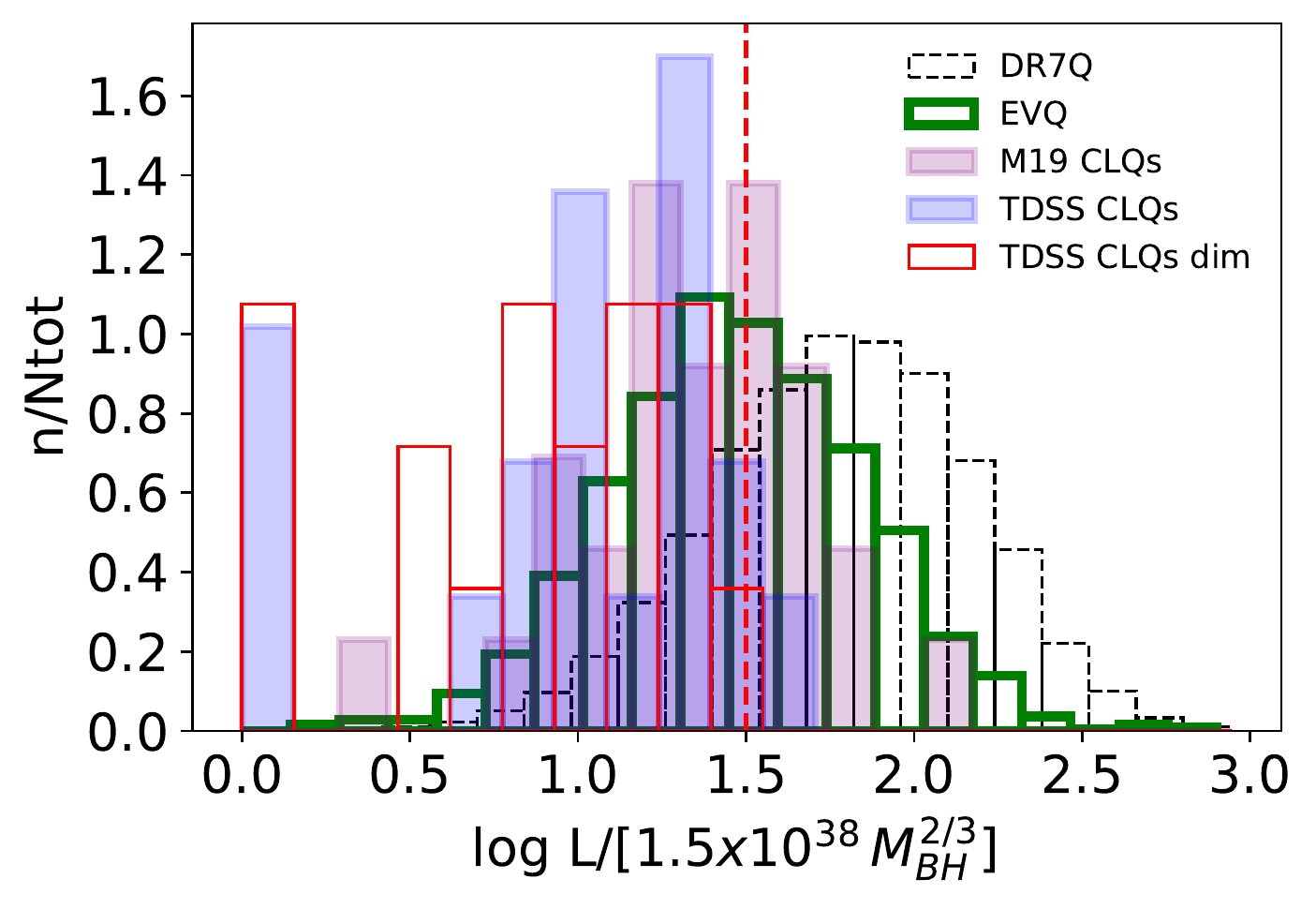}
\caption{
{\em TOP:}
The distribution of Eddington ratios shown as a histogram for the SDSS DR7 quasar sample (\citealt{shen11}; restricted to $z<0.9$; dotted black open bars), $z<0.83$ extremely variable quasars in SDSS/PS1 (EVQs; \citealt{Rumbaugh18}; green open bars), DR7 CLQs from \citet{MacLeod19} (purple filled bars), and our TDSS CLQ sample (blue filled bars). We also show the TDSS CLQ sample dim state Eddington ratios (red open bars).  These latter values are indicative only; they have large errors, because of their very weak,  broad \Hb\ lines.
{\em BOTTOM:}
Histogram of the quantity 
log\,$L / [1.5\times 10^{38}\,M_{BH}^{2/3}]$, where \Mbh\ is in solar masses, for the same samples shown at left.  
The critical parameter $L/M_{BH}^{2/3}$ in the disk-wind model of \citet{Elitzur2014}  predicts when a BLR can form. In these units (similar to \citealt{MacLeod19}), the critical value above which BELs should be observable is shown with a vertical red dashed line.}
\label{f:redd}
\end{figure}

To calculate bolometric luminosity, we use $\Lbol = 8.1 \lambda\,L_{\textrm{ 5100\AA}}$ from \citet{Runnoe12}.  
Figure\,\ref{f:redd} shows the distribution of Eddington ratios as log\,$\REdd$ for several different samples.  
Relative to most previous CLQ samples initially selected explicitly on large amplitude variability (e.g.,\citealt{MacLeod16,MacLeod19}), our own TDSS sample can be thought of as primarily spectroscopically-selected;  candidates are identified by visual inspection of multi-epoch spectra.  While the initial TDSS selection for spectroscopic targeting was indeed by variability, the selection thresholds were at levels associated with typical quasar ($\sim 0.2$mag; \citealt{Morganson15}) rather than more extreme ($\sim 1$\,mag) variability.  However, as demonstrated in Table\,\ref{t:TDSStypes}, the TDSS repeat spectroscopy programs provide the good portion of our CLQ candidates, especially the Disk Emitter (DE) and hypervariable QSO (HypQSO) FES subpgrograms.  The latter objects generally required a variability measure $\gax\,0.5$ mag (\citealt{MacLeod18}, Fig 5.).

Eddington ratio distributions for several quasar samples in Figure\,\ref{f:redd} are shown as histograms, normalized to have the same total area.  To  similar redshifts as our CLQ sample ($z\leq 0.9$), the general quasar population shows a broad distribution peaking at log\,$\REdd \sim -0.8$ or about 10 - 15\% of $\LEdd$.  Extremely variable ($|\Delta g|>1$) quasars from \citet{Rumbaugh18} show an even broader distribution, but centered at lower \REdd\ values near about 4\%.
The bright state \REdd\ values for the sample of (29) CLQs from \citet{MacLeod19} share similar range and mean as the EVQs, and to our own TDSS CLQ sample, all centered  near Eddington ratios of a few percent, about twice as low as \REdd\, for the broader quasar distribution. This may be evidence that the accretion rate is more variable at lower \REdd, yielding stronger variability and occasional darkening of the BLR.  However, it may also be that the FWHM of broad \Hb\ emission, used to derive \Mbh , is typically larger in EVQs and CLQs for other reasons, thereby yielding lower estimates of \REdd.  Indeed, \citet{Ren2021} find that EVQs show a subtle excess in the very broad line component compared with control samples.  
Strong turbulence in the inner accretion disc may be the reason for the continuum variability (e.g., \citealt{Kelly2009, Cai2020}) and may also launch more gas into the inner BLR, where the broadest line component is formed. 
We also plot in Figure\,\ref{f:redd} histograms of these same samples, but mapped to the disk wind parameter of \citet{Elitzur2006} and \citet{Elitzur2014} that determines whether or not a BLR can form. The red
vertical dashed line indicates the predicted critical value above which BELs
should be observable in this model. All the dim state CLQs lie below
it, as the model predicts.

\section{Discussion}\label{s:discussion}

\subsection{Diversity}\label{s:diversity}

The CLQ phenomenon, though simply parameterized here for practical purposes with our \Nsigma$>3$ criterion for broad \Hb, is clearly quite diverse. We note among our sample of 19 CLQs a broad range of changes in restframe optical/UV color, as well as may different timescales and magnitudes of continuum variability. 

There may be subclasses of CLQs caused by different types of activity, different spatial or dynamical configurations.  On the other hand, the apparent variety may also be due to temporal sampling i.e., attributable to observational effects such as where the spectroscopic epochs happen to fall on the light curve. Dense photometric and spectroscopic monitoring of CLQs might reveal a pattern of consistent phases within episodes of dimming or brightening that span the features observed here. 

As noted above, the most common type of spectroscopic variability involves a strong decrease in the restframe UV continuum (the common bluer-when-brighter phenomenon).  
However, there are some CLQs that show strong \Hb\ changes but at most very modest color changes, such as J024508.67+003710.68, J11329.68+531338.78,  J113706.93+481943.68, and J231625.39-002225.50.   Why do these objects not follow the bluer-when-brighter norm? One possibility is that the BLR, a wind, or co-located material itself blocks our sightline to the inner, hotter regions of continuum emission. This might be tested with spectropolarimetry of such objects, especially in comparison to those with more typical continuum changes.

Also, among the confirmed CLQs, we note several different behaviors with regard to the emission lines.
\begin{enumerate}
    \item  All the broad Balmer lines weaken substantially and become narrow, e.g., J021359.79+004226.81, J105513.88+242553.69, J143455.30+572345.10.
    \item Balmer lines other than \Ha\ narrow significantly, e.g., J002311.06+003517.53, J105058.42+241351.18. 
    \item  Broad \Hb\ narrows, but not as dramatically as in other cases, e.g., J224113.54-012108.84.
    \item Some CLQs have double-peaked higher order Balmer lines, e.g., J021259.59-003029.43, J021359.79+004226.81, J105325.40+302419.34, and J135415.54+515925.77.
\end{enumerate}

For the first three cases, we may consider that the drop in continuum flux takes time to propagate outward, with the primary \Ha -emitting region slower to respond than the \Hb\ region.  The response time of \MgII\ is likely even longer than for \Ha\ \citet{Roig14,Guo2019}.  Consistent with this interpretation, case (1) CLQs all have both strong steady dimming in the light curve, and a long time baseline between the bright and dim state spectra. 

A promising interpretation is that an outflowing hydromagnetic disc wind driven by UV and X-ray radiation carries BLR clouds with it, culminating near the radius of the dusty torus (e.g., \citealt{Everett2007, Kollatschny2013}).  The torus and the BLR begin to fade towards lower bolometric luminosities, i.e. low accretion rates \citep{Elitzur2006}, as described further below in \S\,\ref{s:classification}. 

  \subsection{Classification}\label{s:classification}

We find many more turn-off than turn-on CLQs, because we start from a sample identified via their broad emission lines as QSOs by the SDSS pipeline.  However, given the relatively short timescales seen for state transitions in CLQs, the intrinsic rate of turn-on and turn-off activity must be nearly equal, or the space density of broad line AGN would be changing detectably within a few decades.  \citet{Runco2016} obtained repeat spectroscopy within epochs separated by 3 to 9 years (median 6 years) for
a sample of 102 nearby (0.02$\leq z \leq$0.1) Seyfert galaxies with broad \Hb\ and SMBH mass estimates log\Mbh$>7$, and found that 38\% had broad line variability significant enough to change their Seyfert type (i.e., between 1, 1.2, 1.5, 1.8 1.9, or 2).  Twenty-three trended towards weaker broad lines, while seventeen towards stronger broad lines. This slight imbalance is again likely due to their selection criterion requiring initial broad \Hb\ emission.  We note the type changes for individual CLQs in the Appendix.

Most quasars in our candidate sample retain some broad \Ha\ even in the dim state. There are 26 QSOs within our candidate CLQ sample that have \Ha\ within the SDSS spectral range.  We find just 2 of the 19 that seem to lose all broad \Ha\ emission - J013435.90+022839.85 at MJD 56899 and J163620.38+475838.36 at MJD 58257.  The former has \Nsigma$=2.8$, so does not quite qualify as a CLQ by our definition, and has comparatively narrow \Ha\ even in its bright state.

The distinction between Type~1 (broad emission line) and Type~2 (narrow emission line) AGN has mostly been attributed to obscuration of the inner broad line region (BLR) in Type~2s.  The so-called ``unification paradigm'' \citep{antonucci93, urry95, audibert17} holds that the distinction depends on whether our sightline to the BLR is obscured by a dusty molecular torus or perhaps a dusty wind. However, \citet{pen84}, noted that NGC\,4151 and 3C390.3 had transitioned from Type~1 Seyfert (Sy1) to Type~2 (Sy2) over about a decade, and suggested that Sy2s may simply be accreting at low rates, and could potentially return to become broad line Sy1s at higher accretion rates.  Indeed, there are some relatively rare objects deemed true, unobscured, or ``naked'' Sy2s that show no broad emission lines but also no X-ray evidence for absorption or obscuration, even at normal or high Eddington ratios \citep{miniutti13} or at quasar luminosity levels \citep{gallo13,li15}. 

\citet{Hutsemekers19} performed spectropolarimetry of 13 CLQs, of which 7 were in a dim state (narrow line Type 1.9 -- 2) state, and found polarization under 1\% for all but one.  Two of the objects in our sample, 
J002311.06+003517.5 and J100220.18+450927.30 were among those studied.  The low polarization weighs against the hypothesis that dimming and loss of broad line emission in CLQs could be due to obscuration, as does the long timescale expected for dust clouds to move across the sightline to the BLR  ($\gax20$\,yrs; \citealt{MacLeod16,Nenkova2008}).  Especially interesting is their hypothesis that broad line emission seen in polarized spectra of Type~2 AGN may be reflected light from previous bright states, rather than evidence for current obscuration of the BLR.  If true, then in the absence of further brightening episodes, a polarized BLR in a Type~2 AGN should continue to fade with time.

\citet{Elitzur2014} proposed that the intrinsic spectral sequence 1 $\rightarrow$ 1.2/1.5 $\rightarrow$ 1.8/1.9 →$\rightarrow$ 2 is a sequence reflecting decreasing accretion rate onto the SMBH. 
In their disk-wind model for the BLR, a Type\,1 AGN devolves to a “true” Type\,2 AGN below a critical value of L$/M_{BH}^{2/3}$, a parameter similar to the Eddington ratio. CLQs corroborate that the absence of broad emission lines in AGN spectra may be related to a low accretion rate, rather than to obscuration.  The balance of these factors in determining the observed Type~1 vs Type~2 populations remains unclear.\footnote{Another classification paradigm - the division of quasars into ``radio-quiet" and ``radio-loud" has been similarly challenged by the accrual of long term temporal variability data.  ``Radio loudness"  also turns out to be a property of quasars that can change dramatically on human timescales \citep{nyland20}.}

The intrinsic fraction of quasars with Type~2 spectra is controversial, between about a third \citep{Treister2008} and a half \citep{Reyes2008,Lawrence2010} and it may decrease with bolometric luminosity 
\citep{Treister2008,Glikman2018}. This has historically been seen as a decrease in obscuration,
and even part of an evolution as winds from the nucleus clear away dust. However, now that it is clear that 
even luminous quasars can change type (lose or regain their broad line emission) on timescales of months to years,
the relative importance of obscuration versus accretion rate (and subsequent ionization of the BLR) for determining type puts decades of interpretation into question, and also opens new avenues for understanding the life cycles of quasars and the concomitant patterns of growth of their SMBH by accretion.


\subsection{Future Work}
\label{s:future}

As could be said for many astrophysical experiments in the time domain, the ideal campaign to understand the CLQ phenomenon would consist of rapid cadence, multi-epoch, multi-band observations of a large sample of quasars, observed over a long timespan.  Even well short of the ideal, spectroscopy in the time domain is already proving to have profound effects on our interpretation of quasar classes.  Concerted time domain spectroscopic surveys for CLQs have certain advantages over surveys where followup spectroscopy is triggered by photometric or color variability.  Spectroscopy allows a focus directly on the broad line region emission, and so can even detect CLQs where the observed continuum does not change substantially (e.g., in cases where the BLR may be subject to a different ionizing continuum than evident along the observer's sightline) or is mired in host galaxy emission. Repeat spectroscopy surveys, also can find CLQs where the AGN luminosity never dominates the host galaxy emission.

Many more spectroscopically-identified CLQs will be discovered during SDSS-V, which is just beginning, and is expected to last about 5 years.  The Black Hole Mapper program of SDSS-V \citep{Kollmeier2017} includes a significant program of repeat quasar spectroscopy dubbed `AQMES' for All Quasar Multi-epoch Spectroscopy. The AQMES Medium program targets about 2,000 known QSOs to a given magnitude limit (of $i =19.1$) for about 12 epochs of spectroscopy over about 300\,\degsq\, of sky.  AQMES Wide will similarly target about 20,000 known quasars to a similar magnitude limit, but for only a few (1 - 3) epochs over about 3,000\,\degsq\, of sky. 
Combining SDSS-V quasar spectra with prior SDSS archival spectra of the same objects will afford a broad survey of continuum and emission line changes spanning the full parameter space of redshift, luminosity, SMBH mass, Eddington ratios and rest-frame timescales for a variety of emission lines. Many more CLQs will be discovered but, perhaps more importantly, they will be placed in the broader context of quasar spectroscopic variability.  Furthermore, better statistics along with availability of detailed light curves across the sky should allow for enhanced constraints on the distribution of these parameters, to determine whether CLQs represent a uniquely variable species, the extreme tail of a smooth distribution of variability, or simply normal quasars in a highly variable phase.  The first interpretation might require e.g., evidence for distinct properties apart from variability itself, such as distinct host galaxy features, or star formation histories. The power to discriminate between any of these interpretations benefits from increased length and cadence of photometric and spectroscopic monitoring.

\begin{figure}[h!]
\centering
\includegraphics[scale=0.7]{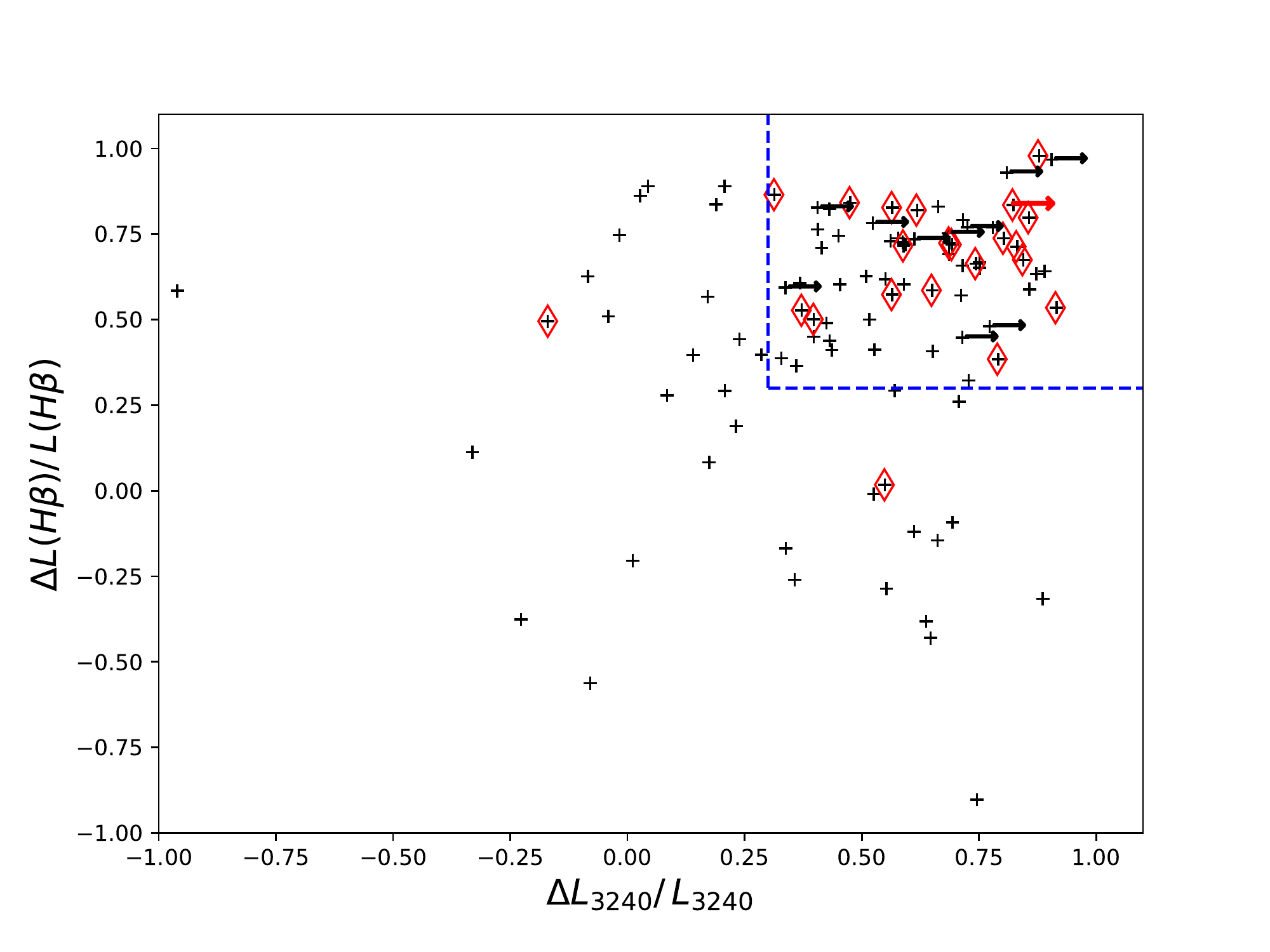}
\caption{Fractional change of broad \Hb\ luminosity plotted against the fractional change in the best-fit power-law. Points show all spectral epochs contrasted against the designated dim epoch for each CLQ candidate.  Contrasted spectral pairs with (\Nsigma\ $>3$) are shown in red.  The blue dashed lines indicate a 30\% fractional change in each axis, which if measured at adequate S/N could serve as reasonable {\emph intrinsic} criteria to declare a quasar to be a bona fide CLQ, for purposes of comparison across different multi-epoch spectral samples or studies. Two CLQs identified only by \Nsigma\ $>3$ in this paper fall outside the box, and are discussed in \S\,\ref{s:future}.}
\label{f:intrCrit}
\end{figure}

Our adoption of the \Nsigma$>3$ criterion has been to be conservative in the presence of large samples of variable S/N spectroscopic data, and to create a numerical criterion that might be adopted by the community to facilitate easy aggregation and/or comparison of samples.  However, if for example we had spectra of S/N$=100$, a 3$\sigma$ change could admittedly represent quite a minor change in broad \Hb\ luminosity. Ideally, criteria to declare identification of a bona fide CLQ should be based on intrinsic properties, measured with adequate significance.
Based on our Fig\,\ref{f:delta_hb} (and similarly, MacLeod Fig 3), among CLQ candidate samples, there does not appear to be a clear lower limit to $\Delta\,L(3240)$ and $\Delta\,L(\Hb)$ that could provide intrinsic criteria. Rather, criteria involving fractional changes might make the most sense.  Figure\,\ref{f:intrCrit} shows the {\emph fractional} change in both $\Delta\,L(3240)$ from the power-law fit and $\Delta\,L(\Hb)$ for our candidate CLQ sample, including all spectral epochs contrasted against the designated dim epoch for each CLQ candidate.  Contrasted spectral pairs with \Nsigma$>3$ are shown in red.  The blue dashed lines indicate 30\% change in each axis.  A sample of typical quasars with multi-epoch spectral analysis would be expected to cluster near the origin of this plot.  Since the full sample shown here was selected to be candidate CLQs by visual inspection, the majority of fractional changes between spectral pairs are found in the upper right corner, where they may meet our suggested intrinsic criteria for CLQ if the S/N of the spectra in the region of \Hb\ is also sufficient.  

Nearly all the CLQs designated as such only by the \Nsigma\ criterion in this paper also meet the other two criteria (fractional change in both $L(3240)$ and broad \Hb\ $>30\%$). One exception (left of the blue dashed box) is J224113.54-012108.84, comparing MJDs 58367 - 55824.  As we remark in the Appendix, this QSO has unusual continuum properties and perhaps a spectroscopic calibration issue.  Another exception  is J100302.62+193251.28, comparing MJDs 57817 - 53762, which appears below the blue dashed box, so a fractional change in broad \Hb\ than prescribed, although we find \Nsigma\ $=4.31$.  We note that both of these exceptions are turn-on CLQs, which remain underrepresented in the literature and deserve further study.

The spectral energy distributions (SEDs) of quasars are known to change significantly with Eddington ratio. More luminous objects have a softer ionizing continuum. The impact of SED changes on line-forming efficiency and even the structure of the BLR is well-documented \citep{korista1988,Zheng1993,Green1998}.
 Clearly, the extremes of accretion rate occurring in CLQs may also correspond to significant changes in SED.  Optical, UV and X-ray monitoring of CLQs during transitions is key to understanding these changes, and their propagation from the central engine to the BLR.  A few nearby Seyferts have been monitored closely throughout state changes e.g., \citet{Krumpe2017} for Mkn1018 ($z=0.043$) and \citet{Trakht2019}, \citet{Ricci2020} for 1ES\,1927+65 ($z=0.019$), but these have 10-100 times lower SMBH masses, and high Eddington ratios compared to our CLQ sample. Some efforts to chart SED changes in CLQs including X-rays have been made, but most have relied on asynchronous observations (e.g., \citealt{Ruan2019}).  A \emph{Chandra} program (P. Green, P.I.) is in progress, catching CLQs in optical and X-ray in both bright and dim states.  We may expect many more CLQs to afford real-time multiwavelength monitoring in coming years via eROSITA \citep{Predehl2021} in X-rays and the Rubin Observatory/LSST in the optical \citep{Ivezic2019}.

\appendix

Here, we discuss the individual CLQ light curves and spectra shown in Figure\,\ref{f:lcspecs}. All the CLQs shown here are turn-offs, dimming with time, unless noted otherwise below.  We include a quantitative estimate of AGN spectral type, an adaptation of the Lick classification scheme \citep{Osterbrock1981}, offered by \citet{Winkler1992}  using the flux ratio $R =$ flux(total \Hb )/flux(total [OIII]$\lambda$5007) to define Type-1, 1.2, 1.5, and 1.8 with R values $R > 5, 2.0 < R < 5.0, 0.333 < R < 2.0, R < 0.333$, respectively. For Type 1, permitted lines are all broad.
Type 1.5 has broad components about equal in flux to narrow. Type-1.9 has no detectable \Hb\ broad emission line, while Type-2 has neither \Hb\ nor \Ha\ broad emission lines detectable i.e., permitted and forbidden lines have similar widths near $\sim 1000$\kms .  Types 1.2 and 1.8 correspond  simply to finer divisions in relative strength of broad and narrow lines. Note that \Ha\ is generally not within the SDSS spectral range for $z>0.4$.


 \emph{002311.06+003517.53}:
This CLQ presents several strong luminosity changes over a nearly
20-year time span in the observed frame.  Brightening of
1.5 mag in $g$ band occurs within a few months.  Later, steady dimming
of 1.5 mag occurs over about a 3 year timeframe.  Six spectral epochs 
confirm strong accompanying changes in the continuum and broad line
emission. From bright to dim state, the spectrum effectively changes from a Type 1.2 to 1.5. 

 \emph{020514.77-045639.74}:
The observed photometric epochs do not coincide well with the dimmest
spectral epoch, but dimming by nearly a magnitude in $g$ band is
evident in the lightcurves. The continuum shape does not change as
strongly as most CLQs, and \Ha\ weakens much less than \Hb.
From bright to dim state, the spectrum effectively changes from a Type 1.0 to 1.5. 

 \emph{021259.59-003029.43}:
This quasar was targeted by TDSS both in the HYPQSO (highly variable)
and DE (disk emitter) samples.  It dimmed by about 1.2 mag in the $r$
band, and likely more than that in $g$ band.  The two latter, dim
state spectra are essentially identical. There appears to have been a
re-brightening by nearly a magnitude after the second dim state
spectrum, on a time scale of about 6 months or less.  In all the
spectra, several of the higher order Balmer lines are double-peaked.
In the bright state spectrum, there is a pronounced red wing to the broad
\Hb\ emission. This quasar's classification remains at Type~1.5.

 \emph{021359.79+004226.81}:
This quasar dimmed in $g$ band by about 1.5 mag, and about half that
in $r$ band.  Again, all the higher order Balmer lines are
double-peaked. In the intermediate state spectrum, the \Hb\ broad
emission has broad shoulders which disappear entirely in the dim state
spectrum.  From bright to dim state, the spectrum effectively changes from a Type 1.5 to 1.8. 

 \emph{024508.67+003710.68}:
Here the $g$ band dims by about 0.5\,mag, while the $r$ band continuum
remains steady.  There are likely flux calibration issues with the
later epoch spectrum.  Nevertheless, it appears that essentially all
broad line emission disappears.  From bright to dim state, the spectrum effectively changes from a Type 1.2 to 1.9. 

 \emph{024932.01+002248.35}:
This quasar suffers dimming of about $\Delta g = 1.5$mag, whereas the
$r$ band continuum dims by about half that amount. The blue continuum
can be seen to almost entirely disappear in the spectra. Some broad
\Hb\ remains. Details of these spectra and model fitting are presented
separately in Figures \ref{f:024932model} and \ref{f:024392DimBright}.
From bright to dim state, the spectrum effectively changes from a Type 1.0 to 1.5. 

 \emph{100302.62+193251.28}:
Judging from the spectra, this CLQ appears to have turned on,
though it is difficult to gauge the change in magnitude,
since the later brighter state spectrum has no nearby phtometric
epoch. The strong \oiii\ lines in this object indicate a very bright narrow
line region, likely lluminated from past bright epochs. Broad
emission from \Hb\ declines markedly, yet \MgII\ remains strong.
From bright to dim state, the spectrum stays at a Type 1.5.

 \emph{105058.42+241351.18}:
Here, the $g$ band likely dims by about half a magnitude. While \Ha\
decreases slightly, all the higher order Balmer lines seem to
disappear.  From bright to dim state, the spectrum effectively changes from a Type 1.2 to 1.5. 

 \emph{105325.40+302419.34}:
The $g$ band dims monotonically by more than a magnitude over
about a dozen years. Broad \Hb\ nearly disappears; broad \Ha\
weakens significantly as well, but the final state may be Type 1.9.
From bright to dim state, the spectrum also effectively changes from a Type 1.2 to 1.5. 

 \emph{105513.88+242553.69}:
The $g$ band dims monotonically by about 2 magnitudes, with an
accompanying striking change in the blue spectroscopic continuum.
\Hb, H$\gamma$, and H$\delta$ all disappear. From bright to dim state, the spectrum effectively changes from a Type 1.0 to 1.9. It may end up as a Type 2 quasar, but at $z=0.496$, \Ha\ is not visible in the spectrum to make that determination.

 \emph{111329.68+531338.78}:
The $g$ band likely dimmed by about a magnitude between bright and dim
states, but spectral epochs fall within gaps in photometric coverage.
All the visible Balmer lines show a marked decrease in flux.
From bright to dim state, the spectrum effectively changes from a Type 1.0 to 1.2.

 \emph{113651.66+445016.48}:
In this turn-on CLQ, the $g$ band magnitude brightens by about
0.5\,mag with a strong upturn in the blue spectral continuum. Broad \Hb\ and \Hg\ strengthen considerably, as does FeII emission, but with little if any change in \Ha .
Changing from dim to bright state, the spectrum effectively changes from a Type 1.5 to 1.2.

 \emph{113706.93+481943.68}:
This CLQ seems to have dimmed steadily by about a magnitude in $g$
band over about a decade, during which broad \Hb\ emission
disappeared. From bright to dim state, the spectrum effectively changes from a Type 1.2 to 1.5.

 \emph{135415.54+515925.77}:
This CLQ seems to show large rapid variability even in the $r$ band
superimposed on a steady overall dimming.  The $g$ band dimmed very
significantly, by about 2 mag.  The higher order Balmer lines appear
to show unusual double narrow peaks in the dim state. The spectral classification remains at Type 1.5.

 \emph{143455.30+572345.10}:
This CLQ dimmed by a full 2 mag in $g$ band, showing a spectacular
disappearance of the BEL components, which began in the bright state
with strong red wings. From bright to dim state, the spectrum effectively changes from a Type 1.2 to 1.8.

 \emph{163620.38+475838.36}:
There is complete disappearance of broad \Hb\ emission in this CLQ,
while the continuum dimming is rather quite modest. Judging from the
photometric points, the early epoch  spectrum would likely have been
much more spectacular had it been obtained about a year earlier. 
From bright to dim state, the spectrum effectively changes from a Type 1.5 to 1.8.

 \emph{224113.54-012108.84}:
This is a turn-on CLQ that brightened by a full magnitude in $g$ band, as is clear in the $< 4000$\AA\ continuum. Four of the five spectra show \Nsigma\ $>3$ relative to the dim state. \Ha\ appears to maintain its broad emission throughout, and retains a Type 1 classification for all its spectral epochs.   The continuum changes are quite unusual, since the reddest spectrum MJD 55824, which we classify as the dim state, is actually brighter in the red than the "bright" bluer states. There may be some calibration issues with the SDSS spectra.

 \emph{231625.39-002225.50}:
For this CLQ, the continuum dimming between spectral epochs is quite
modest - likely no more the $\Delta g\sim 0.2$\,mag.  With
\Nsigma$=3.16$, these epochs barely pass our CLQ criterion, but broad
\Hb\ clearly diminishes significantly by visual inspection.
The spectrum remains a Type 1.5 throughout.

\emph{234623.42+010918.11}
This turn-on CLQ brightened in $g$ band by at leat 1.5\,mag, within
about a year in the observed frame, and remained bright. However, if
the flux calibrations are correct, there were relatively
rapid and significant changes in the spectral continuum blueward of about 4,000\AA in the six later epochs of spectroscopy.
The spectrum remains a Type 1.5 throughout.



\begin{acknowledgments} 
Many thanks to Qian Yang for comments and suggestions.

S.F.A., P.J.G., C.L.M, B.R. and J.J.R. are supported by the National Science
Foundation under Grants Nos. AST- 1715121, and AST-1715763.  C.L.M. and Q.Y. were partially supported for spectroscopic observations and analysis by the National Aeronautics and Space Administration through Chandra Award Numbers GO8-19089X and GO9-20086X, issued by the Chandra X-ray Center, which is operated by the Smithsonian Astrophysical Observatory for and on behalf of the National Aeronautics Space Administration under contract NAS8-03060. 

Photometric light curves used herein are based on observations obtained with the Samuel Oschin 48-inch Telescope at the Palomar Observatory as part of the Zwicky Transient Facility project. ZTF is supported by the National Science Foundation under Grant No. AST-1440341 and a collaboration including Caltech, IPAC, the Weizmann Institute for Science, the Oskar Klein Center at Stockholm University, the University of Maryland, the University of Washington, Deutsches Elektronen-Synchrotron and Humboldt University, Los Alamos National Laboratories, the TANGO Consortium of Taiwan, the University of Wisconsin at Milwaukee, and Lawrence Berkeley National Laboratories. Operations are conducted by COO, IPAC, and UW. 

Funding for the Sloan Digital Sky Survey IV has been provided by the Alfred P. Sloan Foundation, the U.S. Department of Energy Office of Science, and the Participating Institutions. 
SDSS-IV acknowledges support and resources from the Center for High Performance Computing  at the University of Utah. The SDSS website is www.sdss.org.
SDSS-IV is managed by the Astrophysical Research Consortium for the Participating Institutions of the SDSS Collaboration including the Brazilian Participation Group, the Carnegie Institution for Science, Carnegie Mellon University, Center for Astrophysics $|$ Harvard \& Smithsonian, the Chilean Participation Group, the French Participation Group, Instituto de Astrof\'isica de Canarias, The Johns Hopkins University, Kavli Institute for the Physics and Mathematics of the Universe (IPMU) / University of Tokyo, the Korean Participation Group, Lawrence Berkeley National Laboratory, Leibniz Institut f\"ur Astrophysik Potsdam (AIP),  Max-Planck-Institut f\"ur Astronomie (MPIA Heidelberg), Max-Planck-Institut f\"ur Astrophysik (MPA Garching), Max-Planck-Institut f\"ur Extraterrestrische Physik (MPE), National Astronomical Observatories of China, New Mexico State University, New York University, University of Notre Dame, Observat\'ario Nacional / MCTI, The Ohio State University, Pennsylvania State University, Shanghai Astronomical Observatory, United Kingdom Participation Group, Universidad Nacional Aut\'onoma de M\'exico, University of Arizona, University of Colorado Boulder, University of Oxford, University of Portsmouth, University of Utah, University of Virginia, University of Washington, University of Wisconsin, Vanderbilt University, and Yale University.

The authors acknowledge the use of the Catalog Archive Server Jobs System (CasJobs) service at \url{http://casjobs.sdss.org/CasJobs}, developed by the JHU/SDSS team.

This work uses data obtained with the 1.2-m Samuel Oschin Telescope at Palomar Observatory as part of the Palomar Transient Factory project, a scientific collaboration among the California Institute of Technology, Columbia University, Las Cumbres Observatory, the Lawrence Berkeley National Laboratory, the National Energy Research Scientific Computing Center, the University of Oxford, and the Weizmann Institute of Science.  The Zwicky Transient Facility is supported by the NSF under grant AST-1440341 and a collaboration including Caltech, IPAC, the Weizmann Institute for Science,the Oskar Klein Center at Stockholm University, the University of Maryland, the University of Washington, Deutsches Elektronen-Synchrotron and Humboldt University, Los Alamos National Laboratories, the TANGO Consortium of Taiwan, the University of Wisconsin at Milwaukee, and Lawrence Berkeley national Laboratories. Operations are conducted by COO, IPAC and UW. 

The Pan-STARRS1 Surveys (PS1) have been made possible through contributions of the Institute for Astronomy, the University of Hawaii, the Pan-STARRS Project Office, the Max-Planck Society and its participating institutes, the Max Planck Institute for Astronomy, Heidelberg and the Max Planck Institute for Extraterrestrial Physics, Garching, The Johns Hopkins University, Durham University, the University of Edinburgh, Queen's University Belfast, the Harvard-Smithsonian Center for Astrophysics, the Las Cumbres Observatory Global Telescope Network Incorporated, the National Central University of Taiwan, the Space Telescope Science Institute, the National Aeronautics and Space Administration under Grant No. NNX08AR22G issued through the Planetary Science Division of the NASA Science Mission Directorate, the National Science Foundation under Grant No. AST-1238877, the University of Maryland, and Eotvos Lorand University (ELTE).

This work made use of data products from the Catalina Sky Survey.
The CSS survey is funded by the National Aeronautics and Space
Administration under Grant No. NNG05GF22G issued through the Science
Mission Directorate Near-Earth Objects Observations Program.  The CRTS
survey is supported by the U.S.~National Science Foundation under
grants AST-0909182 and AST-1313422.
\end{acknowledgments}

\facilities{Sloan, ING:Herschel, Magellan:Baade (LDSS2 imaging spectrograph, Boller \& Chivens spectrograph), MMT (Blue Channel spectrograph, BinoSpec), PS1, PO:1.2m, SO:Schmidt, SO:1.5m, SO:1m} 

\software{Astropy \citep{astropy1, astropy2},  Matplotlib \citep{matplotlib}, Numpy \citep{numpy}, Scipy \citep{scipy}, TOPCAT \citep{topcat}}

\bibliography{refs}{}
\bibliographystyle{aasjournal}



\end{document}